\documentclass[journal,a4paper,10pt,rotate]{IEEEtran}

\usepackage{amsmath,amsthm,amssymb,amsfonts,upref,cite,epsf,color}
\usepackage{pst-all,pstricks,pst-node,pst-text,pst-3d}

\usepackage{pstricks-add}

\usepackage{multido}
\usepackage{graphicx,epsfig,rotating} 
\usepackage{psfrag,texdraw,bm}
\usepackage{nomencl}
\usepackage{amsmath}
\usepackage{amssymb}
\usepackage{xspace}
\usepackage{dsfont}
\usepackage{pifont}
\usepackage{listings,color}   
\usepackage[T1]{fontenc}
\usepackage{textcomp}
\usepackage{hyperref}

\def \expect[#1] {{\rm E} \left\{ #1 \right\} }
\def \twiddle[#1] {e^{-j \frac{2 \pi}{N}  #1 }}
\def \twiddleneg[#1] {e^{j \frac{2 \pi}{N}  #1 }}
\def \discreteRHS{\bar{R}_{X}}

\def\prefactor{( N \!+\! 1)}

\newcommand{\eq}{\,=\,}
\def\KK{K}
\def\MM{\mathbf{M}}

\newcommand{\be}{\begin{equation}}
\newcommand{\ee}{\end{equation}}
\newcommand{\ist}{\hspace*{.2mm}}
\newcommand{\rmv}{\hspace*{-.2mm}}

\def\lendelay{\Delta M}
\def\lendoppler{\Delta L }
\def\SS{S} 
\def\HH{h}

\allowdisplaybreaks
\renewcommand{\textfraction}{0.01}

\makeatother

\DeclareMathOperator*{\argmin}{argmin}

\allowdisplaybreaks

\begin{document}

\title{Compressive Spectral Estimation for Nonstationary Random Processes\thanks{Copyright (\copyright) 2013 IEEE. Personal use of this material is permitted. 
However, permission to use this material for any other purposes must be obtained from the IEEE by sending a request to pubs-permissions@ieee.org.}
\thanks{Alexander Jung, Georg Taub\"{o}ck, and Franz Hlawatsch are with the Institute of 
Telecommunications, Vienna University of Technology, A-1040 Vienna, Austria (e-mail: \{ajung,$\ist$gtauboec,$\ist$fhlawats\}@nt.tuwien.ac.at). 
This work was supported by the FWF under grant S10603 (Statistical Inference) within the National Research Network SISE
and by the WWTF under grant MA 07-004 (SPORTS). Parts of this work were previously presented at IEEE SSP 2009, Cardiff, Wales, UK,
Aug.--Sept.\ 2009.
}\vspace*{3mm}}

\author{Alexander Jung, Georg Taub\"{o}ck, {\it Member, IEEE}, 
and Franz Hlawatsch, {\it Fellow, IEEE}\\[3mm]
\vspace*{2mm}
 }
 

\maketitle

\vspace*{-5mm}

\begin{abstract}
Estimating the spectral characteristics of a nonstationary random process is an important but challenging
task, which can be facilitated by exploiting structural properties of the process.
In certain applications, the observed processes are underspread, i.e., their time and frequency correlations exhibit a reasonably fast decay, 
and approximately time-frequency sparse, i.e., a reasonably large percentage of the spectral values are small.
For this class of processes, we propose a compressive estimator of the discrete Rihaczek spectrum (RS).
This estimator combines a minimum variance unbiased estimator of the RS (which is a smoothed Rihaczek distribution using an
appropriately designed smoothing kernel) with a compressed sensing technique that exploits the approximate time-frequency sparsity.
As a result of the compression stage, the number of measurements required for good estimation performance can be significantly reduced.
The measurements are values of the ambiguity function of the observed signal at randomly chosen time and frequency lag positions.
We provide bounds on the mean-square estimation error of both the minimum variance unbiased RS estimator and the compressive RS estimator,
and we demonstrate the performance of the compressive estimator by means of simulation results.
The proposed compressive RS estimator can also be used for estimating other time-dependent spectra (e.g., the Wigner-Ville spectrum)
since for an underspread process most spectra are almost equal.
\end{abstract}


\section{{Introduction}}\label{sec.intro}

Estimating the spectral characteristics of a random process is an important task in many signal analysis and processing problems.
Conventional spectral estimation based on the \emph{power spectral density} is restricted to wide-sense stationary and, by extension,
wide-sense cyclostationary processes \cite{Kay88,stoi97}. However, in many applications---including speech and audio, communications, image processing, computer vision, biomedical engineering, and machine monitoring---the signals of interest cannot be well modeled
as wide-sense (cyclo)stationary processes. For example, in cognitive radio systems \cite{CognitiveRadio2005,SpectrumSharingRadios,Sahai09},
the receiver has to infer from the received signal the location of unoccupied frequency bands (``spectral holes'') that can be used for data transmission.
Here, modeling the received signal as a nonstationary process can be advantageous because it potentially
allows a faster estimation of time-varying changes in band occupation \cite{CognitiveRadio2005}.

For a general nonstationary process, a ``power spectral density'' that is nonnegative and extends all the essential properties of the conventional power spectral density
is not available \cite{loynes68,fla-book2,martin-wvs,fla-wigbook,MatzHla06}. Several different definitions of a ``time-dependent (or time-varying) power spectrum''
have been proposed in the literature, see \cite{Page52a,Levin64a,priestley65,loynes68,rihaczek68,mark-physpec,priestley71-physint,tjostheim76,melard78,Gra81,%
martin-wvs,melard89,fla-spectra,detka-TES,fla-wigbook,GM-evolut,fla-book2,GM-phd,CRC_tfrand,MatzHla06,MatzHlaChap08} and references therein.
However, it has been shown \cite{GM-phd,MatzHla06} that in the practically important case of nonstationary processes with fast decaying
time-frequency (TF) correlations---so-called {\em underspread}\/ processes
\cite{koz-corr,koz-hamburg,koz-diss,Koz97b,GM-phd,CRC_tfrand,MatzHlaChap08,MatzHla06}---all
major spectra yield effectively identical results, are (at least approximately) real-valued and nonnegative,
and satisfy several other desirable properties at least approximately. Thus, in the underspread case, the specific choice of
a spectrum is of secondary theoretical importance and can hence be guided by practical considerations such as computational complexity.

Once a specific definition of time-dependent spectrum has been adopted,
an important problem is the estimation of the spectrum from a single observed realization of the process. This nonstationary spectral estimation problem is fundamentally
more difficult than spectral estimation in the (cyclo)stationary case, because long-term averaging cannot be used to reduce the mean-square error (MSE) of the
estimate. Formally, any estimator of a nonparametric time-dependent spectrum can also be viewed as a TF representation
of the observed signal \cite{fla-book2,Cohen95a,hla-magazine,MatzHlaChap08}. Estimators have been previously proposed for several spectra including the Wigner-Ville spectrum 
and the Rihaczek spectrum (RS)
(e.g., \cite{sayeed_est95,ScharfAsilomar2001,ScharfAsilomar1998,scharf_spl05,fla-book2,martin-wvs,fla-spectra,fla-wigbook,amin-boashbook,MatzHlaChap08,koz-riedel,bayram97,BayramChapterFitzgerald,cakrak2001}).

In this paper, extending our work in \cite{jung-ssp09}, we propose a ``compressive'' estimator of the RS that uses the recently
introduced methodology of compressed sensing (CS) \cite{Don06,Can06a}. The proposed estimator is suited to underspread processes that are 
\emph{approximately TF sparse}. The latter property means that only a moderate percentage of the values of
the discrete RS are significantly nonzero. Both assumptions---underspreadness and TF sparsity---are reasonably well satisfied
in many applications, including, e.g., cognitive radio. We consider the RS because it is the simplest time-dependent spectrum from a computational viewpoint,
especially in the discrete setting used. The proposed compressive estimator of the RS is obtained
by augmenting a basic noncompressive estimator (a smoothed version of the Rihaczek distribution (RD), cf.\ \linebreak 
\cite{bayram97,BayramChapterFitzgerald,rihaczek68,fla-book2,hla-magazine,koz-riedel,MatzHlaChap08,cakrak2001,ScharfAsilomar1998,sayeed_est95})
with a CS 
com\-pression-reconstruction stage. Algorithmically, our estimator is similar to the compressive TF representation proposed in
\cite{BorgnatFlandrinICASSP08,BorgnatFlandrinTSP}. In fact, both our estimator and the TF representation of \cite{BorgnatFlandrinICASSP08,BorgnatFlandrinTSP}
are essentially based on a sparsity-regularized inversion of the Fourier transform relationship between a TF distribution and
the values of the ambiguity function (AF) taken at randomly chosen time lag/frequency lag locations.
The sparsity-regularization is achieved by requiring a small $\ell_{1}$-norm of the resulting TF distribution.
However, the setting of \cite{BorgnatFlandrinICASSP08,BorgnatFlandrinTSP}
is that of deterministic TF signal analysis (more specifically, the goal is to improve the TF localization properties of the Wigner distribution),
whereas we consider a stochastic setting, namely, spectral estimation for underspread, approximately TF sparse, nonstationary random processes.

Compressive spectral estimation methods have been proposed previously, also
in the context of cognitive radio \cite{CompWidBandSensThan,CompWidebandSensPolo,Tian2012,Leus2011}.
However, these methods are restricted to the estimation of the power spectral density of stationary or cyclostationary processes.
Furthermore, they perform CS directly on the observed signal (process realization), whereas our method performs CS on
an estimate of a TF autocorrelation function known as the expected ambiguity function (EAF).
This EAF estimate is a quadratic time lag/frequency lag representation of the observed signal that is based on the signal's AF.
It is an intermediate step in the calculation of the spectral estimator,
somewhat similar to a sufficient statistic. In some sense, we perform a twofold compression, first by using only an EAF estimate
(instead of the raw observed signal) for spectral estimation and secondly by ``compressing'' that estimate.
This approach can be advantageous if dedicated hardware units for computing values of the EAF estimate (i.e., AF) 
from an observed continuous-time signal are employed \cite{TuLiu2005,Casasent79,MarksII77,Padin2001}, because fewer such units are required.
It can also be advantageous if the values of the EAF estimate have to be transmitted over low-rate links---e.g., in
wireless sensor networks \cite{zhao2004wsn}---or stored in a memory, because fewer such values need to be transmitted or stored.

The fact that we perform CS in the AF domain and not directly on the signal is a somewhat nonorthodox aspect of our method.
Indeed, the objective of this paper is not to develop a sub-Nyquist sampling scheme in the spirit of, e.g., spectrum-blind sampling
\cite{Bresler2008,XamplingSelSigProc}. Our work is based on the assumption that the original signal of interest is modeled as
a continuous-time random process $X(t)$ that can be (approximately) represented by a finite-length, discrete-time random process $X[n]$.
This discrete-time random process itself is not used in a practical application of our method;
it is only used for the theoretical development of the method. A second assumption is that
values of the AF of a continuous-time process realization $x(t)$ can be computed efficiently.
The computation of the AF values from $x(t)$ using dedicated hardware is described in \cite{TuLiu2005,Casasent79,MarksII77,Padin2001}.

A major focus of our work is an analysis of the estimation accuracy of the proposed compressive estimator.
Because finding a closed-form expression of the MSE is intractable, we derive upper bounds on the MSE.
These bounds depend on two components: the first component is determined by the degree of ``underspreadness,''
corresponding to the concentration of the EAF of the observed process; the second component is related to the TF sparsity properties of the observed process.
As we will see below, there is a tradeoff between these components, since a well concentrated EAF of an underspread process tends to imply a poorly concentrated RS,
which is disadvantageous in terms of TF sparsity.

The remainder of this paper is organized as follows. In Section \ref{sec_EAF-RS},
we state our general setting and review some fundamentals of nonstationary random processes and their TF representation.
In Section \ref{sec_nonstat_est_basic}, we describe a basic noncompressive estimator of the RS.
In Section \ref{sec_nonstat_est_compr}, we develop a compressive estimator by augmenting the noncompressive estimator
with a CS compression-reconstruction stage. Bounds on the MSE of both the noncompressive and compressive estimators
are derived in Section \ref{sec_perf_nonstat}. Finally, numerical results are presented in Section \ref{sec_nonstat_simu}.

\emph{Notation}.\,
The modulus, complex conjugate, real part, and imaginary part of a complex number $a \! \in \! \mathbb{C}$ are denoted by $|a|$, $a^{*}$, $\Re\{a\}$, and $\Im\{a\}$,
respectively. Boldface lowercase letters denote column vectors 
and boldface uppercase letters denote matrices.
The $k$th entry of a vector $\mathbf{a}$ is denoted by ${( \mathbf{a} )}_{k}$, and  the entry
of a matrix $\mathbf{A}$ in the $i$th row and $j$th column by ${( \mathbf{A} )}_{i,j}$.
The superscripts $^{T}$, $^{*}$, and $^{H}$ denote the transpose, conjugate, and Hermitian transpose, respectively, of
a vector or matrix. The $\ell_{1}$-norm of a  vector $\mathbf{a}\in \mathbb{C}^L$ is denoted by
${\| \mathbf{a} \|}_{1} \triangleq \sum_{k=1}^L | {( \mathbf{a} )}_{k} |$, and the $\ell_{2}$-norm 
by ${\| \mathbf{a} \|}_{2} \triangleq \sqrt{ \mathbf{a}^{H} \mathbf{a}}$. The number of nonzero entries is denoted by ${\| \mathbf{a} \|}_{0}$.
The trace of a square matrix $\mathbf{A} \in \mathbb{C}^{M \times M}$ is denoted by ${\rm tr} \{ \mathbf{A} \} \triangleq \sum_{k=1}^{M} {( \mathbf{A} )}_{k,k}$.
Given a matrix $\mathbf{A} \in \mathbb{C}^{M \times N}$, we denote by ${\rm vec} \{ \mathbf{A} \} \in \mathbb{C}^{MN}$ the vector
obtained by stacking all columns of $\mathbf{A}$.
Given two matrices $\mathbf{A} \! \in \! \mathbb{C}^{M_{1} \times N_{1}}$ and $\mathbf{B} \! \in \! \mathbb{C}^{M_{2} \times N_{2}}$, we denote by
$\mathbf{A} \otimes \mathbf{B} \rmv \in \rmv \mathbb{C}^{M_{1}M_{2} \times N_{1}N_{2}}$ their Kronecker product \cite{golub96}.
The inner product of two square matrices $\mathbf{A}, \mathbf{B} \rmv \in \rmv \mathbb{C}^{M \times M}$ is defined as
$\langle \mathbf{A}, \mathbf{B} \rangle \triangleq {\rm tr} \{ \mathbf{A} \mathbf{B}^{H} \}$.
The Kronecker delta is denoted by $\delta[m]$, i.e., $\delta[m] \!=\! 1$ if $m \!=\! 0$ and $\delta[m] \!=\! 0$ otherwise.
Finally, $[N] \triangleq \{0,1,\ldots,N\!-\!1\}$.

\section{EAF and RS}\label{sec_EAF-RS}

In this section, we state our setting and review some fundamentals of the TF representation of nonstationary random processes.
Let $X(t)$ be a bandlimited nonstationary continuous-time random process that can be equivalently represented
by a nonstationary discrete-time random process $X[n]$. We assume that $X[n]$ is zero-mean, circularly symmetric complex, and
defined for $n \!\in\! [N]$.
(As mentioned above, the proposed compressive estimator does not presuppose that the discrete-time samples $X[n]$ are actually computed.)
The autocorrelation function 
of the process $X[n]$ is given by $\gamma_{X}[n_{1},n_{2}] \triangleq {\rm E} \{ X[n_{1}] X^{*}[n_{2}] \}$, where ${\rm E} \{ \cdot \}$ denotes expectation.
Since $X[n]$ is only defined for $n \!\in\! [N]$, we consider
$\gamma_{X}[n_{1},n_{2}]$ only for $n_{1}, n_{2} \!\in\! [N]$.
This is justified for a process that is well concentrated in the interval $[N]$.
An equivalent representation of 
$\gamma_{X}[n_{1},n_{2}]$ is the correlation matrix $\mathbf{\Gamma}_{\!X} \triangleq\, {\rm E} \ist\{ \mathbf{x} \mathbf{x}^H \}$,
where $\mathbf{x} \triangleq (X[0] \; X[1] \,\cdots\, X[N\!-\!1])^T \!\in \mathbb{C}^{N}$; note that ${(\mathbf{\Gamma}_{\!X})}_{n_{1}+1, n_{2}+1} = \gamma_{X}[n_{1},n_{2}]$
for $n_{1}, n_{2} \in [N]$.

We assume that $X[n]$ is an \emph{underspread} process \cite{koz-corr,koz-hamburg,koz-diss,Koz97b,GM-phd,CRC_tfrand,MatzHlaChap08,MatzHla06},
which means that its correlation in time and frequency decays reasonably fast.
The underspread property is phrased mathematically in terms of the discrete EAF,
which is defined as the following discrete Fourier transform (DFT) of the autocorrelation function 
\cite{JachanTSP07,MatzHla06,koz-corr,koz-diss,GM-phd,MatzHlaChap08,CRC_tfrand}:
\vspace{1mm}
\begin{equation}
\label{equ_def_discrete_EAF}
\bar{A}_{X}[m,l] \,\triangleq \sum_{n \in [N]} \rmv\gamma_{X}{[n,n \rmv-\rmv m]}_N \, \twiddle[ln] \, .
\end{equation}
Here, $m$ and $l$ denote discrete time lag and discrete frequency lag, respectively,
and\footnote{It 
will be convenient to consider length-$N$ functions as periodic functions with period
$N$.} 
${[n_1,n_2]}_{N} \triangleq [n_1 \!\! \mod \rmv N, n_2 \!\! \mod \rmv N]$. Note that this definition of $\bar{A}_{X}[m,l]$ is $N$-periodic in both $m$ and $l$.
The EAF $\bar{A}_{X}[m,l]$ is a TF-lag representation of the second-order statistics of $X[n]$ that describes the TF correlation structure of $X[n]$.
A nonstationary process $X[n]$ is said to be underspread if its EAF is well concentrated around the origin in the ($m,l$)-plane, i.e.,
\begin{align}
&\bar{A}_{X}[m,l] \approx 0 \,, \;\; \forall (m,l) \not\in \mathcal{A} \,, \quad\!
  \text{with} \;\,  \mathcal{A} \triangleq {\{- M,\ldots,M \}}_N\nonumber\\[0mm]
&\rule{61mm}{0mm} \times {\{- L,\ldots, L \}}_N \,, \nonumber\\[1mm]
&\rule{5mm}{0mm}\mbox{where} \;\, 0 \le M < \bigg\lfloor \frac{N}{2} \bigg\rfloor \,, \; 0 \le L < \bigg\lfloor \frac{N}{2} \bigg\rfloor \,, \; \mbox{and} \;\, M L \ll\rmv N \ist.
\label{eq_underspread}
\end{align}
Here, e.g., ${\{- M,\ldots,M \}}_N$ denotes the $N$-periodic continuation of the interval $\{- M,\ldots,M \}$.
The concentration of the EAF around the origin can be measured by the \emph{EAF moment} defined in Section \ref{sec_pars}
(see \eqref{equ_nonstat_mom_ell2}). For later reference, we note that the EAF is the expectation of the AF \cite{fla-book2,Cohen95a,hla-magazine}
\be
\label{equ_AF}
A_{X}[m,l] \,\triangleq \sum_{n \in [N]} \! X{[n]} \ist\ist X^{*}{[n\!-\!m]}_N \, \twiddle[ln] \ist ,
\ee
i.e., $\bar{A}_{X}[m,l] = {\rm E} \big\{ A_{X}[m,l] \big\}$.

Nonstationary spectral estimation is the problem of estimating a ``time-dependent power spectrum'' of the nonstationary process
$X[n]$ from a single realization $x[n]$ observed for $n \in [N]$. As mentioned earlier, there is no
definition of a ``time-dependent power spectrum'' that satisfies all desirable properties \cite{loynes68,fla-book2,martin-wvs,fla-wigbook,MatzHla06}.
However, in the underspread case considered, most reasonable definitions of a time-dependent power spectrum
are approximately equal, represent the mean energy distribution of the process over time and frequency,
and approximately satisfy all desirable properties \cite{GM-phd,MatzHla06}. Therefore, we use the simplest such definition, which is the RS
\cite{rihaczek68,fla-book2,scharf_spl05,fla-wigbook}. The discrete RS is defined as the following DFT of the autocorrelation function:
\begin{equation}
\label{def_discrete_RHS}
\bar{R}_X[n,k] \,\triangleq \sum_{m \in [N]} \!\gamma_{X}{[n,n \rmv-\rmv m]}_N \, \twiddle[km] \ist .
\end{equation}
Just as the EAF and AF, the RS $\bar{R}_X[n,k]$ is $N$-periodic in both its variables.
Furthermore, the RS is complex-valued in general, but it is approximately real-valued and nonnegative in the underspread case \cite{GM-phd,MatzHla06}.
The RS is related to the EAF via a symplectic two-dimensional (2D) 
\pagebreak 
DFT:
\begin{align}
\bar{R}_X[n,k] &\eq \frac{1}{N} \!\sum_{m,l \in [N]} \!\bar{A}_{X}[m,l] \, \twiddle[(km-nl)] \,, \label{equ_fourier_eaf_rhs}\\[0mm]
\bar{A}_{X}[m,l] &\eq \frac{1}{N} \!\sum_{n,k \in [N]} \!\bar{R}_X[n,k] \, \twiddleneg[(mk-ln)] \,.
\label{equ_inv_fourier_eaf_rhs}
\end{align}
Relation \eqref{equ_fourier_eaf_rhs} extends the 
Fourier transform relation between the power spectral density and the autocorrelation function of a stationary process \cite{Kay88,stoi97} to the nonstationary case.
It follows from \eqref{equ_fourier_eaf_rhs} that the RS of an underspread process is a smooth function.
Furthermore, the RS is the expectation of the RD defined as \cite{rihaczek68,fla-book2,Cohen95a,hla-magazine,scharf_spl05}
\begin{align*}
R_{X}[n,k] &\,\triangleq \sum_{m \in [N]} \!X{[n]}_N \ist X^{*}{[n\!-\!m]}_N \, \twiddle[km] \\[.5mm] 
&\eq X{[n]}_N \ist \hat{X}^*{[k]}_N \, e^{-j \frac{2 \pi}{N} nk} \,,
\end{align*}
where $\hat{X}[k] \triangleq \sum_{n \in [N]} X[n] \ist e^{- j \frac{2 \pi}{N} kn}$ is the DFT of $X[n]$. That is, $\bar{R}_{X}[n,k]  = {\rm E} \big\{ R_{X}[n,k] \big\}$.
The 2D DFT relations \eqref{equ_fourier_eaf_rhs}, \eqref{equ_inv_fourier_eaf_rhs} hold also for the RD and AF, i.e.,
\begin{align}
R_X[n,k] &\eq \frac{1}{N} \!\sum_{m,l \in [N]} \!A_{X}[m,l] \, \twiddle[(km-nl)] \,, \label{RD_AF_fourier}\\[0mm]
A_{X}[m,l] &\eq \frac{1}{N} \!\sum_{n,k \in [N]} \!R_X[n,k] \, \twiddleneg[(mk-ln)] \,. \nonumber
\end{align}

Our central assumption, besides the underspread property, is that the nonstationary process $X[n]$ is ``approximately TF sparse'' in the sense that only
a moderate percentage of the RS values $\bar{R}_X[n,k]$ within the fundamental $(n,k)$-region $[N]^2 \!= [N] \times [N]$ are significantly nonzero.
For such approximately TF sparse processes, we will develop a compressive estimator of the RS by augmenting a basic RS
estimator with a compression-reconstruction stage. We present the basic estimator first.

\section{Basic RS Estimator}\label{sec_nonstat_est_basic}

In analogy to well-known estimators of the Wigner-Ville spectrum \cite{fla-book2,martin-wvs,fla-spectra,fla-wigbook,amin-boashbook,MatzHlaChap08,koz-riedel},
a basic (noncompressive)
estimator of the RS $\bar{R}_X[n,k]$ is given by the following smoothed version of the RD \cite{scharf_spl05,koz-riedel}:
\be
\label{RD_convolve}
\hat{R}_{X}[n,k] \,\triangleq\, \frac{1}{N} \!\sum_{n'\!,k' \in [N]} \!\Phi{[n\!-\!n'\!,k\!-\!k']} \, R_{X}[n'\!,k'] \,.
\ee
Here, $\Phi[n,k]$ is a smoothing function that is $N$-periodic in both arguments. Because of \eqref{equ_inv_fourier_eaf_rhs}, the symplectic 2D inverse DFT of
$\hat{R}_{X}[n,k]$,
\be
\label{EAF_est}
\hat{A}_{X}[m,l] \,\triangleq\ist \frac{1}{N} \!\sum_{n,k \in [N]} \!\hat{R}_{X}[n,k] \, e^{j\frac{2\pi}{N} (mk - ln)} \ist,
\ee
can be viewed as an estimator of the EAF $\bar{A}_{X}[m,l]$. Using \eqref{RD_convolve} and \eqref{RD_AF_fourier} in \eqref{EAF_est}, we obtain
\be
\label{AF_weight}
\hat{A}_{X}[m,l] \eq \phi[m,l] \, A_X[m,l] \, ,
\ee
where the 2D window (weighting, taper) function $\phi[m,l]$ is related to the smoothing function $\Phi[n,k]$ through a 
\pagebreak 
2D DFT, i.e.,
\be
\label{eq:phi}
\phi[m,l] \,\triangleq\, \frac{1}{N} \!\sum_{n,k \in [N]} \!\Phi[n,k] \, e^{j\frac{2\pi}{N} (mk - ln)}\,.
\ee
Note that $\phi[m,l]$ and $\hat{A}_{X}[m,l]$ are $N$-periodic in both $m$ and $l$.

We now consider the choice of the smoothing function $\Phi[n,k]$ or, equivalently, of the window function $\phi[m,l]$.
Our performance criterion is the 
\vspace{1mm}
MSE
\[
\varepsilon \,\ist\triangleq\,\ist  {\rm E} \big\{ \big\| \hat{R}_{X} \rmv-\rmv \bar{R}_{X} \big\|^{2}_{2} \big\}
  \eq  \!\sum_{n,k \in [N]} \!{\rm E} \big\{ \big| \hat{R}_{X}[n,k] \rmv-\rmv \bar{R}_{X}[n,k] \big|^{2} \big\} \,.
\]
The MSE can be decomposed as
$\varepsilon = B^{2}\rmv + V$ with the squared bias term $B^2 \triangleq \big\| \ist{\rm E} \{\hat{R}_{X} \} \rmv-\rmv \bar{R}_{X} \big\|_2^{2}$
and the variance $V \triangleq {\rm E} \big\{ \big\| \hat{R}_{X} \rmv- {\rm E} \{ \hat{R}_{X} \} \big\|_2^{2} \big\}$.
We will consider a \emph{minimum variance unbiased} (MVU)
design\footnote{The 
MVU design is analytically tractable and well established in TF spectrum estimation \cite{MatzHlaChap08,koz-riedel}.
An alternative design of $\Phi[n,k]$ could be based on the minimax rationale \cite{LC};
however, there does not seem to exist a simple solution to the minimax design
problem.} 
of $\Phi[n,k]$.
This means that $\hat{R}_{X}[n,k]$ is required to be unbiased, i.e., $B \rmv=\rmv 0$, and the variance $V$ is minimized under this constraint.
More specifically, we will adopt the MVU design proposed in \cite{koz-riedel,MatzHlaChap08}, which is based on the idealizing assumption that the
EAF $\bar{A}_{X}[m,l]$ is supported on a periodized rectangular region
$\mathcal{A} = {\{- M,\ldots,M \}}_N\rmv \times {\{- L,\ldots, L \}}_N$, i.e.,
$\bar{A}_{X}[m,l] = 0$ for all $(m,l) \not\in \mathcal{A}$, with $0 \le M < \lfloor N/2 \rfloor$ and $0 \le L < \lfloor N/2 \rfloor$. This is somewhat similar to
the underspread property \eqref{eq_underspread}; however, it is an exact, rather than approximate, support constraint. As a further
difference from the underspread property, we do not require that $M L \ll\rmv N$.
We note that this idealizing exact support constraint is only needed for the MVU interpretation of our design of $\Phi[n,k]$; in particular,
it will not be used for our performance analysis in Section \ref{sec_perf_nonstat}.
The size of $\mathcal{A}$---i.e., the choice of $L$ and $M$---is a design parameter that can be chosen freely in principle.
The resulting estimator $\hat{R}_{X,\text{MVU}}[n,k]$ (cf.\ \eqref{equ_MVU_nonlin_nonstat_eaf}) can
be applied to any process $X[n]$, including, in particular, processes whose EAF $\bar{A}_{X}[m,l]$ is not exactly supported on $\mathcal{A}$.

We briefly review the derivation of the MVU smoothing function presented in \cite{koz-riedel,MatzHlaChap08}.
Using \eqref{AF_weight} and ${\rm E} \big\{ A_{X}[m,l] \big\} = \bar{A}_{X}[m,l]$, the bias term 
$B^2 = \big\| \ist{\rm E} \{\hat{R}_{X} \} - \bar{R}_{X} \big\|_2^{2} = \big\| \ist{\rm E} \{\hat{A}_{X} \} \rmv-\rmv \bar{A}_{X} \big\|_2^{2}$ can be expressed as
\begin{equation}
B^2 \rmv\eq \!\sum_{m,l \in [N]} \rmv\big| (\phi[m,l] \rmv-\rmv 1) \, \bar{A}_{X}[m,l] \big|^{2} .
  \label{equ_exact_bias_non_11}
\end{equation}
Thus, $B^2 \!=\rmv 0$ if and only if $\phi[m,l] \rmv=\rmv 1$ on the support of $\bar{A}_{X}[m,l]$, i.e., for all $(m,l) \!\in\! \mathcal{A}$.
Under the constraint $B^2 \!=\rmv 0$, minimizing the variance of $\hat{R}_{X}[n,k]$ is equivalent to minimizing the mean power
\begin{align*}
P &\,\triangleq\,\ist {\rm E} \big\{ \big\| \hat{R}_{X} \big\|^{2}_{2} \big\}\\[.5mm]
&\stackrel{\eqref{EAF_est}}{\eq}\ist {\rm E} \big\{ \big\| \hat{A}_{X} \big\|^{2}_{2} \big\}\\[.5mm]
&\stackrel{\eqref{AF_weight}}{\eq}\ist {\rm E} \big\{ \big\| \phi[m,l] \, A_X[m,l] \ist\big\|^{2}_{2} \big\}\\[1mm]
&\eq\! \sum_{m,l \in [N]}  \! |\phi[m,l]|^{2} \, {\rm E} \big\{ \big|A_X [m,l]\big|^{2} \big\} \,.
\end{align*}
Splitting this sum into a sum over ${[N]}^2 \cap \mathcal{A}$ (where $\phi[m,l] \rmv=\rmv 1$) and a sum over
${[N]}^2 \cap \overline{\mathcal{A}}$ (here, $\overline{\mathcal{A}}$ denotes the complement of $\mathcal{A}$), it is clear that
$P$ is minimized if and only if the latter sum is zero. This means that $\phi[m,l]$ must be zero for
$(m,l) \in {[N]}^2 \cap \overline{\mathcal{A}}$, and further, due to the periodicity of $\phi[m,l]$, for $(m,l) \in \overline{\mathcal{A}}$.
Thus, we conclude that the MVU window function (DFT of the MVU smoothing function) is the indicator function $I_{\rmv \mathcal{A}}[m,l]$ of the
EAF support $\mathcal{A} = {\{- M,\ldots,M \}}_N\rmv \times {\{- L,\ldots, L \}}_N$:
\begin{equation}
\label{equ_def_indicator_function_MVU_EAF}
 \phi_{\text{MVU}}[m,l] \eq I_{\rmv \mathcal{A}}[m,l]\,\triangleq\ist \begin{cases}
   1, & (m,l) \!\in\! \mathcal{A} \\[0mm]
   0, & \text{otherwise} \ist .
   \end{cases}
\end{equation}
The corresponding EAF estimator in \eqref{AF_weight} is obtained as
\begin{align}
\hat{A}_{X,\text{MVU}}[m,l] 
  &\eq \phi_{\text{MVU}}[m,l] \, A_X[m,l] \nonumber\\[1mm]
  &\eq I_{\rmv \mathcal{A}}[m,l] \, A_X[m,l] \label{AF_weight_MVU_0}\\[1mm]
  &\,=\ist \begin{cases} A_X[m,l], & (m,l) \!\in\! \mathcal{A} \\[0mm]
   0, & \text{otherwise} \ist.
   \end{cases}
\label{AF_weight_MVU}
\end{align}
Therefore, the MVU estimator of the RS is given by (see \eqref{EAF_est})
\begin{align}
\hat{R}_{X,\text{MVU}}[n,k] &\eq  \frac{1}{N} \!\sum_{m,l \in [N]} \!\hat{A}_{X,\text{MVU}}[m,l] \, e^{-j \frac{2 \pi}{N}(km - nl)} \nonumber\\[-4mm]
&  \label{equ_MVU_nonlin_nonstat_eaf_0}\\[-1mm]
  &\eq  \frac{1}{N} \!\sum_{m=-M}^{M} \sum_{l=-L}^{L} \rmv A_{X} {[m,l]} \, e^{-j \frac{2 \pi}{N}(km - nl)} \,,\nonumber\\[-3mm]
& \label{equ_MVU_nonlin_nonstat_eaf}\\[-7mm]
& \nonumber
\end{align}
where the periodicity of the summand with respect to $m$ and $l$ has been exploited in the last step.

\section{Compressive RS Estimator}\label{sec_nonstat_est_compr}

Next, we will augment the basic RS estimator presented in the previous section with a compression-reconstruction stage.

\vspace{-1mm}

\subsection{Basic DFT Relation}\label{sec_DFT-relation}

The proposed compressive RS estimator is based on a 2D DFT relation that will now be derived.
We recall from \eqref{AF_weight_MVU} that the EAF estimate $\hat{A}_{X,\text{MVU}}[m,l]$ is exactly zero outside the effective EAF support
$\mathcal{A} = {\{- M,\ldots,M \}}_N\rmv \times {\{- L,\ldots, L \}}_N$, where $0 \le M < \lfloor N/2 \rfloor$ and $0 \le L < \lfloor N/2 \rfloor$. 
In what follows, we will denote by
\begin{equation}
\label{equ_def_mathcal_S}
\SS \,\triangleq\, \big| {[N]}^2 \rmv\cap \mathcal{A} \big| \eq (2M \rmv+\! 1)(2L\rmv+\! 1)
\end{equation}
the size of one period of $\mathcal{A}$.
Because $2M \rmv+\! 1$ and $2L\rmv+\! 1$ do not necessarily divide $N$, we furthermore define an ``extended effective EAF support''
as the periodized rectangular region $\mathcal{A}' \triangleq \{-M,\ldots,-M + \Delta M -1 {\}}_{N} \times \{-L,\ldots, -L + \Delta L -1 {\}}_{N}$.
Here, $\Delta M$ and $\Delta L$ are chosen as the smallest integers such that $\Delta M \geq 2M+1$ and $\Delta L \geq 2L+1$ and, moreover, 
$\Delta M$ and $\Delta L$ divide $N$, i.e, there are integers $\Delta n$, $\Delta k$ such that $\Delta n \,\Delta L=\Delta k \,\Delta M= N$ or, equivalently,
\be
\Delta n \ist= \frac{N}{\lendoppler} \,, \quad\; \Delta k \ist= \frac{N}{\lendelay} \,.
\label{equ_def_delta_n_k}
\ee
The size of one period of $\mathcal{A}'$ is
\[
\SS' \,\triangleq\, \big| {[N]}^2 \rmv\cap \mathcal{A}' \big| \eq \Delta M \ist\Delta L \,.
\]
Note that
\begin{equation}
\label{equ_ineq_S_S_prime}
\mathcal{A}\^{E}\subseteq \mathcal{A}' \ist, \quad\; \SS \leq \SS',
\end{equation}
although typically $\SS \approx \SS'$. Let us arrange the values of one period of $\hat{A}_{X,\text{MVU}}[m,l]$ that are located within $\mathcal{A}'$ into
a matrix $\mathbf{A} \in \mathbb{C}^{\lendelay \times \lendoppler}$, i.e.,
\begin{align}
{( \mathbf{A} )}_{m+1,\ist l+1} &\,\triangleq\,  \hat{A}_{X,\text{MVU}}[m - M, l\rmv-\rmv L] \,,\nonumber\\[.5mm] 
& \rule{20mm}{0mm} m \in [\lendelay] \,, \;\, l\in [\lendoppler] \,.
\label{eq:A_A}
\end{align}
Alternatively, we can represent $\hat{A}_{X,\text{MVU}}[m,l]$ by the matrix $\mathbf{R} \in \mathbb{C}^{\lendoppler \times \lendelay}$ whose entries are
given by the following 2D DFT of dimension 
\vspace{1mm}
$\lendoppler \times \lendelay$:
\begin{align}
{( \mathbf{R} )}_{p+1,\ist q+1} &\,\triangleq \sum_{m \in [\Delta M]} \sum_{l \in [\Delta L]} \!{( \mathbf{A} )}_{m+1,\ist l+1} \nonumber\\[-1mm] 
& \rule{22mm}{0mm} \times e^{-j 2 \pi \big( \frac{q(m-M)}{\lendelay} \ist-\ist \frac{p(l-L)}{\lendoppler} \big)} \,
  \label{eq:R_A}\\[1mm]
&\stackrel{\eqref{eq:A_A}}{=} \sum_{m=-M}^{-M + \Delta M - 1} \,\sum_{l=-L}^{-L + \Delta L- 1} \! \hat{A}_{X,\text{MVU}}[m,l] \nonumber\\[-1.5mm] 
& \rule{35mm}{0mm} \times e^{-j 2 \pi \big( \frac{qm}{\lendelay} \ist-\ist \frac{pl}{\lendoppler} \big)} \nonumber\\[-1mm]
  &\stackrel{\eqref{AF_weight_MVU}}{\eq} \! \sum_{m=-M}^{M} \sum_{l=-L}^{L}
  \! A_X[m,l] \, e^{-j 2 \pi \big( \frac{qm}{\lendelay} \ist-\ist \frac{pl}{\lendoppler} \big)}\ist,\nonumber\\[-1.5mm] 
& \rule{27mm}{0mm}p \in [\lendoppler] \ist, \; q\in [ \lendelay] \,.
\label{equ_S_A} \\[-4mm]
&\nonumber
\end{align}
It can be seen by comparing \eqref{equ_S_A} and \eqref{equ_MVU_nonlin_nonstat_eaf} that
the matrix entries ${( \mathbf{R} )}_{p+1,\ist q+1}$ equal (up to a constant factor) a subsampled version of $\hat{R}_{X,\text{MVU}}[n,k]$, i.e.,
\begin{align}
&{( \mathbf{R} )}_{p+1,\ist q+1} \ist=\ist N \ist \hat{R}_{X,\text{MVU}}[p \ist\Delta n, q \ist\Delta k] \,,\nonumber\\[.5mm] 
& \rule{35mm}{0mm} p \in [\lendoppler] \ist, \; q\in [\lendelay] \,,
\label{equ_R_def}
\end{align}
with $\Delta n = N/\lendoppler$ and $\Delta k = N/\lendelay$ as in \eqref{equ_def_delta_n_k}. This subsampling does not cause a loss of information because
$\hat{A}_{X,\text{MVU}}[m,l]$ is supported in $\mathcal{A}$, and therefore, by \eqref{equ_ineq_S_S_prime}, also in
$\mathcal{A}' = \{-M,\ldots,-M + \Delta M -1 {\}}_{N} \times \{-L,\ldots, -L + \Delta L -1 {\}}_{N}$.

Inverting \eqref{eq:R_A}, we obtain
\begin{align}
&{( \mathbf{A} )}_{m+1,\ist l+1} \nonumber\\[1mm] 
& \rule{.2mm}{0mm}\eq \frac{1}{\SS'} \!\sum_{p \in [\lendoppler]} \sum_{q \in [\lendelay]} \rmv\!{( \mathbf{R} )}_{p+1,\ist q+1} 
  \, e^{j 2 \pi \big( \frac{(m-M)q}{\lendelay} \ist-\ist \frac{(l-L)p}{\lendoppler} \big)} \ist, \quad\;\;\nonumber\\[-1mm] 
& \rule{45mm}{0mm} m\in [\lendelay] \ist, \; l\in [\lendoppler] \,.
\label{equ_S_A_inv}
\end{align}
This 2D DFT relation will constitute an important basis for our compressive RS estimator. It can be compactly written as
\begin{equation}
\label{equ_data_model_non}
\mathbf{U} \mathbf{r} \ist=\ist \mathbf{a} \,,
\end{equation}
where $\mathbf{r} \ist\triangleq\ist {\rm vec} \{ \mathbf{R} \} \in \mathbb{C}^{\SS'}$,
$\mathbf{a} \ist\triangleq\ist {\rm vec} \{ \mathbf{A}^{\!T} \} \in \mathbb{C}^{\SS'}$, and
\be
\label{equ_def_U}
\mathbf{U} \,\triangleq\, \frac{1}{\SS' } \, \mathbf{F}^{*}_{\rmv  \lendelay} \rmv\otimes \mathbf{F}_{\rmv \lendoppler}
  \in\, \mathbb{C}^{\SS'  \times \SS' } ,
\ee
with $\mathbf{F}_{\rmv \lendelay}$ defined as ${( \mathbf{F}_{\rmv \lendelay} )}_{q+1,m+1} \triangleq e^{-j2 \pi \frac{q(m-M)}{\lendelay}}$, $q,m \in [\lendelay]$
and $\mathbf{F}_{\rmv \lendoppler}$ defined as ${( \mathbf{F}_{\rmv \lendoppler} )}_{p+1,l+1} \triangleq e^{-j2 \pi \frac{p(l-L)}{\lendoppler}}$, $p,l \in [\lendoppler]$.

\pagebreak 

Furthermore, using \eqref{eq:A_A} in \eqref{equ_MVU_nonlin_nonstat_eaf_0}, we obtain
\begin{align*}
\hat{R}_{X,\text{MVU}}[n,k] &\eq \frac{1}{N} \!\sum_{m \in [\Delta M]} \sum_{l \in [\Delta L]} \! {( \mathbf{A} )}_{m+1,\ist l+1} \nonumber\\[.5mm] 
& \rule{22mm}{0mm} \times e^{-j \frac{2 \pi}{N}[k(m-M) - n(l-L)]} \,.
\end{align*}
Inserting \eqref{equ_S_A_inv}, we see that
the basic RS estimate $\hat{R}_{X,\text{MVU}}[n,k]$ can be calculated from $\mathbf{r}$ (or, equivalently, from $\mathbf{R}$) according to
\begin{align}
&\hspace{-3mm}\hat{R}_{X,\text{MVU}}[n,k] \eq \mathcal{L} \{ \mathbf{r} \} [n,k] \nonumber\\[0mm]
&\hspace{-2mm}\triangleq\, \frac{1}{N\SS' } \!\rmv\sum_{m \in [\Delta M]} \sum_{l \in [\Delta L]}
  \Bigg[ \sum_{p \in [\lendoppler]} \sum_{q \in [\lendelay]} \rmv\! {( \mathbf{R} )}_{p+1,\ist q+1} \nonumber\\[.5mm] 
& \rule{4mm}{0mm} \times e^{j 2 \pi \big( \frac{(m-M)q}{\lendelay} \ist-\ist \frac{(l-L)p}{\lendoppler} \big)} \Bigg]
  \ist e^{-j \frac{2 \pi}{N}[k(m-M) - n(l-L)]} \,. \label{equ_reconstr_operator_non} 
\end{align}

\vspace{-6mm}

\subsection{Measurement Equation and Sparse Reconstruction}\label{sec_meas-eq}

The compressive RS estimator can be obtained by combining the results of the previous subsection
with standard results from CS theory \cite{Can06a,rud06}. To motivate our development, we assume that $\hat{R}_{X,\text{MVU}}[\ist p \ist\Delta n, q \ist\Delta k]$
is approximately $K$-sparse for some $K \!<\rmv \SS'\!$, i.e.,
at most $K$ of the $\SS'$ values of the basic RS estimator $\hat{R}_{X,\text{MVU}}[n,k]$ on the subsampled grid $(n,k) = (p \ist\Delta n, q \ist\Delta k)$ are significantly
nonzero. (Because $\hat{R}_{X,\text{MVU}}[n,k]$ is an estimator of the RS, this assumption is consistent with our basic assumption
that the RS $\bar{R}_X[n,k]$ itself is approximately sparse.)
Due to \eqref{equ_R_def}, it follows that the matrix $\mathbf{R}$ and, equivalently, the vector $\mathbf{r} \equiv {\rm vec} \{ \mathbf{R} \}$
are approximately $K$-sparse. Furthermore, according to \eqref{equ_data_model_non}, $\mathbf{r} \rmv\in\rmv \mathbb{C}^{\SS'}$ is related to the EAF
estimate $\mathbf{a} \equiv {\rm vec} \{ \mathbf{A}^{\!T} \} \rmv\in\rmv \mathbb{C}^{\SS'}$
as $\mathbf{U} \mathbf{r} = \mathbf{a}$, where $\mathbf{U}$ (see \eqref{equ_def_U}) is an orthogonal (up to a factor) and equimodular matrix
of size $\SS' \!\times\rmv \SS'\rmv$, i.e., $\mathbf{U}^{H} \mathbf{U} = \frac{1}{\SS'} \ist \mathbf{I}$ and $| ( \mathbf{U} )_{i,j} | = \frac{1}{\SS'}$.
Let us define $\mathbf{a}^{(P)}\! \in \mathbb{C}^{P}$ as the vector made up of $P$ randomly selected entries of $\mathbf{a}$, for some $P \!<\rmv \SS'$
(typically, $P \!\ll\rmv \SS'$). Thus, recalling \eqref{eq:A_A} and \eqref{AF_weight_MVU_0}, the entries of $\mathbf{a}^{(P)}$ are $P$ values of the masked AF
$I_{\rmv \mathcal{A}}[m,l] \ist A_X[m,l]$ randomly located within the region ${[N]}^2 \ist\cap\ist\ist \mathcal{A}'$ or,
equivalently,\footnote{Typically, 
the region ${[N]}^2 \cap \mathcal{A}'$ is only slightly larger than the effective EAF support ${[N]}^2 \cap \mathcal{A}$.
Thus, most of the $P$ entries of $\mathbf{a}^{(P)}$ are values of $A_X[m,l]$ randomly located within ${[N]}^2 \cap \mathcal{A}$
or, equivalently, within $\{- M,\ldots,M \} \times \{- L,\ldots, L \}$. The remaining entries of $\mathbf{a}^{(P)}$ are
zero.} 
the values of $I_{\rmv \mathcal{A}}[m,l] \ist A_X[m,l]$ at $P$ randomly chosen TF lag positions
$(m,l) \in \{- M,\ldots, -M + \Delta M -1 \} \times \{  - L ,\ldots, -L + \Delta L-1 \}$. We have then from \eqref{equ_data_model_non}
\be
\MM\mathbf{r} \ist=\ist \mathbf{a}^{(P)} \ist,
\label{equ_meas-eq_1}
\ee
where the matrix $\MM \in \mathbb{C}^{P \times \SS'}$ is obtained by randomly selecting $P$ rows from
$\mathbf{U}$; the indices of these rows equal the indices of the entries selected from $\mathbf{a}$.

Equation \eqref{equ_meas-eq_1} is an instance of a \emph{measurement equation} as considered in CS theory.
Because the ``measurement matrix'' $\MM$ is formed by randomly selecting $P$ rows from $\mathbf{U}$,
and $\mathbf{U}$ is a unitary (up to a factor) and equimodular matrix, CS theory \cite{Can06a,rud06} provides the following result:
For
\begin{equation}
\label{equ_P_cond_nonstat}
P \,\geq\, C \, (\log \SS'  )^4 \ist \KK \ist=\ist C \ist \big[ \log (\lendelay) + \log (\lendoppler) \big]^4 \ist \KK \ist,
\end{equation}
where
$C$ is a positive constant that does not depend on $\mathbf{r}$,
the result of Basis Pursuit \cite{Chen98atomicdecomposition} operating on $\mathbf{a}^{(P)}$, i.e.,
\begin{equation}
\label{equ_def_bp_nonstat}
\hat{\mathbf{r}} \,\triangleq\, \argmin_{\mathbf{r}' \rmv:\ist \MM\mathbf{r}' \ist=\ist \mathbf{a}^{(P)} } {\|\mathbf{r}'\|}_{1}  \,,
\end{equation}
satisfies with overwhelming 
probability\footnote{That 
is, the probability of \eqref{equ_err_bound_bp_nonstat} not being true decreases exponentially with
$P\rmv$.}
\begin{equation}
\label{equ_err_bound_bp_nonstat}
{\| \hat{\mathbf{r}} \rmv-\rmv \mathbf{r} \|}_{2}  \,\leq\,  \frac{D}{\sqrt{\KK}} \, {\| \mathbf{r} \rmv-\rmv \mathbf{r}^{\mathcal{G}} \|}_{1} \,.
\end{equation}
Here, $D$ is another positive constant that does not depend on $\mathbf{r}$, and $\mathbf{r}^{\mathcal{G}}$ denotes the vector that is obtained
by zeroing all entries of $\mathbf{r}$ except the $K$ entries whose indices are in a given index set $\mathcal{G} \subseteq \{1,\ldots, \SS' \}$ of size $|\mathcal{G}| \!=\! K$.
Since $\mathbf{r}$ is approximately $\KK$-sparse, the index set $\mathcal{G}$ can be chosen such that the corresponding entries
${\{ {(\mathbf{r})}_{k} \}}_{k \in \mathcal{G}}$ comprise, with high
probability,\footnote{Note 
that the index set $\mathcal{G}$ is deterministic and fixed, whereas the indices of the largest entries of $\mathbf{r}$ may vary with each realization
of the random process. However, for the performance analysis in Section \ref{sec_perf_nonstat},
it is sufficient to assume that the index set $\mathcal{G}$ \emph{approximately} contains the indices of the largest entries of $\mathbf{r}$ for each
realization.} 
the significantly nonzero entries of $\mathbf{r}$, implying a small norm ${\| \mathbf{r} \rmv-\rmv \mathbf{r}^{\mathcal{G}} \|}_{1}$.
The bound \eqref{equ_err_bound_bp_nonstat} then shows that the Basis Pursuit is capable of reconstructing $\mathbf{r}$---and, thus, the subsampled basic RS estimator
$\hat{R}_{X,\text{MVU}}[p \ist\Delta n, q \ist\Delta k]$---from the compressed AF vector $\mathbf{a}^{(P)}$ with a small
reconstruction error ${\| \hat{\mathbf{r}} \rmv-\rmv \mathbf{r} \|}_{2}$.
(We recall, at this point, that the entries of $\mathbf{r}$ equal the values of $\hat{R}_{X,\text{MVU}}[p \ist\Delta n, q \ist\Delta k]$.)
The minimization in \eqref{equ_def_bp_nonstat} can be implemented numerically using standard tools, e.g., the MATLAB toolbox \rm{CVX} \cite{GrantBoydCVX}.

\subsection{The Compressive RS Estimator}\label{sec_sub_nonstat_est_compr}

From the Basis Pursuit reconstruction result $\hat{\mathbf{r}}$ in \eqref{equ_def_bp_nonstat}, a compressive approximation of
the basic RS estimator $\hat{R}_{X,\text{MVU}}[n,k]$ in \eqref{equ_MVU_nonlin_nonstat_eaf} is finally obtained by substituting $\hat{\mathbf{r}}$ for
$\mathbf{r}$ in \eqref{equ_reconstr_operator_non}:
\begin{align}
&\hspace{-3mm}\hat{R}_{X,\text{CS}}[n,k] \eq \mathcal{L} \{ \hat{\mathbf{r}} \} [n,k] \nonumber\\[0mm]
&\hspace{-2mm}=\, \frac{1}{N\SS' } \!\rmv\sum_{m \in [\Delta M]} \sum_{l \in [\Delta L]}
  \Bigg[ \sum_{p \in [\lendoppler]} \sum_{q \in [\lendelay]} \rmv\! {( \hat{\mathbf{R}} )}_{p+1,\ist q+1} \nonumber\\[.5mm] 
& \rule{4mm}{0mm} \times e^{j 2 \pi \big( \frac{(m-M)q}{\lendelay} \ist-\ist \frac{(l-L)p}{\lendoppler} \big)} \Bigg]
  \ist e^{-j \frac{2 \pi}{N}[k(m-M) - n(l-L)]} \,, \label{equ_MVU_nonstat_est_CS} 
\end{align}
where $\hat{\mathbf{R}} = \mathrm{unvec} \{ \hat{\mathbf{r}} \} \in \mathbb{C}^{\lendoppler \times \lendelay}$ is the matrix corresponding to $\hat{\mathbf{r}}$.
This defines the compressive RS estimator.

To summarize, the proposed compressive RS estimator $\hat{R}_{X,\text{CS}}[n,k]$ is calculated by the following steps.

\begin{enumerate}

\vspace{1.5mm}

\item Choose $K \!<\! \SS'$ such that it reflects the prior intuition about the effective sparsity of the subsampled RS $\bar{R}_X[\ist p \ist\Delta n, q \ist\Delta k]$,
$(p,q) \in [\lendoppler] \times [\lendelay]$.
(Equivalently, $KN^{2}/\SS'$ reflects the prior intuition about the effective sparsity of the RS $\bar{R}_X[n,k]$, $(n,k) \in [N]^2$.)

\vspace{1.5mm}

\item
Acquire $P \geq C \ist \big[ \log (\lendelay) + \log (\lendoppler) \big]^4 \ist \KK$
values of the masked AF $I_{\rmv \mathcal{A}}[m,l] \, A_X[m,l]$ at randomly chosen TF lag
positions\footnote{More 
precisely, we choose uniformly at random a size-$P$ subset of $\{- M,\ldots, -M + \Delta M -1 \} \times \{  - L ,\ldots, -L + \Delta L-1 \}$, containing $P$ different TF lag positions
$(m,l)$.}
$(m,l) \in \{- M,\ldots,-M +\Delta M -1 \} \times \{- L,\ldots, -L+ \Delta L -1 \}$.
Let $\mathbf{a}^{(P)}\rmv$ denote the vector containing these ``compressive measurements.''
A compression has been achieved if $P < \SS'  \equiv \lendelay \ist \lendoppler$; the ``compression factor'' is $\SS'/P \ge 1$.
It is important to note that the AF values $A_X[m,l]$ can be equivalently obtained (up to small aliasing errors that are typically negligible)
from the continuous-TF-lag AF of the underlying continuous-time process
$X(t)$.\footnote{The 
continuous-TF-lag AF is defined as $A_{X}(\tau,\nu) \triangleq \int_{- \infty}^{\infty} X(t) \ist X^{*}(t - \tau) \ist e^{- j 2 \pi \nu t} \ist dt$.
If the process $X(t)$ is bandlimited to the frequency band $[0,1/(2T_{\text{s}})]$ and effectively localized within the time interval $[0,NT_{\text{s}}/2]$, we can use the
approximation
\begin{align*}
A_{X}[m,l] &\,\stackrel{\eqref{equ_AF}}{=}\!  \sum_{n\in [N]} \! X[n] \ist X^{*}{[n \rmv-\rmv m]}_{N} \ist e^{- j \frac{2 \pi}{N}  l n }\\
  &\,\ist\approx\!  \sum_{n\in [N]} \! X(nT_{\text{s}}) \ist X^{*}((n \rmv-\rmv m)T_{\text{s}}) \ist\ist e^{- j \frac{2 \pi}{N}  l n }\\
  &\,\ist\approx \frac{1}{T_{\text{s}}} \, A_{X}\bigg(mT_{\text{s}}, \frac{l}{NT_{\text{s}}}\bigg) \ist , \quad \mbox{for} \;\, m,l \in [ \lfloor N/2 \rfloor ] \ist\ist.
\end{align*}
Here, $X[n]$ is obtained from the continuous-time process $X(t)$ by regular sampling with period $T_{\text{s}}$, i.e., $X[n] = X(nT_{\text{s}})$ for $n \rmv\in\rmv [N]$.
Thus, $A_{X}[m,l]$ can be approximately calculated from the AF $A_{X}(\tau,\nu)$ of the continuous-time process 
$X(t)$.}

\vspace{1.5mm}

\item
Form the ``measurement matrix'' $\MM \in \mathbb{C}^{P \times \SS'}$ comprising those rows of 
$\mathbf{U} \in \mathbb{C}^{\SS' \times \SS'}$ (see \eqref{equ_def_U}) whose indices correspond to the TF lag positions $(m,l)$
chosen in Step 2.

\vspace{1.5mm}

\item
Compute an estimate $\hat{\mathbf{r}}$ of $\mathbf{r}$ from $\mathbf{a}^{(P)}\rmv$ by means of the Basis Pursuit \eqref{equ_def_bp_nonstat}, i.e.,
$\hat{\mathbf{r}} = \argmin_{\mathbf{r}' \rmv:\ist \MM\mathbf{r}' \ist=\ist \mathbf{a}^{(P)} } {\|\mathbf{r}'\|}_{1}$.

\vspace{1.5mm}

\item
From $\hat{\mathbf{r}}$, calculate $\hat{R}_{X,\text{CS}}[n,k] = \mathcal{L} \{ \hat{\mathbf{r}} \}[n,k]$ according to
\eqref{equ_MVU_nonstat_est_CS}. This step can be implemented efficiently by two successive 2D FFT operations.

\vspace{1.5mm}

\end{enumerate}

Based on the error bound \eqref{equ_err_bound_bp_nonstat} (with the index set $\mathcal{G}$ chosen as described below \eqref{equ_err_bound_bp_nonstat}), 
the compressive RS estimator $\hat{R}_{X,\text{CS}}[n,k]$
can be expected to be close to the noncompressive basic RS estimator $\hat{R}_{X,\text{MVU}}[n,k]$
in \eqref{equ_MVU_nonlin_nonstat_eaf} if the subsampled RS estimate $\hat{R}_{X,\text{MVU}}[\ist p \ist\Delta n, q \ist\Delta k]$
is approximately $\KK$-sparse. In Section \ref{sec_perf_nonstat}, we will derive an upper bound on the approximation error (MSE) 
that is formulated in terms of certain parameters depending on second-order statistics of the process $X[n]$, including the RS, $\bar{R}_X[n,k]$.

As previously mentioned in Section \ref{sec.intro}, from an algorithmic viewpoint, our compressive RS estimator $\hat{R}_{X,\text{CS}}[n,k]$ is
similar to the compressive TF representation proposed in \cite{BorgnatFlandrinICASSP08,BorgnatFlandrinTSP}.
However, the setting of \cite{BorgnatFlandrinICASSP08,BorgnatFlandrinTSP}
is that of deterministic TF signal analysis (improving the TF localization of the Wigner distribution),
rather than spectral estimation for nonstationary random processes.

\vspace{-2mm}

\subsection{An Improved Compressive RS Estimator}\label{sec_sub_nonstat_est_compr_impr}

The compressive RS estimator $\hat{R}_{X,\text{CS}}[n,k]$ in \eqref{equ_MVU_nonstat_est_CS} is related to the compressive EAF estimator
$\hat{A}_{X,\text{CS}}[m,l]$ defined 
\vspace*{-2mm}
as
\be
\hat{A}_{X,\text{CS}}[m,l] \,\triangleq \begin{cases}
& \hspace*{-3.5mm}\displaystyle\frac{1}{\SS'} \!\!\sum\limits_{p \in [\lendoppler]} \sum\limits_{q \in [\lendelay]} \! {( \hat{\mathbf{R}} )}_{p+1,\ist q+1} \, 
  e^{j 2 \pi \big( \frac{mq}{\lendelay} \ist-\ist \frac{lp}{\lendoppler} \big)},\\[5mm]
&\hspace*{3mm}(m,l) \rmv\in\rmv {\{- M,\ldots, -M + \Delta M \rmv-\rmv 1 \}}_{N} \\
& \hspace*{17mm}\times {\{  - L ,\ldots, -L + \Delta L \rmv-\rmv 1 \}}_{N}\\[0mm]
&\hspace*{-3.5mm}0 \,, \hspace*{2.9mm} \text{otherwise.}
\end{cases}
\label{equ_def_compr_EAF_estimator}
\ee
This relation is given by the 2D DFT
\begin{align}
\hat{R}_{X,\text{CS}}[n,k] &\eq \frac{1}{N} \!\sum_{m =-M}^{-M+\Delta M -1} \,\sum_{l = - L}^{-L+ \Delta L-1} \!\hat{A}_{X,\text{CS}}[m,l] \nonumber\\[0mm]
&\hspace*{36mm} \times e^{-j \frac{2 \pi}{N}(km - nl)} \ist.
\label{equ_accompanying_RS_est_AF_est_compressive}
\end{align}
Now, although the AF and EAF satisfy the following symmetry property:
\begin{subequations}
\begin{align}
A_{X}^*[-m,-l] \, e^{-j \frac{2 \pi}{N} ml} &\ist=\ist A_{X}[m,l] \label{equ_symmetry_discrete_AF}\\[1.5mm]
\bar{A}_{X}^{*}[-m,-l] \, e^{-j \frac{2 \pi}{N} ml} &\ist=\ist \bar{A}_{X}[m,l] \ist ,\label{equ_symmetry_discrete_EAF} 
\end{align}
\label{equ_symmetry}
\end{subequations}
$\!\rmv$the EAF estimator $\hat{A}_{X,\text{CS}}[m,l]$ does not exhibit this symmetry property in general.
This fact suggests the following simple symmetrization modification (postprocessing) of the EAF estimator:
\begin{equation}
\label{equ_modified_compressive_est_EAF}
\hat{A}_{X,\text{CS}}^{(\text{s})}[m,l] \,\triangleq\, \frac{1}{2} \ist \big[ \hat{A}_{X,\text{CS}}[m,l] + \hat{A}_{X,\text{CS}}^{*}[-m,-l] \,e^{-j \frac{2 \pi}{N} ml} \ist \big] \,.
\end{equation}
This, in turn, naturally leads to the definition of a ``symmetrized'' RS estimator $\hat{R}_{X,\text{CS}}^{(\text{s})}[n,k]$ via the 2D DFT transform in
\eqref{equ_accompanying_RS_est_AF_est_compressive}, i.e.,
\begin{align*}
&\hat{R}_{X,\text{CS}}^{(\text{s})}[n,k] \,\triangleq\, \frac{1}{N} \!\sum_{m =-M}^{-M+\Delta M -1} \,\sum_{l = - L}^{-L+ \Delta L-1} \!\hat{A}_{X,\text{CS}}^{(\text{s})}[m,l] \nonumber\\[-1mm]
&\rule{55mm}{0mm} \times e^{-j \frac{2 \pi}{N}(km - nl)} \ist.
\end{align*}
The following explicit expression of the symmetrized RS estimator is easily shown:
\begin{align}
&\hspace*{-1mm}\hat{R}_{X,\text{CS}}^{(\text{s})}[n,k]\nonumber\\[0mm]
&\rule{-1mm}{0mm}\eq \frac{1}{2NS'}  \!\rmv\sum_{m \in [\Delta M] } \,\sum_{l \in [\Delta L] }
  \Bigg[ \sum_{p \in [\lendoppler]} \sum_{q \in [\lendelay]} \rmv\!  \Big[  {( \hat{\mathbf{R}} )}_{p+1,\ist q+1}\nonumber\\[0mm]
&\rule{4mm}{0mm} +\ist {( \hat{\mathbf{R}} )}^{*}_{p +1,\ist q +1} \ist e^{-j \frac{2 \pi}{N} (m-M)(l-L)} \Big]
  \ist e^{j 2 \pi \big( \frac{(m-M)q}{\lendelay} \ist-\ist \frac{(l-L)p}{\lendoppler} \big)} \Bigg] \;\nonumber\\[0mm]
&\rule{40mm}{0mm} \times e^{-j \frac{2 \pi}{N}[k(m-M) - n(l-L)]} \, .
\label{equ_modified_compressive_RS_estimator_explicit_expr} 
\end{align}
This expression replaces \eqref{equ_MVU_nonstat_est_CS}. In Appendix A, we show that the MSE of the symmetrized 
RS estimator $\hat{R}_{X,\text{CS}}^{(\text{s})}[n,k]$ is always smaller than (or equal to) that of the original RS estimator $\hat{R}_{X,\text{CS}}[n,k]$, i.e.,
\be
{\rm E} \big\{ \big\| \hat{R}_{X,\text{CS}}^{(\text{s})} \rmv-\rmv \bar{R}_X \big\|^{2}_{2} \big\} \,\leq\, {\rm E} \big\{ \big\| \hat{R}_{X,\text{CS}} \rmv-\rmv \bar{R}_X \big\|^{2}_{2} \big\} \,.  \label{equ_lower_mse_modified_compr_RS_est}
\ee
Thus, the upper bound on the MSE of $\hat{R}_{X,\text{CS}}[n,k]$ to be derived in Section \ref{sec_perf_nonstat} also applies to the MSE 
\pagebreak 
of $\hat{R}_{X,\text{CS}}^{(\text{s})}[n,k]$.
To summarize, by using instead of the compressive RS estimator in \eqref{equ_MVU_nonstat_est_CS} the symmetrized compressive RS estimator
$\hat{R}_{X,\text{CS}}^{(\text{s})}[n,k]$ given by \eqref{equ_modified_compressive_RS_estimator_explicit_expr}, we can typically reduce the MSE.

Finally, we mention that in the case where no compression is performed, i.e., $S'/P \!=\!1$,
the basic (noncompressive) estimator $\hat{R}_{X,\text{MVU}}[n,k]$,
the compressive estimator $\hat{R}_{X,\text{CS}}[n,k]$,
and the symmetrized compressive estimator $\hat{R}_{X,\text{CS}}^{(\text{s})}[n,k]$ all coincide,
i.e., $\hat{R}_{X,\text{MVU}}[n,k] \equiv \hat{R}_{X,\text{CS}}[n,k] \equiv \hat{R}_{X,\text{CS}}^{(\text{s})}[n,k]$.
The equivalence $\hat{R}_{X,\text{MVU}}[n,k] \equiv \hat{R}_{X,\text{CS}}[n,k]$ can be verified by observing that for $S'/P =1$,
the measurement matrix $\mathbf{M}$ in \eqref{equ_meas-eq_1} coincides with the invertible matrix $\mathbf{U}$ in \eqref{equ_data_model_non}.
Therefore, the vectors $\mathbf{r} =  {\rm vec} \{ \mathbf{R} \}$ in \eqref{equ_data_model_non} and $\hat{\mathbf{r}} = {\rm vec} \{ \hat{\mathbf{R}} \}$ 
in \eqref{equ_def_bp_nonstat} coincide, and so do the corresponding RS estimators $\hat{R}_{X,\text{MVU}}[n,k]$ and $\hat{R}_{X,\text{CS}}[n,k]$ (cf.\ \eqref{equ_reconstr_operator_non} and \eqref{equ_MVU_nonstat_est_CS}).
To verify that $\hat{R}_{X,\text{CS}}^{(\text{s})}[n,k] \equiv \hat{R}_{X,\text{MVU}}[n,k]$ for $S'/P \!=\!1$,
note that because of \eqref{equ_S_A_inv} and \eqref{equ_def_compr_EAF_estimator},
$\hat{R}_{X,\text{CS}}[n,k] \equiv \hat{R}_{X,\text{MVU}}[n,k]$ is equivalent to $\hat{A}_{X,\text{CS}}[m,l] \equiv \hat{A}_{X,\text{MVU}}[m,l]$.
Since $\mathcal{A} = {\{- M,\ldots,M \}}_N\rmv \times {\{- L,\ldots, L \}}_N$ is symmetric,
it follows from expression \eqref{AF_weight_MVU} that the basic EAF estimator $\hat{A}_{X,\text{MVU}}[m,l]$
satisfies the symmetry relation \eqref{equ_symmetry}, and hence
$\hat{A}_{X,\text{MVU}}^{(\text{s})}[m,l] \triangleq \frac{1}{2} \ist \big[ \hat{A}_{X,\text{MVU}}[m,l] + \hat{A}_{X,\text{MVU}}^{*}[-m,-l] \,e^{-j \frac{2 \pi}{N} ml} \ist \big]
\!=\! \hat{A}_{X,\text{MVU}}[m,l]$.
Thus, for $S'/P$\linebreak 
$=\!1$, we have
$\hat{A}_{X,\text{CS}}^{(\text{s})}[m,l] = \hat{A}_{X,\text{MVU}}^{(\text{s})}[m,l] = \hat{A}_{X,\text{MVU}}[m,l]$, and in turn
$\hat{R}_{X,\text{CS}}^{(\text{s})}[n,k] = \hat{R}_{X,\text{MVU}}[n,k]$.

\section{MSE Bounds}\label{sec_perf_nonstat}

In this section, we derive an upper bound on the MSE of the proposed compressive RS estimator $\hat{R}_{X,\text{CS}}[n,k]$,
\begin{align*}
\varepsilon_{\text{CS}} &\,\triangleq\, {\rm E} \big\{ \big\| \hat{R}_{X,\text{CS}} \rmv-\rmv \bar{R}_{X} \big\|^{2}_{2} \big\}\\[1mm]
  &\eq \! \sum_{n,k \in [N]} \!{\rm E} \big\{ \big| \hat{R}_{X,\text{CS}}[n,k] \rmv-\rmv \bar{R}_{X}[n,k] \big|^{2} \big\} \,,
\end{align*}
under the assumption that $X[n]$ is a circularly symmetric complex Gaussian nonstationary process. We do \emph{not} assume that the EAF $\bar{A}_{X}[m,l]$ is exactly
supported on some periodic lag rectangle $\mathcal{A} = {\{- M,\ldots,M \}}_N\rmv \times {\{- L,\ldots, L \}}_N$
with $0 \le M < \lfloor N/2 \rfloor$ and $0 \le L < \lfloor N/2 \rfloor$.

\vspace{-1mm}

\subsection{Parameters}\label{sec_pars}

Our MSE bound depends on three parameters of the second-order statistics of the process $X[n]$, which will be defined first.

\vspace{1.5mm}

\begin{enumerate}

\item
As a measure (in the broad sense) of the sparsity of $\discreteRHS[n,k]$,
we define the \emph{TF sparsity moment}
\be
\label{eq_TF-spars-moment}
\sigma_{\rmv\rmv X}^{(w)} \triangleq\, \frac{1}{{\|\bar{R}_{X}\|}^{2}_{2} } \ist
  \Bigg[ \sum_{n,k \in [N]} \!w[n,k] \, \big|\bar{R}_{X}[n,k]\big| \Bigg]^{2} \rmv,
\ee
where $w[n,k] \ge 0$ is a suitably chosen weighting function and ${\|\bar{R}_{X}\|}^{2}_{2} \triangleq \sum_{n,k \in [N]} \big|\bar{R}_{X}[n,k]\big|^2$
(i.e., the norm is taken over one period of $\bar{R}_{X}[n,k]$).
In particular, for $w[n,k] \equiv 1$, $\sigma_{\rmv\rmv X}^{(w)} \rmv= {\|\bar{R}_{X}\|}^{2}_{1} / {\|\bar{R}_{X}\|}^{2}_{2}\ist$.

\vspace{2mm}

\item
For another way to measure the TF sparsity, let us first denote by
\be
\label{equ_mean_function_RS_MVU_def}
\widetilde{R}_{X,\text{MVU}}[n,k] \,\triangleq\, {\rm E} \big\{ \hat{R}_{X,\text{MVU}}[n,k] \big\}
\ee
the expectation of the basic RS estimator $\hat{R}_{X,\text{MVU}}[n,k]$ in \eqref{equ_MVU_nonlin_nonstat_eaf}.
It follows from \eqref{RD_convolve} that $\widetilde{R}_{X,\text{MVU}}[n,k]$ is a smoothed version of the RS, i.e.,
\begin{align}
&\widetilde{R}_{X,\text{MVU}}[n,k]\nonumber\\[1mm]
&\rule{2mm}{0mm}\eq \frac{1}{N} \rmv\!\sum_{n'\!,k' \in [N]} \rmv\!\Phi_{\text{MVU}}{[n\!-\!n'\!,k\!-\!k']} \,\ist {\rm E} \big\{ R_{X}[n'\!,k'] \big\}\nonumber\\[0mm]
&\rule{2mm}{0mm}\eq \frac{1}{N} \rmv\!\sum_{n'\!,k' \in [N]} \rmv\!\Phi_{\text{MVU}}{[n\!-\!n'\!,k\!-\!k']} \,\bar{R}_{X}[n'\!,k'] \,,
\label{equ_mean_function_RS_MVU}
\end{align}
where ${\rm E} \big\{ R_{X}[n,k] \big\} = \bar{R}_{X}[n,k]$ has been used in the last step. Due to \eqref{eq:phi}, the smoothing kernel is given by
\begin{align}
\Phi_{\text{MVU}}[n,k] &\,\triangleq\, \frac{1}{N} \!\sum_{m,l \in [N]} \!\phi_{\text{MVU}}[m,l] \, e^{-j\frac{2\pi}{N} (km-nl)} \nonumber\\[.5mm]
&\stackrel{\eqref{equ_def_indicator_function_MVU_EAF}}{\eq} \frac{1}{N} \!\sum_{m,l \in [N]} \!I_{\rmv \mathcal{A}}[m,l] \, e^{-j\frac{2\pi}{N} (km-nl)}
\label{eq:Phi_MVU}\\[.5mm]
&\eq \frac{1}{N} \!\sum_{m=-M}^M \sum_{m=-L}^L \!e^{-j\frac{2\pi}{N} (km-nl)} \ist. \nonumber
\end{align}
Because of the smoothing, the number of significantly nonzero values of $\widetilde{R}_{X,\text{MVU}}[n,k]$ may be larger than the
number of significantly nonzero values of the RS $\bar{R}_{X}[n,k]$. However, for an underspread process, the RS is inherently smooth, 
which implies that the smoothed RS is close to the RS. Therefore, for an underspread process with a small number of significantly nonzero RS values, we can expect that also
the smoothed RS consists of only a small number of significantly nonzero values.
Let us denote by $\mathcal{G}(\KK)$ the set of indices $(p,q) \in [\lendoppler] \!\times\! [\lendelay]$ of the $\KK$ largest (in magnitude)
values of the subsampled expected RS estimator, $\widetilde{R}_{X,\text{MVU}}[\ist p \ist\Delta n, q \ist\Delta k]$.
Let $\overline{\mathcal{G}(\KK)} \triangleq ([\lendoppler] \!\times\! [\lendelay]) \setminus \mathcal{G}(\KK)$,
and note that $\big|\overline{\mathcal{G}(\KK)} \big| = \SS' \!-\!\KK$. We then define the \emph{TF sparsity
profile}\footnote{We 
note that this definition is different from that
in \cite{jung-icassp09}.} 
\be
\tilde\sigma_{X}(\KK) \,\triangleq\, \frac{1}{{\|\bar{R}_{X}\|}^{2}_{2} } \ist \sum_{(p,q) \in\ist \overline{\mathcal{G}(\KK)}} \! \HH_ {p,q} \,,
\label{eq_sigma-tilde_nonstat_0}
\vspace{-2mm}
\ee
with
\begin{align}
\HH_ {p,q} &\,\triangleq\, {\rm E} \big\{ \big| { (\mathbf{R}) }_{p+1,\ist q+1} \big|^{2} \big\} \nonumber\\[1mm]
  &\stackrel{\eqref{equ_R_def}}{=} N^2 \, {\rm E} \big\{ \big| \hat{R}_{X,\text{MVU}}[\ist p \ist\Delta n, q \ist\Delta k] \big|^{2} \big\}\,.
\label{eq_sigma-tilde_nonstat} 
\end{align}
For later use, we note that
\be
\sum_{(p,q) \in\ist \overline{\mathcal{G}(\KK)} } \! \HH_ {p,q} \eq {\rm E} \big\{ \big\| \mathbf{r}^{\overline{\mathcal{G}(\KK)}} \big\|_2^2 \big\}
  \eq {\rm E} \big\{ \big\| \mathbf{r} \rmv-\rmv \mathbf{r}^{\mathcal{G}(K)} \big\|_2^2 \big\} \,,
\label{eq:r_rK_P}
\pagebreak 
\ee
where $\mathbf{r}^{ \overline{\mathcal{G}(\KK)}}$ (resp.\ $\mathbf{r}^{\mathcal{G}(K)}$) denotes the vector that is obtained from
$\mathbf{r} \equiv {\rm vec} \{ \mathbf{R} \}$ by zeroing all entries except the $\SS' \!-\! \KK$ (resp.\ $\KK$) entries
whose indices correspond to the
indices\footnote{For 
convenience, though with an abuse of notation, we denote by $\mathcal{G}(K)$ both a set of indices $k$ of ${(\mathbf{r})}_k$
and the corresponding set of 2D indices $(p,q)$ of ${( \mathbf{R} )}_{p+1,\ist q+1} = ({\rm unvec} \{ \mathbf{r} \})_{p+1,\ist q+1}$ or equivalently
of $\hat{R}_{X,\text{MVU}}[p \ist\Delta n, q \ist\Delta k]$. Thus, depending on the context, we will write $k \rmv\in \mathcal{G}(\KK)$ or
$(p,q) \rmv\in\rmv \mathcal{G}(\KK)$.} 
$(p,q) \rmv\in \overline{\mathcal{G}(\KK)}$ (resp.\ $(p,q) \rmv\in \mathcal{G}(\KK)$).

\vspace{2mm}

\item The ``TF correlation width'' of $X[n]$ can be measured by the \emph{EAF moment} \cite{GM-phd,MatzHla06}
\begin{equation}
\label{equ_nonstat_mom_ell2}
m_{X}^{(\psi)} \triangleq\, \frac{1}{{\|\bar{A}_{X}\|}^{2}_{2} } \, \sum_{m,l \in [N]} \!\psi [m,l] \, \big|\bar{A}_{X}[m,l]\big|^{2} \ist,
\end{equation}
where $\psi [m,l]$ is some weighting function that is generally zero or small at the origin $(0,0)$ and increases with increasing $|m|$ and $|l|$,
and ${\|\bar{A}_{X}\|}^{2}_{2} \triangleq \sum_{m,l \in [N]} \big|\bar{A}_{X}[m,l]\big|^2\rmv\rmv = {\|\bar{R}_{X}\|}^{2}_{2}$.
For an underspread process $X[n]$ and a reasonable choice of $\psi[m,l]$, $m_{X}^{(\psi)}$
is small ($\ll\rmv\rmv 1$).

\end{enumerate}

\subsection{Bound on the MSE of the Basic RS Estimator}\label{sec_nonstat_est_basic_msebound}

Our bound on the MSE $\varepsilon_{\text{CS}} = {\rm E} \big\{ \big\| \hat{R}_{X,\text{CS}} \rmv-\rmv \discreteRHS \big\|^{2}_{2} \big\}$ is
a combination of a bound on the MSE of the basic (noncompressive) RS estimator $\hat{R}_{X,\text{MVU}}[n,k]$
and a bound on the excess MSE introduced by the compression.
First, we derive the bound on the MSE of the basic RS estimator,
\[
\varepsilon \,\triangleq\, {\rm E} \big\{ \big\| \hat{R}_{X,\text{MVU}} \rmv-\rmv \discreteRHS \big\|^{2}_{2} \big\} \,.
\]
As in Section \ref{sec_nonstat_est_basic}, we use the decomposition
\be
\varepsilon \ist\ist=\ist\ist B^{2}\rmv + V \ist,
\label{equ_MSE-B-V}
\ee
with the squared bias term $B^2 = \big\| \ist{\rm E} \{\hat{R}_{X,\text{MVU}} \} \rmv-\rmv \bar{R}_{X} \big\|_2^{2}$
and the variance $V = {\rm E} \big\{ \big\| \hat{R}_{X,\text{MVU}} \rmv- {\rm E} \{ \hat{R}_{X,\text{MVU}} \} \big\|_2^{2} \big\}$.

\vspace{2mm}

\subsubsection{Bias}\label{sec_bias}
An expression of the bias term is obtained by setting $\phi[m,l] = \phi_{\text{MVU}}[m,l] = I_{\rmv \mathcal{A}}[m,l]$ in \eqref{equ_exact_bias_non_11}:
\begin{align*}
B^2 &\rmv\eq \!\sum_{m,l \in [N]} \rmv\big| (I_{\rmv \mathcal{A}}[m,l] \rmv-\rmv 1) \, \bar{A}_{X}[m,l] \big|^{2}\\[.5mm]
&\rmv\eq \!\sum_{m,l \in [N]} \! I_{\overline{\mathcal{A}} }[m,l] \, \big| \bar{A}_{X}[m,l] \big|^{2} ,
\end{align*}
where $I_{\overline{\mathcal{A}} }[m,l] = 1 - I_{\mathcal{A}}[m,l]$ is the indicator function of the complement $\overline{\mathcal{A}}$
of the effective EAF support region $\mathcal{A} = {\{- M,\ldots,M \}}_N\rmv \times {\{- L,\ldots, L \}}_N$, 
\vspace{-1.5mm}
i.e.,
\[
 I_{\overline{\mathcal{A}} }[m,l]\,=\ist \begin{cases} 1, & (m,l) \rmv\not\in\rmv \mathcal{A} \\[0mm]
   0, & \text{otherwise} \ist.
   \end{cases}
\]
We can write $B^2$ in terms of the EAF moment \eqref{equ_nonstat_mom_ell2} with weighting function $\psi [m,l] = I_{\overline{\mathcal{A}} }[m,l]$:
\begin{equation}
B^2
\rmv\eq\rmv {\| \bar{A}_{X} \|}^{2}_{2} \,\ist m_{X}^{(I_{\overline{\mathcal{A}} })}
  \rmv\eq {\|\bar{R}_{X}\|}^{2}_{2} \,\ist m_{X}^{(I_{\overline{\mathcal{A}} })}.
  \label{equ_exact_bias_non_MVU}
\pagebreak 
\end{equation}
Note that $m_{X}^{(I_{\overline{\mathcal{A}} })}\! =\rmv 0$, and thus $B^2 \!=\rmv 0$, if and only if the EAF $\bar{A}_{X}[m,l]$ is exactly
supported on $\mathcal{A}$.

\vspace{2mm}

\subsubsection{Variance}\label{sec_var}
In what follows, we will use the (scaled) discrete TF shift matrices $\mathbf{J}_{m,l}$ of size $N \!\times\! N$
whose action on $\mathbf{x} \!\in\! \mathbb{C}^N$ is given by
\[
{(\mathbf{J}_{m,l} \ist\ist \mathbf{x})}_{n+1} \eq \frac{1}{\sqrt{N}} \, ( \mathbf{x} )_{{(n-m)}_{N} \ist+\ist 1} \, e^{ j \frac{2 \pi}{N} l n} \ist , \quad n \in [N] \,,
\]
with ${(n)}_{N} \triangleq n \,\ist {\rm mod} \, N$. Basic properties of the family of TF shift matrices ${\{ \mathbf{J}_{m,l} \}}_{m,l \in [N]}$ are considered in Appendix B.
Using $\mathbf{J}_{m,l}$, $\hat{R}_{X,\text{MVU}}[n,k]$ can be written as a quadratic form in $\mathbf{x} = (X[0] \,\cdots\, X[N\!-\!1])^T$. In fact,
starting from \eqref{equ_MVU_nonlin_nonstat_eaf} and using \eqref{equ_AF_J}, we can develop $\hat{R}_{X,\text{MVU}}[n,k]$ as follows:
\begin{align*}
\hat{R}_{X,\text{MVU}}[n,k]  & \,\stackrel{\eqref{equ_MVU_nonlin_nonstat_eaf}}{\eq} \frac{1}{N} \!\sum_{m=-M}^{M} \sum_{l=-L}^{L} \rmv A_{X} [m,l] \, e^{-j \frac{2 \pi}{N}(km - nl)} \nonumber \\[1mm]
&\,\stackrel{\eqref{equ_AF_J}}{\eq} \frac{1}{\sqrt{N}} \!\sum_{m=-M}^{M} \sum_{l=-L}^{L}
  \rmv \langle \mathbf{x}\ist \mathbf{x}^{H} \!, \mathbf{J}_{m,l} \rangle \,e^{-j \frac{2 \pi}{N}(km - nl)} \nonumber \\[1mm]
& \,\eq
\Bigg\langle \mathbf{x}\ist \mathbf{x}^{H}  ,\, \frac{1}{\sqrt{N}} \!\sum_{m=-M}^{M} \sum_{l=-L}^{L} \rmv
  e^{j \frac{2 \pi}{N}(km - nl)} \, \mathbf{J}_{m,l} \Bigg\rangle \,. 
\end{align*}
Setting
\begin{equation}
\label{equ_def_C_n_k}
\mathbf{C}_{n,k} \,\triangleq\, \frac{1}{\sqrt{N}} \!\sum_{m=-M}^{M} \sum_{l=-L}^{L} \rmv e^{j \frac{2 \pi}{N}(km - nl)} \, \mathbf{J}_{m,l} \,,
\end{equation}
this becomes
\begin{align}
\hat{R}_{X,\text{MVU}}[n,k] &\eq \big\langle \mathbf{x}\ist \mathbf{x}^{H} \rmv, \mathbf{C}_{n,k} \big\rangle \nonumber\\[.8mm]
&\eq {\rm tr} \big \{  \mathbf{x}\ist \mathbf{x}^{H}  \mathbf{C}_{n,k}^{H} \big \}\nonumber\\[.8mm]
&\eq \mathbf{x}^{H} \mathbf{C}^{H}_{n,k} \ist\ist \mathbf{x} \, .
\label{equ_est_quad_form_non}
\end{align}
Note that the matrix $\mathbf{C}_{n,k}$ is not Hermitian in general.

Splitting $\hat{R}_{X,\text{MVU}}[n,k]$ into its real and imaginary parts, we have
\begin{align}
{\rm var} \big\{\hat{R}_{X,\text{MVU}}[n,k] \big\} &\eq {\rm var} \big\{ \Re \{ \hat{R}_{X,\text{MVU}}[n,k] \} \big\} \nonumber\\[.5mm]  
&\rule{8mm}{0mm}\ist+\ist  {\rm var} \big\{ \Im \{ \hat{R}_{X,\text{MVU}}[n,k] \} \big\} \,.
\label{equ_var_Re_Im}
\end{align}
It is easily shown that
\begin{align}
\Re \{ \hat{R}_{X,\text{MVU}}[n,k] \} &\eq \mathbf{x}^{H} \mathbf{C}^{(\text{R})}_{n,k} \ist \mathbf{x} \,, \label{equ_R_Re}\\[.5mm]  
\Im \{ \hat{R}_{X,\text{MVU}}[n,k] \} &\eq \mathbf{x}^{H}  \mathbf{C}^{(\text{I})}_{n,k} \ist \mathbf{x} \,, \label{equ_R_Im}
\end{align}
with the Hermitian matrices
\be
\mathbf{C}^{(\text{R})}_{n,k} \ist\triangleq\ist \frac{1}{2} \ist \big( \mathbf{C}^{H}_{n,k} + \mathbf{C}_{n,k} \big) \,, \quad
  \mathbf{C}^{(\text{I})}_{n,k} \ist\triangleq\ist \frac{1}{2j} \ist \big( \mathbf{C}^{H}_{n,k} \rmv\rmv-\rmv \mathbf{C}_{n,k} \big) \,.
\label{eq:C_Re-Im}
\ee
Inserting \eqref{equ_R_Re} and \eqref{equ_R_Im} into \eqref{equ_var_Re_Im} and using a standard result for the
variance of a Hermitian form of a circularly symmetric complex Gaussian random vector \cite{Tziritas87}, we obtain
\begin{align}
{\rm var} \big\{\hat{R}_{X,\text{MVU}}[n,k] \big\} &\eq {\rm tr} \big\{  \mathbf{C}^{(\text{R})}_{n,k} \mathbf{\Gamma}_{\!X}  \mathbf{C}^{(\text{R})}_{n,k} \mathbf{\Gamma}_{\!X} \rmv\big\} 
  \nonumber\\[.5mm]  
&\rule{10mm}{0mm}\ist+\ist {\rm tr} \big\{  \mathbf{C}^{(\text{I})}_{n,k}\mathbf{\Gamma}_{\!X}   \mathbf{C}^{(\text{I})}_{n,k}\mathbf{\Gamma}_{\!X} \rmv\big\} \,,
\label{equ_var_mean_non}\\[-7.5mm]
& \nonumber
\end{align}
with $\mathbf{\Gamma}_{\!X} \triangleq\, {\rm E} \ist\{ \mathbf{x} \mathbf{x}^H \}$.

Using this expression, we next derive an upper bound on $V = {\rm E} \big\{ \big\| \hat{R}_{X,\text{MVU}} \rmv- {\rm E} \{ \hat{R}_{X,\text{MVU}} \} \big\|_2^{2} \big\}$.
We have
\begin{align}
V &= \!\sum_{n,k \in [N]} \!{\rm E} \big\{ \big| \hat{R}_{X,\text{MVU}}[n,k] \rmv-\rmv {\rm E} \{ \hat{R}_{X,\text{MVU}}[n,k] \} \big|^{2} \big\} \nonumber \\[.5mm]
  &= \!\sum_{n,k \in [N]} \rmv\! {\rm var} \big\{\hat{R}_{X,\text{MVU}}[n,k] \big\} \nonumber \\[.5mm]
  &\!\stackrel{\eqref{equ_var_mean_non}}{=} \sum_{n,k \in [N]} \rmv\! 
    {\rm tr} \big\{ \mathbf{C}^{(\text{R})}_{n,k} \mathbf{\Gamma}_{\!X}  \mathbf{C}^{(\text{R})}_{n,k}\mathbf{\Gamma}_{\!X} \rmv\big\}\nonumber \\[-1mm]
  &\rule{20mm}{0mm} \ist+\rmv \sum_{n,k \in [N]} \rmv\! {\rm tr} \big\{ \mathbf{C}^{(\text{I})}_{n,k} \mathbf{\Gamma}_{\!X}  \mathbf{C}^{(\text{I})}_{n,k} \mathbf{\Gamma}_{\!X} \rmv\big\} \,.
  \label{equ_variance_nonstat_equ_s_x_mvu} 
\end{align}
It is then shown in Appendix C that
\be
V = \!\sum_{m,l \in [N]} \rmv \big| \bar{A}_{X}[m,l] \big|^{2} \, \chi[m,l] \,,
\label{equ_var_A_non}
\vspace{-2mm}
\ee
with
\begin{align}
\chi[m,l] &\eq \frac{1}{N} \!\sum_{m'\!,l' \in [N]} \!I_{\mathcal{A}}[m'\!,l'] \, e^{j\frac{2 \pi}{N} (lm' - ml')}\label{equ_chi_def_1_0} \\[.5mm]
  &\eq \frac{1}{N} \! \sum_{m'=-M}^{M} \sum_{l'=-L}^{L} \! e^{j\frac{2 \pi}{N} (lm' - ml')}\,.
\label{equ_chi_def_1}
\end{align}
We can bound the magnitude of $\chi[m,l]$ according to
\begin{align*}
|\chi[m,l]| &\,\leq\, \frac{1}{N} \! \sum_{m'=-M}^{M} \sum_{l'=-L}^{L} \! \big| e^{j\frac{2 \pi}{N} (lm' - ml')} \big|\nonumber \\[.5mm]
  &\eq \frac{1}{N} \, (2M+1) (2L+1) \nonumber \\[1.5mm]
  &\eq \frac{\SS}{N} \,\ist .\\[-5mm]
\end{align*}
Combining with \eqref{equ_var_A_non} leads to the following bound on $V$:
\begin{align}
V &\,\leq \rmv \sum_{m,l \in [N]} \rmv \big| \bar{A}_{X}[m,l] \big|^{2} \, \big| \chi[m,l] \big|\nonumber \\[.5mm]
  &\,\leq\, \frac{\SS }{N} \! \sum_{m,l \in [N]} \rmv\big| \bar{A}_{X}[m,l] \big|^{2}\nonumber \\[.5mm]
  &\stackrel{\eqref{equ_inv_fourier_eaf_rhs}}{\eq} \frac{\SS }{N} \, {\|\bar{R}_{X}\|}^{2}_{2}  \,.
\label{equ_bound_var_non_11}
\end{align}

\vspace{1mm}

\subsubsection{MSE}\label{sec_nonstat_est_basic_msebound_mse}
Finally, the desired bound on the MSE $\varepsilon = {\rm E} \big\{ \big\| \hat{R}_{X,\text{MVU}} \rmv-\rmv \discreteRHS \big\|^{2}_{2} \big\}$
is obtained by inserting \eqref{equ_exact_bias_non_MVU} and \eqref{equ_bound_var_non_11} into the expansion \eqref{equ_MSE-B-V}:
\begin{align}
\varepsilon &\eq B^{2}\rmv + V\nonumber \\[.5mm]
  &\,\leq\, {\|\bar{R}_{X}\|}^{2}_{2} \,\ist m_{X}^{(I_{\overline{\mathcal{A}} })} + \frac{\SS }{N} \, {\|\bar{R}_{X}\|}^{2}_{2}\nonumber \\[.5mm]
  &\eq {\|\bar{R}_{X}\|}^{2}_{2} \, \bigg( m_{X}^{(I_{\overline{\mathcal{A}} })} \rmv+\ist \frac{\SS}{N} \bigg) \,.
\label{bound_basic_nonstat_nonlin}\\[-6.5mm]
&\nonumber
\end{align}
This bound is small if $X[n]$ is underspread, i.e., if $m_{X}^{(I_{\overline{\mathcal{A}} })} \!\ll \rmv 1$ and $\SS \rmv\ll\! N$.


\subsection{Bound on the Excess MSE Due to Compression}\label{sec_nonstat_est_excess_msebound}

\vspace{1mm}

The excess MSE caused by the compression is given by
\[
\Delta \varepsilon \,\triangleq\, {\rm E} \big\{ \big\| \hat{R}_{X,\text{CS}} \rmv-\rmv \hat{R}_{X,\text{MVU}} \big\|^{2}_{2} \big\} \,.
\]
Because of the Fourier transform relations \eqref{equ_reconstr_operator_non} and \eqref{equ_MVU_nonstat_est_CS}, we have
\be
\Delta \varepsilon \eq \frac{1}{\SS'}  \, {\rm E} \big\{ {\|\hat{\mathbf{r}} \rmv-\rmv \mathbf{r} \|}_2^{2} \big\} \,.
\label{equ_Deltaeps_r}
\ee
As in Section \ref{sec_meas-eq}, let $K$ denote a nominal sparsity degree that is chosen according to
our intuition about the approximate sparsity of $\hat{R}_{X,\text{MVU}}[\ist p \ist\Delta n, q \ist\Delta k]$
and, equivalently, $\mathbf{r}$. We assume that the number $P$ of randomly selected AF samples is sufficiently large so that \eqref{equ_err_bound_bp_nonstat}
is satisfied, i.e.,
\be
{\| \hat{\mathbf{r}} \rmv-\rmv \mathbf{r} \|}^2_{2} \,\leq\,  \frac{D^2}{\KK} \, {\| \mathbf{r} \rmv-\rmv \mathbf{r}^{\mathcal{G}}\|}_{1}^2 \,,
\label{equ_err_bound_bp_nonstat_recall}
\ee
for any index set $\mathcal{G}$ of size $|\mathcal{G}| \!=\! K$. (A sufficient condition is \eqref{equ_P_cond_nonstat}.) 
An intuitively reasonable choice of $K$ and $\mathcal{G}$ can be based on the smoothed
RS $\widetilde{R}_{X,\text{MVU}}[n,k] = {\rm E} \big\{ \hat{R}_{X,\text{MVU}}[n,k] \big\}$ in \eqref{equ_mean_function_RS_MVU_def},
\eqref{equ_mean_function_RS_MVU}: we choose
$K$ as the number of significantly nonzero values $\widetilde{R}_{X,\text{MVU}}[p \Delta n,q \Delta k]$,
and $\mathcal{G} = \mathcal{G}(\KK)$ of size $K$ as the set of those indices of $\mathbf{r}$
that correspond to these significant values---equivalently, to the $K$ largest (in magnitude) values $\widetilde{R}_{X,\text{MVU}}[p \Delta n,q \Delta k]$.
Thus, $\mathbf{r}^{\mathcal{G}(\KK)}$ comprises those $K$ values
$\hat{R}_{X,\text{MVU}}[p \Delta n,q \Delta k]$ for which the corresponding values
$\widetilde{R}_{X,\text{MVU}}[p \Delta n,q \Delta k]$ are largest (in magnitude).

Based on this choice, we will now derive an approximate upper bound on the excess MSE $\Delta\varepsilon$.
Inserting \eqref{equ_err_bound_bp_nonstat_recall} into \eqref{equ_Deltaeps_r}, we obtain
\be
\Delta\varepsilon \,\leq\, \frac{D^2}{\SS'  \KK} \, {\rm E} \big\{ \big\| \mathbf{r} \rmv-\rmv \mathbf{r}^{\mathcal{G}(K)} \big\|_{1}^2 \big\} \,.
\label{eq:bound_excess_nonstat_nonlin_0}
\ee
Using the
inequality\footnote{Indeed, 
the $\ell_{1}$-norm of an arbitrary vector $\mathbf{z}$ can be expressed as
${\| \mathbf{z} \|}_{1} = \mathbf{z}^{H}\mathbf{a}(\mathbf{z})$, where $\mathbf{a}(\mathbf{z})$
is given elementwise by ${(\mathbf{a}(\mathbf{z}))}_{k} \triangleq z_{k}/|z_{k}|$ for $z_{k} \rmv\not=\rmv 0$ and
${(\mathbf{a}(\mathbf{z}))}_{k} \triangleq 0$ for $z_{k} \rmv=\rmv 0$. Clearly, ${\|\mathbf{a}(\mathbf{z}) \|}^{2}_{2} = {\| \mathbf{z} \|}_{0}$,
and thus ${\| \mathbf{z} \|}_{1}^{2}  = ( \mathbf{z}^{H} \mathbf{a}(\mathbf{z}) )^{2} \leq {\| \mathbf{z} \|}^{2}_{2} \, {\| \mathbf{a}(\mathbf{z}) \|}^{2}_{2}
= {\| \mathbf{z} \|}^{2}_{2} \, {\| \mathbf{z} \|}_{0}$, where the Cauchy-Schwarz inequality has been
used.} 
${\| \cdot \|}_{1}^{2} \leq {\| \cdot \|}_{0} \ist\ist {\| \cdot \|}_{2}^{2}$, we have
$\big\| \mathbf{r} \!-\! \mathbf{r}^{\mathcal{G}(K)} \big\|_{1}^2 \leq
\big\| \mathbf{r} \!-\! \mathbf{r}^{\mathcal{G}(K)} \big\|_{0} \ist\ist \big\| \mathbf{r} \!-\! \mathbf{r}^{\mathcal{G}(K)} \big\|_{2}^{2}
\leq (\SS'  \!-\! \KK) \ist\ist \big\| \mathbf{r} \!-\! \mathbf{r}^{\mathcal{G}(K)} \big\|_{2}^{2}$,
and thus \eqref{eq:bound_excess_nonstat_nonlin_0} becomes further
\begin{align}
\Delta\varepsilon &\,\leq\, \frac{(\SS'  \!-\! \KK) \ist\ist D^2}{\SS'  \KK} \, {\rm E} \big\{ \big\| \mathbf{r} \rmv-\rmv \mathbf{r}^{\mathcal{G}(K)} \big\|_2^2 \big\}\nonumber \\[1mm]
  &\ist\stackrel{\eqref{eq:r_rK_P}}{\eq}\ist \frac{(\SS'  \!-\! \KK) \ist\ist D^2}{\SS'  \KK} \!\!
                           \sum_{(p,q) \in\ist \overline{\mathcal{G}(\KK)} } \! \HH_ {p,q} \label{eq:bound_excess_nonstat_nonlin_1} \\[.5mm]
  &\ist\stackrel{\eqref{eq_sigma-tilde_nonstat_0}}{\eq}\ist \frac{(\SS'  \!-\! \KK) \ist\ist D^2}{\SS' \KK} \,{\|\bar{R}_{X}\|}^{2}_{2} \,\, \tilde{\sigma}_{X}(\KK) \,.
\label{eq:bound_excess_nonstat_nonlin}
\end{align}

In what follows, we will derive an approximate expression of $\HH_ {p,q} = {\rm E} \big\{ \big| { (\mathbf{R}) }_{p+1,\ist q+1} \big|^{2} \big\}$
in terms of $\bar{R}_{X}[n,k]$; this expression will show under which condition $\tilde{\sigma}_{X}(\KK) \propto \sum_{(p,q) \in\ist \overline{\mathcal{G}(\KK)} } \HH_ {p,q}$ is small.
We have
\begin{align}
\HH_ {p,q} &\eq {\rm E} \big\{ \big| { (\mathbf{R}) }_{p+1,\ist q+1} \big|^{2} \big\} \nonumber\\[.7mm]
&\eq {\rm var} \big\{ { (\mathbf{R}) }_{p+1,\ist q+1} \big\} \ist+\ist\ist \big| {\rm E} \big\{ { (\mathbf{R}) }_{p+1,\ist q+1} \big\} \big|^{2} \nonumber \\[1.5mm]
&\eq {\rm var} \big\{ \Re\big\{{(\mathbf{R}) }_{p+1,\ist q+1}\big\}  \big\}
  \ist+\ist {\rm var} \big\{ \Im\big\{{(\mathbf{R}) }_{p+1,\ist q+1}\big\}  \big\}\nonumber\\[.5mm]
&\rule{37mm}{0mm}\ist+\ist\ist \big| {\rm E} \big\{ { (\mathbf{R}) }_{p+1,\ist q+1} \big\} \big|^{2} .
  \label{eq:h_sum} 
\end{align}
Using \eqref{equ_S_A} and \eqref{equ_AF_J}, we can express ${ (\mathbf{R}) }_{p+1,\ist q+1}$ as a quadratic form:
\begin{align}
{ (\mathbf{R}) }_{p+1,\ist q+1}  & \rmv\stackrel{\eqref{equ_S_A}}{\eq} \sum_{m=-M}^{M} \sum_{l=-L}^{L} \! A_{X}[m,l] \,
   e^{-j 2 \pi \big(\frac{qm} {\lendelay} \ist-\ist \frac{pl}{\lendoppler} \big)} \nonumber \\[1mm]
&\rmv\stackrel{\eqref{equ_AF_J}}{\eq} \ist\sqrt{N} \!\rmv \sum_{m=-M}^{M} \sum_{l=-L}^{L}
  \rmv \langle \mathbf{x}\ist \mathbf{x}^{H} \!, \mathbf{J}_{m,l} \rangle \,e^{-j 2 \pi \big(\frac{qm} {\lendelay} \ist-\ist \frac{pl}{\lendoppler} \big)} \nonumber \\[1mm]
& \eq \langle \mathbf{x} \ist \mathbf{x}^{H} \!, \mathbf{T}_{\rmv p,q}\rangle \nonumber \\[1mm]
&\eq {\rm tr} \big\{\mathbf{x} \ist \mathbf{x}^{H} \mathbf{T}^{H}_{\rmv p,q} \big\} \nonumber \\[1mm]
&\eq \mathbf{x}^{H} \mathbf{T}^{H}_{\rmv p,q} \ist \mathbf{x}
\label{equ_R_T} \,,\\[-6.5mm]
&\nonumber
\end{align}
with
\vspace{-1mm}
\be
\label{equ_def_T_p_q}
\mathbf{T}_{\rmv p,q} \ist\triangleq\ist \sqrt{N} \!\rmv \sum_{m=-M}^{M} \sum_{l=-L}^{L} \! e^{j 2 \pi \big(\frac{qm} {\lendelay} \ist-\ist \frac{pl}{\lendoppler} \big)} \,\mathbf{J}_{m,l} \,.
\ee
Note that the matrix $\mathbf{T}_{\rmv p,q}$ is not Hermitian in general.
Inserting \eqref{equ_R_T} into \eqref{eq:h_sum} then yields
\[
\HH_ {p,q} =\ist {\rm var} \big\{ \mathbf{x}^{H} \mathbf{T}^{(\text{R})}_{\rmv p,q} \ist \mathbf{x} \big\}
  \ist+\ist {\rm var} \big\{ \mathbf{x}^{H} \mathbf{T}^{(\text{I})}_{\rmv p,q} \ist \mathbf{x} \big\}
  \ist+\ist\ist \big| {\rm E} \big\{ \mathbf{x}^{H} \mathbf{T}^{H}_{\rmv p,q} \ist \mathbf{x} \big\} \big|^{2} ,
\]
with the Hermitian matrices
\be
\mathbf{T}^{(\text{R})}_{\rmv p,q} \ist\ist\triangleq\ist\ist \frac{1}{2} \ist \big( \mathbf{T}_{\rmv p,q}^{H} + \mathbf{T}_{\rmv p,q} \big) \,, \quad\!
\mathbf{T}^{(\text{I})}_{\rmv p,q} \ist\ist\triangleq\ist\ist \frac{1}{2j} \ist \big( \mathbf{T}^{H}_{\rmv p,q} \rmv-\rmv \mathbf{T}_{\rmv p,q} \big) \,.
\label{equ_def_T_p_q_RI}
\ee
Using standard results for the variance and mean of a Hermitian form of a
circularly symmetric complex Gaussian vector \cite{Tziritas87}, we obtain further
\begin{align}
\hspace*{-1mm}\HH_ {p,q} &=\ist {\rm tr} \big\{\mathbf{T}^{(\text{R})}_{\rmv p,q} \mathbf{\Gamma}_{\!X} \rmv \mathbf{T}^{(\text{R})}_{\rmv p,q}  \mathbf{\Gamma}_{\!X} \big\}
    \ist+\ist {\rm tr} \big\{ \mathbf{T}^{(\text{I})}_{\rmv p,q} \mathbf{\Gamma}_{\!X} \rmv \mathbf{T}^{(\text{I})}_{\rmv p,q}  \mathbf{\Gamma}_{\!X} \big\}\nonumber\\[.8mm]
&\rule{46mm}{0mm}\ist+\ist \big| {\rm tr} \big\{ \mathbf{\Gamma}_{\!X} \rmv \mathbf{T}^{H}_{\rmv p,q} \big\} \big|^{2} .
\label{equ_power_non_exact}
\end{align}

There does not seem to exist a simple closed-form expression of \eqref{equ_power_non_exact} in terms of the EAF $\bar{A}_{X}[m,l]$ or the RS $\bar{R}_{X}[n,k]$.
However, under the assumption that the process $X[n]$ is underspread and the effective EAF support dimensions $M$, $L$ (cf.\ \eqref{equ_def_mathcal_S}) 
are accordingly chosen to be small, the following approximation is derived in Appendix D:
\begin{align}
\hspace*{-2mm}\HH_ {p,q} &\approx\ist N \!\! \sum_{n,k \in [N]} \! \big| \bar{R}_{X}[n,k] \, \Phi_{\text{MVU}}[n\!-\!p \ist\Delta n,k\!-\!q \ist\Delta k] \big|^{2}\nonumber\\[-.2mm]
  &\rule{5mm}{0mm}\ist+\,  \ist\ist \Bigg| \sum_{n,k \in [N]} \!\! \bar{R}_{X}[n,k] \, \Phi_{\text{MVU}}[n\!-\!p \ist\Delta n,k\!-\!q \ist\Delta k] \ist\ist \Bigg|^{2} \rmv,
\label{equ_approx_P_p_q_sum_quad} 
\end{align}
where, as before, $\Delta n = N/\lendoppler$ and $\Delta k = N/\lendelay$.
Comparing with \eqref{equ_mean_function_RS_MVU} and noting that $\Phi_{\text{MVU}}[-n,-k] = \Phi_{\text{MVU}}[n,k]$,
it is seen that the second term on the right hand side of \eqref{equ_approx_P_p_q_sum_quad} is
$N^2 \ist\ist \big| \widetilde{R}_{X,\text{MVU}}[p \ist\Delta n, q \ist\Delta k] \big|^{2}\rmv$.
Using the inequality ${\|\cdot\|}_2^2 \le {\|\cdot\|}_1^2$ \cite{golub96} to bound the first term on the right-hand side of \eqref{equ_approx_P_p_q_sum_quad},
and using a trivial upper bound on the second term, we obtain
\begin{align}
\HH_ {p,q} &\ist\lessapprox\ist N \ist \Bigg[ \sum_{n,k \in [N]} \!\big| \bar{R}_{X}[n,k] \, \Phi_{\text{MVU}}[n\!-\!p \ist\Delta n,k\!-\!q \ist\Delta k] \big| \Bigg]^{2}\nonumber\\[-.5mm]
  &\rule{8mm}{0mm}\!+\ist  \ist \Bigg[ \sum_{n,k \in [N]} \!\big|\bar{R}_{X}[n,k] \, \Phi_{\text{MVU}}[n\!-\!p \ist\Delta n,k\!-\!q \ist\Delta k] \big| \Bigg]^{2} \nonumber \\[0mm]
     &\ist=\ist \prefactor \ist \Bigg[ \sum_{n,k \in [N]} \!\big| \bar{R}_{X}[n,k] \, \Phi_{\text{MVU}}[n\!-\!p \ist\Delta n,k\!-\!q \ist\Delta k] \big| \Bigg]^{2}\rmv\rmv.
     \label{equ_approx_power_non_4}
\end{align}
Here, $\sum_{n,k \in [N]} \!\big| \bar{R}_{X}[n,k] \, \Phi_{\text{MVU}}[n\!-\!p \ist\Delta n,k\!-\!q \ist\Delta k] \big|$
can be interpreted as a local average of the RS modulus $\big| \discreteRHS[n,k] \big|$
about the TF point $(p \ist\Delta n, q \ist\Delta k)$. Thus, the (approximate) upper bound \eqref{equ_approx_power_non_4} shows that
$\HH_ {p,q}$ is small if $\discreteRHS[n,k]$ is small within a neighborhood of $(p \ist\Delta n, q \ist\Delta k)$ or, said differently, if $(p \ist\Delta n, q \ist\Delta k)$ is located outside
a broadened version of the effective support of $\discreteRHS[n,k]$. The broadening is stronger for a larger spread of $\Phi_{\text{MVU}}[n,k]$.
According to \eqref{eq:Phi_MVU}, $\Phi_{\text{MVU}}[n,k]$ is the 2D DFT of the indicator function $I_{\rmv \mathcal{A}}[m,l]$,
and thus the broadening depends on the size of the effective EAF support $\mathcal{A}$; it will be stronger if $\mathcal{A}$ is smaller,
i.e., if the process $X[n]$ is more underspread. Since a stronger broadening implies a poorer sparsity, this demonstrates an intrinsic tradeoff between the
underspreadness and the TF sparsity of $X[n]$: better underspreadness implies a smaller effective EAF support $\mathcal{A}$,
whereas better TF sparsity requires a larger $\mathcal{A}$.

With this ``broadening'' interpretation in mind, we reconsider $\tilde{\sigma}_{X}(\KK) \propto \sum_{(p,q) \in\ist \overline{\mathcal{G}(\KK)} } \HH_ {p,q}$ in the bound
\eqref{eq:bound_excess_nonstat_nonlin}. Recall that $\mathcal{G}(\KK)$ was defined as the set of those indices of $\mathbf{r}$
such that the corresponding values $\widetilde{R}_{X,\text{MVU}}[p \ist\Delta n, q \ist\Delta k]$ are the $K$ largest (in magnitude). Therefore, a small $\tilde{\sigma}_{X}(\KK)$
requires that $\KK$ is chosen such that $\KK \ist \Delta n \ist\ist \Delta k$ is approximately equal to the area of the broadened effective support of $\discreteRHS[n,k]$,
because then $\sum_{n,k \in [N]} \!\big| \bar{R}_{X}[n,k] \, \Phi_{\text{MVU}}[n\!-\!p \ist\Delta n,k\!-\!q \ist\Delta k] \big| \approx 0$
for $(p,q) \rmv\in\rmv \overline{\mathcal{G}(\KK)}$ and thus, using \eqref{equ_approx_power_non_4}, 
$\tilde{\sigma}_{X}(\KK) \propto \sum_{(p,q) \in\ist \overline{\mathcal{G}(\KK)} } \HH_ {p,q} \approx 0$.

\vspace{.5mm}

Using \eqref{equ_approx_power_non_4}, we can upper-bound the MSE bound in \eqref{eq:bound_excess_nonstat_nonlin_1},
$\Delta\varepsilon \leq \frac{(\SS'  - \KK) \ist\ist D^2}{\SS'  \KK} \sum_{(p,q) \in\ist \overline{\mathcal{G}(\KK)} } \HH_ {p,q}$,
which results in a simpler (but generally looser) upper bound. Indeed, we have
\begin{align}
\sum_{(p,q) \in\ist \overline{\mathcal{G}(\KK)} } \! \HH_ {p,q} & \ist\stackrel{\eqref{equ_approx_power_non_4}}{\ist\lessapprox}\ist
 \prefactor \!\!\sum_{(p,q) \in\ist \overline{\mathcal{G}(\KK)} } \Bigg[ \sum_{n,k \in [N]} \! \big| \bar{R}_{X}[n,k] \nonumber\\[-1.5mm]
  &\rule{21mm}{0mm} \times \Phi_{\text{MVU}}[n\!-\!p \ist\Delta n,k\!-\!q \ist\Delta k] \big| \Bigg]^{2} \nonumber \\[-.5mm]
& \ist\ist\stackrel{(*)}{\leq}\, \prefactor \ist \Bigg[ \sum_{(p,q) \in\ist \overline{\mathcal{G}(\KK)} } \, \Bigg| \!\sum_{n,k \in [N]} \! \big| \bar{R}_{X}[n,k] \nonumber\\[-1.5mm]
  &\rule{21mm}{0mm} \times \Phi_{\text{MVU}}[n\!-\!p \ist\Delta n,k\!-\!q \ist\Delta k] \big| \ist\ist \Bigg| \, \Bigg]^{2} \nonumber \\[-.5mm]
& \ist\eq\, \prefactor \ist \Bigg[ \sum_{(p,q) \in\ist \overline{\mathcal{G}(\KK)} } \ist \sum_{n,k \in [N]} \! \big| \bar{R}_{X}[n,k]\big| \nonumber\\[-1.5mm]
  &\rule{21mm}{0mm} \times \big| \Phi_{\text{MVU}}[n\!-\!p \ist\Delta n,k\!-\!q \ist\Delta k] \big| \ist \Bigg]^{2} \nonumber \\[-.5mm]
& \ist\eq\, \prefactor \ist \Bigg[ \sum_{n,k \in [N]} \! \big| \bar{R}_{X}[n,k]\big|  \nonumber\\[-1.5mm]
  &\rule{7.5mm}{0mm} \times\!\! \sum_{(p,q) \in\ist \overline{\mathcal{G}(\KK)} }\big| \Phi_{\text{MVU}}[n\!-\!p \ist\Delta n,k\!-\!q \ist\Delta k] \big| \ist \Bigg]^{2} \nonumber \\[1mm]
& \ist\eq\, \prefactor \ist \Bigg[ \sum_{n,k \in [N]} \! \big| \bar{R}_{X}[n,k]\big| \, w_{\Phi}[n,k] \ist \Bigg]^{2} \rmv ,
\label{eq:sum_P_bound_0}
\end{align}
where ${\|\cdot\|}_{2}^{2} \leq {\|\cdot \|}_{1}^{2}$ was used in the step labeled with $(*)$ and
\be
w_{\Phi}[n,k] \,\triangleq\! \sum_{(p,q) \in\ist \overline{\mathcal{G}(\KK)} } \! \big| \Phi_{\text{MVU}}[n\!-\!p \ist\Delta n,k\!-\!q \ist\Delta k] \big| \,.
\label{eq:w_s_def}
\vspace{-1mm}
\ee
Comparing with the definition of the TF sparsity moment $\sigma_{\rmv\rmv X}^{(w)}$ in \eqref{eq_TF-spars-moment},
it is seen that the approximate bound \eqref{eq:sum_P_bound_0} can be written as
\be
\sum_{(p,q) \in\ist \overline{\mathcal{G}(\KK)} } \! \HH_ {p,q} \ist\,\lessapprox\,\ist
  \prefactor \, {\|\bar{R}_{X}\|}^{2}_{2} \,\ist\ist \sigma^{(w_{\Phi})}_{\rmv\rmv X} \ist.
\label{eq:sum_P_bound}
\ee
Inserting \eqref{eq:sum_P_bound} into \eqref{eq:bound_excess_nonstat_nonlin_1} then gives the approximate MSE bound
\begin{equation}
\Delta\varepsilon \,\lessapprox\, \frac{(\SS'  \!-\! \KK) \ist\ist D^2}{\SS' \KK} \ist \prefactor \, {\|\bar{R}_{X}\|}^{2}_{2} \,\ist\ist \sigma^{(w_{\Phi})}_{\rmv\rmv X} \ist.
\label{eq:bound_excess_nonstat_nonlin_2}
\end{equation}
A small excess MSE $\Delta\varepsilon$ can be achieved if the TF sparsity moment
$\sigma^{(w_{\Phi})}_{\rmv\rmv X}\! \propto \big[ \sum_{n,k \in [N]} \big| \bar{R}_{X}[n,k]\big| \, w_{\Phi}[n,k] \big]^2$ is small. This, in turn, is the case
if the RS $\bar{R}_{X}[n,k]$ is negligible within the effective support of the TF weighting function $w_{\Phi}[n,k]$. Due to
\eqref{eq:w_s_def}, the size of the effective support of $w_{\Phi}[n,k]$, which is concentrated around the points
$\big\{ (p \ist\Delta n, q \ist\Delta k) \big\}_{(p,q) \in\ist \overline{\mathcal{G}(\KK)}}$,  is not larger than $\SS' \!-\!K$ times the size of the effective support of
$\Phi_{\text{MVU}}[n,k]$ (recall that $\big| \overline{\mathcal{G}(\KK)} \big| = \SS' \!-\! \KK$).
Because of the DFT expression \eqref{eq:Phi_MVU} and the fact that $\big| {[N]}^2 \rmv\cap \mathcal{A} \big| = \SS$ (see \eqref{equ_def_mathcal_S}),
the size of the effective support of $\Phi_{\text{MVU}}[n,k]$ within one period ${[N]}^2$ can be estimated by $N^{2}\rmv\rmv /S$.
Thus, for a small TF sparsity moment $\sigma^{(w_{\Phi})}_{\rmv\rmv X}\rmv$, the RS $\bar{R}_{X}[n,k]$ should effectively vanish (within $[N]^2$) on a region of size at least
$(\SS' \!-\rmv K) \ist N^{2}\rmv\rmv /S \stackrel{\eqref{equ_ineq_S_S_prime}}{\geq} (\SS -\rmv K) \ist N^{2}\rmv\rmv /S = N^{2} - K N^{2}\rmv\rmv /S$.
Since typically $S' \rmv\approx S$, implying that $(\SS' \!-\rmv K) \ist N^{2}\rmv\rmv /S \approx N^{2} - K N^{2}\rmv\rmv /S$, it follows
that the size of the effective support (within $[N]^2$) of the RS $\bar{R}_{X}[n,k]$ should not be larger than $K N^{2}\rmv\rmv /S$.
Note that $K$ was defined as our prior intuition about the number of significantly nonzero values $\widetilde{R}_{X,\text{MVU}}[p \Delta n,q \Delta k]$;
furthermore, $N^{2}\rmv\rmv /S$ is related to the TF undersampling in $\widetilde{R}_{X,\text{MVU}}[p \Delta n,q \Delta k]$
because (for $S' \rmv\approx S$) it is approximately equal to the ratio of the number of samples $\big\{ \widetilde{R}_{X,\text{MVU}}[n,k] \big\}_{n,k\in [N]}$ 
(which is $N^2$) to the number of samples $\big\{ \widetilde{R}_{X,\text{MVU}}[p \Delta n,q \Delta k] \big\}_{p\in [\lendoppler],q \in [\lendelay]}$ (which is $S'$).

\vspace{-1mm}

\subsection{Combining the Two MSE Bounds}\label{sec_nonstat_est_comb_msebound}

We will now combine the bound \eqref{bound_basic_nonstat_nonlin} on $\varepsilon =  {\rm E} \big\{ \big\| \hat{R}_{X,\text{MVU}} \rmv-\rmv \discreteRHS \big\|^{2}_{2} \big\}$
and the bound \eqref{eq:bound_excess_nonstat_nonlin} or \eqref{eq:bound_excess_nonstat_nonlin_2} on
$\Delta \varepsilon = {\rm E} \big\{ \big\| \hat{R}_{X,\text{CS}} \rmv-\rmv \hat{R}_{X,\text{MVU}} \big\|^{2}_{2} \big\}$
into a bound on the MSE $\varepsilon_{\text{CS}} = {\rm E} \big\{ \big\| \hat{R}_{X,\text{CS}} \rmv-\rmv \bar{R}_{X} \big\|^{2}_{2} \big\}$
of the proposed compressive RS estimator $\hat{R}_{X,\text{CS}}[n,k]$. To this end,
let us define the \emph{norm} of a random process $Y[n,k]$ that is $N$-periodic in $n$ and $k$ as
\[
{\| Y \|}_{\text{R}} \,\triangleq\, \sqrt{ {\rm E} \big\{ {\|Y \|}_2^{2} \big\} }
  \eq \sqrt{ \sum_{n,k \in [N]} \!\rmv {\rm E} \big\{ {| Y[n,k] |}^{2} \big\} } \,.
\]

The estimation error of the compressive RS estimator can be expanded as
\begin{align*}
&\hat{R}_{X,\text{CS}}[n,k] \rmv-\rmv \bar{R}_{X}[n,k]\nonumber\\[1mm]
&\rule{2mm}{0mm}\eq \hat{R}_{X,\text{CS}}[n,k] \rmv-\rmv \hat{R}_{X,\text{MVU}}[n,k] + \hat{R}_{X,\text{MVU}}[n,k] \rmv-\rmv \discreteRHS[n,k] \nonumber\\[1mm]
&\rule{2mm}{0mm}\eq Y_1[n,k] + Y_2[n,k] \,,
\end{align*}
where we have set $Y_1[n,k] \triangleq \hat{R}_{X,\text{MVU}}[n,k] \rmv-\rmv \discreteRHS[n,k]$ and
$Y_2[n,k] \triangleq \hat{R}_{X,\text{CS}}[n,k] \rmv-\rmv \hat{R}_{X,\text{MVU}}[n,k]$. Hence, the MSE of the compressive RS estimator can be rewritten as
\[
\varepsilon_{\text{CS}} \ist=\ist {\rm E} \big\{ \big\| \hat{R}_{X,\text{CS}} - \bar{R}_{X} \big\|^{2}_{2} \big\}
  \ist=\ist {\rm E} \big\{ {\| Y_1 \rmv+\rmv Y_2 \|}^{2}_{2} \big\}
  \ist=\ist {\| Y_1 \rmv+\rmv Y_2 \|}_{\text{R}}^2 \,.
\]
Using the triangle inequality \cite{golub96} ${\| Y_1 + Y_2 \|}_{\text{R}} \ist\le\ist {\| Y_1 \|}_{\text{R}} + {\| Y_2 \|}_{\text{R}}$, we obtain the bound
$\varepsilon_{\text{CS}} \le \big( {\| Y_1 \|}_{\text{R}} + {\| Y_2 \|}_{\text{R}} \big)^{2}$. Recognizing that
${\| Y_1 \|}_{\text{R}} = \sqrt{ {\rm E} \big\{ \big\| \hat{R}_{X,\text{MVU}} \rmv-\rmv \discreteRHS \big\|_2^{2} \big\} } = \sqrt{ \varepsilon }$ and
${\| Y_2 \|}_{\text{R}} = \sqrt{ {\rm E} \big\{ \big\| \hat{R}_{X,\text{CS}} \rmv-\rmv \hat{R}_{X,\text{MVU}} \big\|_2^{2} \big\} } = \sqrt{ \Delta \varepsilon }$,
this bound can be rewritten as
\[
\varepsilon_{\text{CS}} \ist\ist\leq\ist\ist \big( \sqrt{\varepsilon} + \sqrt{\Delta \varepsilon } \ist\big)^{2} \ist.
\]
Inserting the bounds \eqref{bound_basic_nonstat_nonlin} on $\varepsilon$ and \eqref{eq:bound_excess_nonstat_nonlin}
on $\Delta \varepsilon$ then results in the following bound on $\varepsilon_{\text{CS}}$:
\[
\varepsilon_{\text{CS}} \,\leq\, {\|\bar{R}_{X}\|}^{2}_{2} \Bigg[ \sqrt{ m_{X}^{(I_{\overline{\mathcal{A}} })} \rmv+ \frac{\SS }{N}}
  \ist+\ist \sqrt{ \frac{(\SS'  \!-\! \KK) \ist\ist D^2}{\SS' \KK} \,\tilde{\sigma}_{X}(\KK)} \,\ist \Bigg]^{2} .
\]
Alternatively, using the approximate bound \eqref{eq:bound_excess_nonstat_nonlin_2} on $\Delta \varepsilon$ instead of \eqref{eq:bound_excess_nonstat_nonlin},
we obtain the simpler (but looser) approximate bound
\begin{align*}
\varepsilon_{\text{CS}} &\,\lessapprox\, {\|\bar{R}_{X}\|}^{2}_{2} \Bigg[ \sqrt{ m_{X}^{(I_{\overline{\mathcal{A}} })} \rmv+ \frac{\SS }{N}} \\[-1mm]
&\rule{21mm}{0mm} \ist+\ist \sqrt{\frac{(\SS'  \!-\! \KK) \ist\ist D^2}{\SS' \KK} \ist \prefactor \, \sigma^{(w_{\Phi})}_{\rmv\rmv X}} \,\ist\Bigg]^{2} .
\end{align*}

We note that our bounds on $\Delta \varepsilon$ are based on the CS bound \eqref{equ_err_bound_bp_nonstat} together with \eqref{equ_P_cond_nonstat},
which is known to be very loose \cite{rud06}.
Thus, for a given nominal sparsity degree $K$ and a given number of measurements $P$ satisfying \eqref{equ_P_cond_nonstat},
our upper bounds on $\Delta\varepsilon$ and, in turn, on $\varepsilon_\text{CS}$ will generally be quite pessimistic, i.e., too high.
However, the bounds are still valuable theoretically in the sense of an asymptotic analysis, because they show that the MSE decreases with increasing underspreadness
(expressed by a smaller moment $m_{X}^{(I_{\overline{\mathcal{A}} })}$ and a smaller ratio $\SS/N$)
and with increasing TF sparsity (expressed by a smaller moment $\sigma^{(w_{\Phi})}_{\rmv\rmv X}$).

\section{Numerical Study}\label{sec_nonstat_simu}

\renewcommand{\textfraction}{0.01}

We will assess the performance of our compressive spectral estimator for two simple examples. 
The first example is inspired by a cognitive radio application; the second example concerns the analysis of chirp-like signals.

\vspace{-1mm}

\subsection{Orthogonal Frequency Division Multiplexing Symbol Process}  \label{sec_simu_ofdm}

\begin{figure*}[t]
\vspace{-2mm}
\psfrag{(a)}[c][c][0.9]{\uput*{2mm}[270]{0}{{\normalsize (a)}}}
\psfrag{(b)}[c][c][0.9]{\uput*{2mm}[270]{0}{(b)}}
\psfrag{RHS}[c][c][0.9]{\uput*{2mm}[90]{0}{}} 
\psfrag{EAF}[c][c][0.9]{\uput*{2mm}[90]{0}{}} 
\psfrag{x_m_10}[c][c][0.9]{\uput*{2mm}[270]{0}{$-10$}}
\psfrag{x_m_5}[c][c][0.9]{\uput*{2mm}[270]{0}{$-5$}}
\psfrag{x_10}[c][c][0.9]{\uput*{2mm}[270]{0}{$10$}}
\psfrag{m}[c][c][0.9]{\uput*{4.5mm}[270]{0}{$m$}}
\psfrag{n}[c][c][0.9]{\uput*{4.5mm}[270]{0}{$n$}}
\psfrag{l}[c][c][0.9]{\uput*{3mm}[90]{0}{$l$}}
\psfrag{k}[c][c][0.9]{\uput*{3mm}[90]{0}{$k$}}
\psfrag{n_0}[c][c][0.9]{\uput*{.5mm}[270]{0}{\small{$0$}}}
\psfrag{n_100}[c][c][0.9]{\uput*{.5mm}[270]{0}{\small{$100$}}}
\psfrag{n_200}[c][c][0.9]{\uput*{.5mm}[270]{0}{\small{$200$}}}
\psfrag{n_300}[c][c][0.9]{\uput*{.5mm}[270]{0}{\small{$300$}}}
\psfrag{n_400}[c][c][0.9]{\uput*{.5mm}[270]{0}{\small{$400$}}}
\psfrag{n_500}[c][c][0.9]{\uput*{.5mm}[270]{0}{\small{$500$}}}
\psfrag{n0_m_Ncp}[c][c][0.9]{\uput*{2mm}[270]{0}{\small{$0$}}}
\psfrag{n0}[c][c][0.9]{\uput*{0mm}[270]{0}{\small{$0$}}}
\psfrag{n0_p_N_s_m_1}[c][c][0.9]{\uput*{0mm}[270]{0}{\small{$0$}}}
\psfrag{x_0}[c][c][0.9]{\uput*{.5mm}[270]{0}{\small{$0$}}}
\psfrag{x_100}[c][c][0.9]{\uput*{.5mm}[270]{0}{\small{$100$}}}
\psfrag{x_200}[c][c][0.9]{\uput*{.4mm}[270]{0}{\small{$200$}}}
\psfrag{x_m_100}[c][c][0.9]{\uput*{.5mm}[270]{0}{\small{$\!\!\!-100$}}}
\psfrag{x_m_200}[c][c][0.9]{\uput*{.45mm}[270]{0}{\small{$\!\!\!-200$}}}
\psfrag{y_m_50}[r][r][0.9]{\uput*{1mm}[180]{0}{$$}}
\psfrag{y_m_100}[r][r][0.9]{\uput*{1mm}[180]{0}{\small{$-100$}}}
\psfrag{y_m_150}[r][r][0.9]{\uput*{1mm}[180]{0}{$$}}
\psfrag{y_150}[r][r][0.9]{\uput*{1mm}[180]{0}{$$}}
\psfrag{y_m_200}[r][r][0.9]{\uput*{1mm}[180]{0}{\small{$-200$}}}
\psfrag{y_200}[r][r][0.9]{\uput*{1mm}[180]{0}{\small{$200$}}}
\psfrag{y_50}[r][r][0.9]{\uput*{1mm}[180]{0}{$$}}
\psfrag{y_100}[r][r][0.9]{\uput*{1mm}[180]{0}{\small{$100$}}}
\psfrag{y_15}[r][r][0.9]{\uput*{1mm}[180]{0}{$15$}}
\psfrag{y_m_20}[r][r][0.9]{\uput*{1mm}[180]{0}{$-20$}}
\psfrag{y_m_15}[r][r][0.9]{\uput*{1mm}[180]{0}{$-15$}}
\psfrag{y_m_10}[r][r][0.9]{\uput*{1mm}[180]{0}{$-10$}}
\psfrag{y_m_5}[r][r][0.9]{\uput*{1mm}[180]{0}{$-5$}}
\psfrag{y_10}[r][r][0.9]{\uput*{1mm}[180]{0}{$10$}}
\psfrag{y_5}[r][r][0.9]{\uput*{1mm}[180]{0}{$5$}}
\psfrag{y_0}[r][r][0.9]{\uput*{1mm}[180]{0}{\small{$0$}}}
\psfrag{0}[r][r][0.9]{\uput*{1mm}[180]{0}{\small{$0$}}}
\centering
\includegraphics[height=5cm,width=11cm]{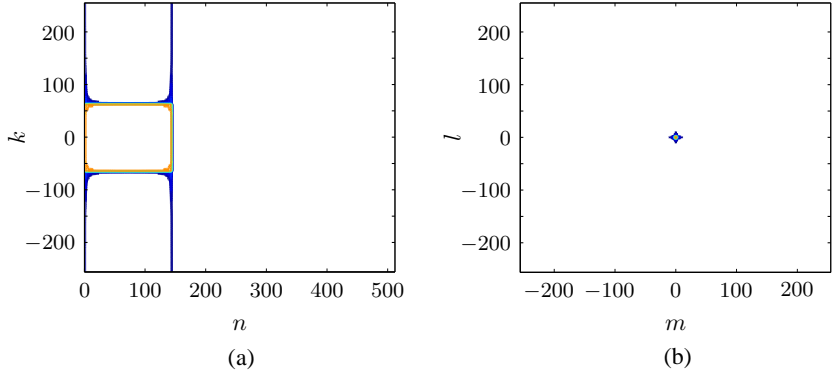}
\psfrag{x_250}[c][c][0.9]{\uput*{1mm}[270]{0}{$250$}}
\renewcommand{\baselinestretch}{1.2}\small\normalsize
\vspace{2mm}
\caption{TF representation of the OFDM process $X[n]$:
(a) Real part of RS $\bar{R}_X[n,k]$, displayed for $(n,k) \in [N] \times \{-N/2,\ldots,N/2-1\}$,
with $N \!= \rmv 512$; (b) magnitude of EAF $\bar{A}_{X}[m,l]$, displayed for $(m,l) \in \{-N/2,\ldots,N/2-1\}^2$.}
\label{fig_EAF}
\vspace{0mm}
\end{figure*}

\subsubsection{Simulation Setting}  \label{sec_simu_ofdm_setting}
In a cognitive radio system, a given transmitter/receiver node has to monitor a large overall frequency band and determine the unoccupied
bands that it can use for its own transmission \cite{CognitiveRadio2005,SpectrumSharingRadios,Sahai09}.
In our simulation, we consider a single active transmitter employing orthogonal frequency division multiplexing (OFDM) \cite{Bingham1990,hanzo03}, which is
a modulation scheme employed, e.g., for wireless local area networks \cite{hanzo03,80211},
digital video broadcasting \cite{Reimers98,DVBTSpec,dvb01},
and long term evolution cellular systems \cite{holma09}. 
We use $Q \!=\! 64$ subcarriers and a cyclic prefix whose length is $1/8$ of the symbol length.
Each subcarrier $i \!\in\! [Q]$ transmits a symbol $s_{i}$ that is randomly selected from a quadrature phase-shift keying (QPSK) 
constellation with normalized symbol energy $|s_i|^2 = 1$.
All QPSK symbols are equally likely, and the different subcarrier symbols $s_{i}$ are statistically independent.
The OFDM modulator uses an inverse DFT of length $Q \!=\! 64$ to map the frequency-domain transmit symbols $s_{i}$ into the (discrete)
time domain; this is followed by insertion of a cyclic prefix.
Assuming an idealized, noise-free channel for simplicity, the resulting transmit signal is also observed by the receiver. However, we assume that our receiver
monitors an overall bandwidth that is twice the nominal OFDM bandwidth, $B$. This corresponds to a twofold oversampling, i.e.,
a sampling period of $1/(2B)$, and can be easily realized by using an inverse DFT of length $N_\text{s} = 2\ist Q = 128$.
The lengths of an OFDM symbol and of the cyclic prefix are then given by $N_\text{s} = 128$ and $N_\text{cp} = 128/8 = 16$ samples, respectively.
To keep the simulation complexity low, we assume that a single OFDM symbol is transmitted, with silent periods before and afterwards.
Thus, the received time-domain signal (discrete-time baseband representation) is given by
\begin{equation} 
\label{equ_def_OFDM_process}
X[n] \eq \begin{cases}
   \displaystyle \ist\sum_{i \in [Q]} \rmv s_{i} \, e^{j \frac{2 \pi}{2Q} (n-n_{0})i}, & n \in \{n_{0} \!-\! N_\text{cp}, \ldots,\\[-3mm]
     &\rule{13mm}{0mm}{n_{0} \rmv+\rmv N_\text{s} \!-\! 1 \}}_{N}\\[0mm]
   \ist0 \,, &\text{otherwise}.
\end{cases}
\vspace{1mm}
\end{equation}
Here, $n_{0}$ denotes an arbitrary but fixed time offset. In our simulation, we used $n_{0} = N_\text{cp}$ and considered $X[n]$ for $n \in [N]$ with $N \!=\! 512$.

Because of the random $s_{i}$, $X[n]$ is a nonstationary random process. The RS and EAF of $X[n]$ are easily obtained from, respectively,
\eqref{def_discrete_RHS} and \eqref{equ_def_discrete_EAF} as
\begin{align*}
\bar{R}_X[n,k] &\eq \begin{cases}
   \displaystyle \ist \sum_{i \in [Q]}  {\rm dir}\bigg(\rmv N_\text{s} \rmv+\rmv N_\text{cp} \ist, \frac{k}{N} \rmv-\rmv \frac{i}{2Q} \bigg) \, e^{-j\frac{2 \pi}{N}nk},&\\[-1.5mm]
     &\rule{-45mm}{0mm}n \in {\{n_{0} \!-\! N_\text{cp}, \ldots, n_{0} \rmv+\rmv N_\text{s} \!-\! 1 \}}_N\\[.5mm]
   \ist0 \,, &\rule{-45mm}{0mm}\text{otherwise}\ist ;
\end{cases}\\[1.5mm]
\bar{A}_{X}[m,l] &\eq \begin{cases}
   \displaystyle \ist\sum_{i \in [Q]}  {\rm dir}\bigg(\rmv N_\text{s} \rmv+\rmv N_\text{cp} \!-\rmv m \ist,-\frac{l}{N} \bigg) \, e^{-j\frac{2 \pi}{2Q}mi},&\\[-1.5mm]
     &\rule{-41mm}{0mm}m \in {\{-N_{\text{cp}} \!-\! N_{\text{s}}  \rmv+\rmv 1,\ldots,0 \}}_{N} \\[1.5mm]
   \displaystyle \ist {\bar{A}^{*}_{X}[-m,-l]}_{N} \, e^{-j \frac{2 \pi}{N}ml} ,&\\[0mm]
     &\rule{-41mm}{0mm} m \in {\{1,\ldots,N_{\text{cp}} \rmv+\rmv N_{\text{s}}  \!-\! 1 \}}_{N} \\[1mm]
   \ist 0 \,, &\rule{-41mm}{0mm}\text{otherwise} \ist ,
\end{cases}
\end{align*}
where ${\rm dir}(n,\theta) \triangleq \sum_{n'=0}^{n-1} e^{j  \pi \theta n'}=e^{j\pi \theta (n-1)} \frac{\sin(\pi \theta n)}{\sin( \pi \theta)}$.
Note that the expression for $\bar{A}_{X}[m,l]$ requires that $N_{\text{s}}+N_\text{cp} < N/2$, a condition that is fulfilled in our simulation since $128+16<512/2$.
The RS and EAF are shown in Fig.\ \ref{fig_EAF}. From this figure, we can conclude that the process $X[n]$ is reasonably TF sparse but only
moderately underspread (the latter observation follows from the fact that $\bar{R}_X[n,k]$ is not very smooth).
Note that the TF sparsity could be further improved if we considered longer silent periods before and/or after the OFDM symbol,
and if we considered a wider band (i.e., if we used an oversampling factor larger than $2$).

\begin{figure}[b]
\vspace{-2mm}
\psfrag{y_0}[r][r][0.9]{\uput*{1mm}[180]{0}{\small{$0$}}}
\psfrag{y_0_2}[r][r][0.9]{\uput*{1mm}[180]{0}{\small{$0.2$}}}
\psfrag{y_0_4}[r][r][0.9]{\uput*{1mm}[180]{0}{\small{$0.4$}}}
\psfrag{y_0_6}[r][r][0.9]{\uput*{1mm}[180]{0}{\small{$0.6$}}}
\psfrag{y_0_8}[r][r][0.9]{\uput*{1mm}[180]{0}{\small{$0.8$}}}
\psfrag{y_1}[r][r][0.9]{\uput*{1mm}[180]{0}{\small{$1$}}}
\psfrag{x_1}[c][c][0.9]{\uput*{.5mm}[270]{0}{}}
\psfrag{x_10}[c][c][0.9]{\uput*{.5mm}[270]{0}{\small{$10$}}}
\psfrag{x_20}[c][c][0.9]{\uput*{.5mm}[270]{0}{\small{$20$}}}
\psfrag{x_30}[c][c][0.9]{\uput*{.5mm}[270]{0}{\small{$30$}}}
\psfrag{x_40}[c][c][0.9]{\uput*{.5mm}[270]{0}{\small{$40$}}}
\psfrag{x_50}[c][c][0.9]{\uput*{.5mm}[270]{0}{\small{$50$}}}
\psfrag{x_60}[c][c][0.9]{\uput*{.5mm}[270]{0}{\small{$60$}}}
\psfrag{x_70}[c][c][0.9]{\uput*{.5mm}[270]{0}{\small{$70$}}}
\psfrag{ylabel}[c][c][0.9]{\uput{10mm}[90]{270}{}}
\psfrag{exact}[c][c][0.9]{\uput{0mm}[0]{0}{$\!\HH_ {{(p,q)}_{r}}/\HH_ {{(p,q)}_{1}}$}}
\psfrag{approx}[c][c][0.9]{\uput{0mm}[0]{0}{$\!\!\rmv\tilde{\HH}_ {{(p,q)}_{r}}/\HH_ {{(p,q)}_{1}}$}}
\psfrag{title}[c][c][1]{\uput{2mm}[90]{0}{}} 
\psfrag{xlabel}[c][c][1]{\uput{4mm}[270]{0}{\vspace*{3mm} \small{$r$}}}
\centering
\includegraphics[height=3.4cm,width=7.2cm]{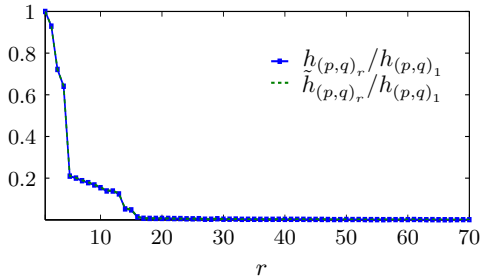}
\psfrag{x_250}[c][c][0.9]{\uput*{1mm}[270]{0}{$250$}}
\renewcommand{\baselinestretch}{1.25}\small\normalsize
\vspace{1mm}
\caption{$\HH_ {{(p,q)}_{r}}\!$ normalized by $\HH_ {{(p,q)}_{1}}\!$ and the corresponding normalized approximation $\tilde{\HH}_ {{(p,q)}_{r}}/\HH_ {{(p,q)}_{1}}\!$
according to \eqref{equ_approx_P_p_q_sum_quad} versus $r$.}
\label{fig_P_r_approx}
\end{figure}

\begin{figure*}[t]
\vspace{3mm}
\centering
\psfrag{f}[c][c][1]{\uput{0.1mm}[160]{0}{$k$}}
\psfrag{t}[c][c][1]{\uput{3mm}[340]{0}{$n$}}
\psfrag{(a)}[c][c][1]{\uput{4.5mm}[270]{0}{\,\quad\quad{\small (a)}}}
\psfrag{(b)}[c][c][1]{\uput{4.5mm}[270]{0}{\,\quad\,{\small (b)}}}
\psfrag{(c)}[c][c][1]{\uput{4.5mm}[270]{0}{\quad\;\,{\small (c)}}}
\psfrag{(d)}[c][c][1]{\uput{4.5mm}[270]{0}{\quad$\rmv${\small (d)}}}
\psfrag{(e)}[c][c][1]{\uput{4.5mm}[270]{0}{\,\;\;\;\;{\small (e)}}}
\psfrag{(f)}[c][c][1]{\uput{4.5mm}[270]{0}{\;\;\;\;\;\,{\small (f)}}}
\psfrag{(g)}[c][c][1]{\uput{4.5mm}[270]{0}{\,\;\;\,{\small (g)}}}
\centering
\hspace*{.5mm}\includegraphics[height=6cm,width=17cm]{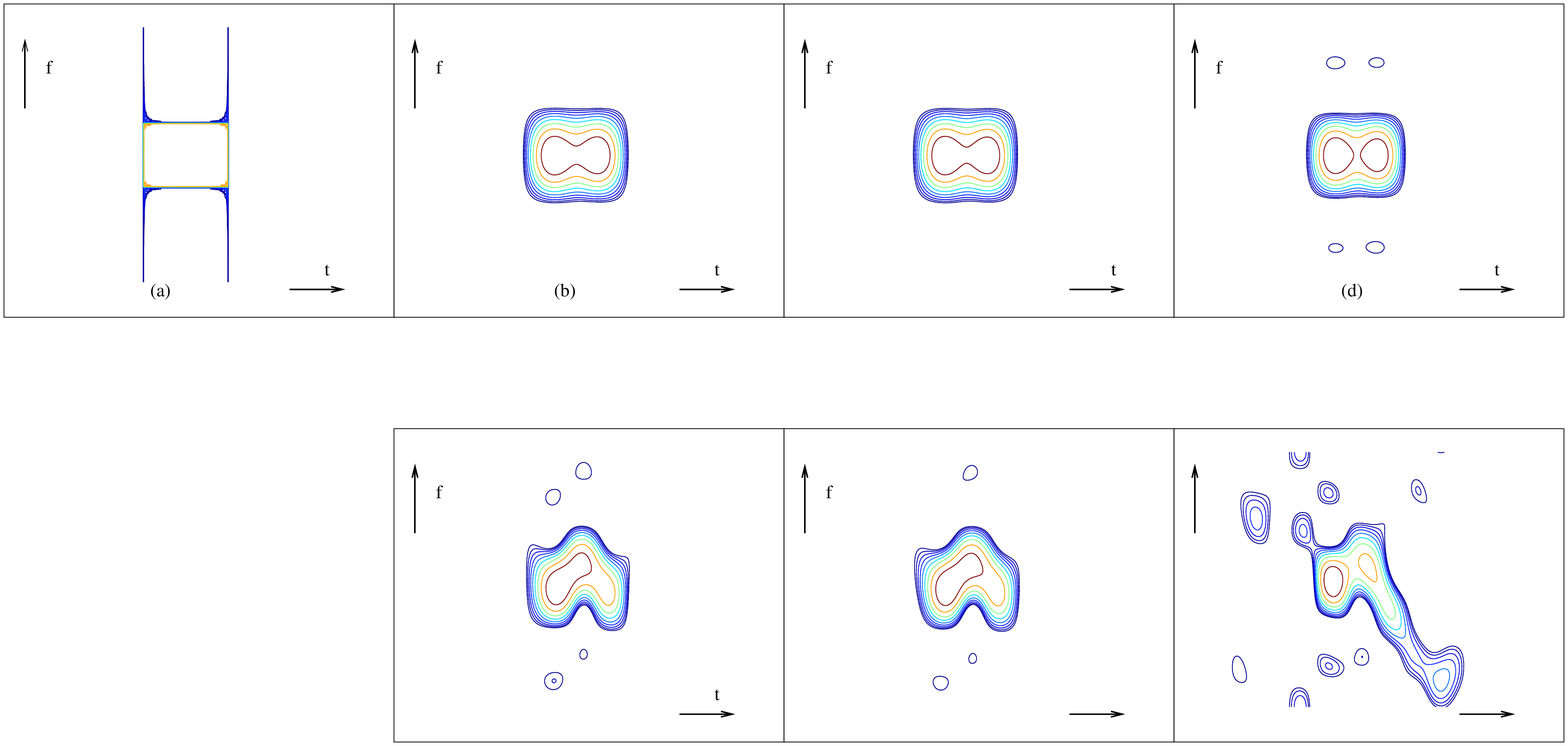}
\vspace*{3.5mm}
\renewcommand{\baselinestretch}{1.20}\small\normalsize
\caption{Averages and single realizations of RS estimators:
(a) RS of the OFDM process $X[n]$;
(b) average of the noncompressive estimator $\hat{R}_{X,\text{MVU}}[n,k]$ (compression factor $\SS'\! /P\rmv=\!1$);
(c) and (d) average of the compressive estimator $\hat{R}_{X,\text{CS}}[n,k]$ for $\SS'\! /P\rmv=\rmv 2$ and $\SS'\! /P\rmv\approx\rmv 5$, respectively;
(e) realization of $\hat{R}_{X,\text{MVU}}[n,k]$;
(f) and (g) realization of $\hat{R}_{X,\text{CS}}[n,k]$ for $\SS'\! /P\rmv=\rmv 2$ and $\SS'\! /P\rmv\approx\rmv 5$, respectively.
The real parts of all TF functions are shown for $(n,k) \in  \{-150,\ldots,361\} \times \{-N/2,\ldots,N/2-1\}$, with $N \!= \rmv 512$.
}
\label{fig_TF}
\vspace{-1mm}
\end{figure*}

For the design of the compressive RS estimator $\hat{R}_{X,\text{CS}}[n,k]$ in \eqref{equ_MVU_nonstat_est_CS}, we
used $M = 3$, $L=7$, $\Delta M = 8$, and $\Delta L =16$. This corresponds to choosing the effective EAF support (see \eqref{eq_underspread}) as
$\mathcal{A} = {\{- 3,\ldots,3 \}}_{512}\rmv \times {\{- 7,\ldots, 7 \}}_{512}$, of size
$\SS \equiv (2M \rmv+\! 1)(2L\rmv+\! 1) = 105$; furthermore, the size of the extended effective EAF support $\mathcal{A}'$ is $\SS' \equiv \Delta M \ist\Delta L = 128$.
For an assessment of the TF sparsity of $X[n]$, we consider
$\HH_ {p,q} = N^2 \, {\rm E} \big\{ \big| \hat{R}_{X,\text{MVU}}[\ist p \ist\Delta n, q \ist\Delta k] \big|^{2} \big\}$, which underlies the TF sparsity
profile $\tilde\sigma_{X}(\KK)$ in \eqref{eq_sigma-tilde_nonstat_0}. Let ${(p,q)}_{r}$ with $r \in \{1,\ldots,\SS'\}$ be the TF index of the $r$th largest (in magnitude) 
value of the set $\big\{ \widetilde{R}_{X,\text{MVU}}[p \Delta n, q \Delta k]\^{E}\big\}_{(p,q) \in [\lendoppler] \times [\lendelay]}$,
where, as before, $\widetilde{R}_{X,\text{MVU}}[n,k] = {\rm E} \big\{ \hat{R}_{X,\text{MVU}}[n,k] \big\}
= \frac{1}{N} \sum_{n'\!,k' \in [N]} \Phi_{\text{MVU}}{[n\!-\!n'\!,k\!-\!k']} \,\bar{R}_{X}[n'\!,k']$ (see \eqref{equ_mean_function_RS_MVU_def}, \eqref{equ_mean_function_RS_MVU}).
In Fig.\ \ref{fig_P_r_approx}, we show the values $\HH_ {{(p,q)}_{r}}$ along with the corresponding approximations \eqref{equ_approx_P_p_q_sum_quad}---here
denoted $\tilde{\HH}_ {{(p,q)}_{r}}$---as a function of the index $r$. It is seen that $\HH_ {{(p,q)}_{r}}$ is close to zero for $r$ larger than 15.
Furthermore, we can conclude that the ordering of the values $\widetilde{R}_{X,\text{MVU}}[p \Delta n, q \Delta k]$
according to decreasing magnitude matches the ordering of the values $\HH_ {p,q}$ very well. Thus, for TF positions $(p \Delta n, q \Delta k)$ for which
$\big| \widetilde{R}_{X,\text{MVU}}[p \Delta n, q \Delta k] \big|$ is large, we can expect that also $\HH_ {p,q}$ is large.
Finally, it is seen that the curves representing $\tilde{\HH}_ {{(p,q)}_{r}}$ and $\HH_ {{(p,q)}_{r}}$ coincide,
which shows that the approximation \eqref{equ_approx_P_p_q_sum_quad} is very accurate.

\vspace{.5mm}

\subsubsection{Simulation Results}  \label{sec_simu_ofdm_results}
We now consider the estimation of the RS $\bar{R}_X[n,k]$ from a single realization of $X[n]$ that is observed for $n \in [512]$.
To evaluate the estimation performance, we generated 1000 realizations of the QPSK symbols ${\{s_{i}\}}_{i \in [64]}$
and computed the corresponding realizations of $X[n]$. In Fig.\ \ref{fig_TF}, we show the average of 1000 realizations of the compressive RS estimator 
$\hat{R}_{X,\text{CS}}[n,k]$ (obtained for the 1000 realizations of $X[n]$) as well as a single realization of $\hat{R}_{X,\text{CS}}[n,k]$
for compression factors $\SS'\! /P\rmv=\rmv 1$, $2$, and approximately $5$ or, equivalently, $P\rmv=\rmv 128$, $64$, and $25$ randomly located AF measurements.
The optimization in \eqref{equ_def_bp_nonstat}, which is required for the computation of $\hat{R}_{X,\text{CS}}[n,k]$ in \eqref{equ_MVU_nonstat_est_CS},
was carried out using the \rm{MATLAB} library \rm{CVX} \cite{GrantBoydCVX}. The true RS is also re-displayed for easy comparison with the estimates.

The case $\SS'\! /P \!=\! 1$ corresponds to the basic RS estimator $\hat{R}_{X,\text{MVU}}[n,k]$ in \eqref{equ_MVU_nonlin_nonstat_eaf} (cf.\
the discussion at the end of Section \ref{sec_sub_nonstat_est_compr_impr}).
We see that already in this case, even for the average $\hat{R}_{X,\text{CS}}[n,k]$, there are noticeable deviations from the true RS.
In fact, the average of the 1000 basic RS estimates $\hat{R}_{X,\text{MVU}}[n,k]$ closely approximates the expected basic RS estimator
$\widetilde{R}_{X,\text{MVU}}[n,k] = {\rm E} \big\{ \hat{R}_{X,\text{MVU}}[n,k] \big\}$, which according to \eqref{equ_mean_function_RS_MVU} 
is a smoothed version of the RS. This smoothing leads to a noticeable deviation from
the RS, because the RS itself is not very smooth. The limited smoothness of the RS corresponds to the fact that the process $X[n]$ is only moderately underspread.
For compression factor $\SS'\! /P=2$, there is no visible degradation of the average estimate relative to the basic estimator. For $\SS'\! /P \approx 5$,
a small degradation is visible. The results obtained for the individual realizations suggest a random variation and deviation from the true TF support of the RS that are 
higher for compression factor $\SS'\! /P \approx 5$. The results of the symmetrized compressive estimator $\hat{R}_{X,\text{CS}}^{(\text{s})}[n,k]$ 
in \eqref{equ_modified_compressive_RS_estimator_explicit_expr} are not shown in Fig.\ \ref{fig_TF} because they can hardly be distinguished visually 
from those of $\hat{R}_{X,\text{CS}}[n,k]$.

\begin{figure}[b!]
\vspace{-2mm}
\centering
\psfrag{NMSE}[c][c][0.8]{\uput{0.1mm}[90]{0}{NMSE [dB]}}
\psfrag{N/M}[c][c][0.8]{\uput{0.0005mm}[270]{0}{$\SS /P$}}
\psfrag{y_0}[r][r][0.9]{\uput*{1mm}[175]{0}{$0\ist$}}
\psfrag{y_0_05}[r][r][0.9]{\uput*{1mm}[180]{0}{}}
\psfrag{y_0_1}[r][r][0.9]{\uput*{1mm}[175]{0}{\small{$0.1$}}}
\psfrag{y_0_15}[r][r][0.9]{\uput*{1mm}[180]{0}{}}
\psfrag{y_0_2}[r][r][0.9]{\uput*{1mm}[175]{0}{\small{$0.2$}}}
\psfrag{y_0_25}[r][r][0.9]{\uput*{1mm}[180]{0}{}}
\psfrag{y_0_3}[r][r][0.9]{\uput*{1mm}[175]{0}{\small{$0.3$}}}
\psfrag{y_0_35}[r][r][0.9]{\uput*{1mm}[180]{0}{}}
\psfrag{y_0_4}[r][r][0.9]{\uput*{1mm}[175]{0}{\small{$0.4$}}}
\psfrag{y_0_45}[r][r][0.9]{\uput*{1mm}[180]{0}{}}
\psfrag{y_0_5}[r][r][0.9]{\uput*{1mm}[175]{0}{\small{$0.5$}}}
\psfrag{y_0_55}[r][r][0.9]{\uput*{1mm}[175]{0}{}}
\psfrag{y_0_6}[r][r][0.9]{\uput*{1mm}[180]{0}{\small{$0.6$}}}
\psfrag{y_0_8}[r][r][0.9]{\uput*{1mm}[180]{0}{$0.8$}}
\psfrag{data1}[c][c][0.61]{\uput*{0mm}[1]{0}{\hspace*{-3.0mm}{\small $\varepsilon_{\text{CS}}/{\| \bar{R}_{X} \|}^{2}_{2}$}}}
\psfrag{data3}[c][c][0.61]{\uput*{0mm}[-1]{0}{\hspace*{-4mm}{\small $B_{\text{CS}}^{2}/{\| \bar{R}_{X} \|}^{2}_{2}$}}}
\psfrag{data2}[c][c][0.61]{\uput*{0mm}[0]{0}{\hspace*{-3.8mm}{\small $V_{\text{CS}} /{\| \bar{R}_{X} \|}^{2}_{2}$}}}
\psfrag{data4}[c][c][0.61]{\uput*{0mm}[5]{0}{\hspace*{-3.5mm}{\small $\varepsilon_{\text{CS}^{(\text{s})}}/{\| \bar{R}_{X} \|}^{2}_{2}$}}}
\psfrag{data6}[c][c][0.61]{\uput*{0mm}[-5]{0}{\hspace*{-4.0mm}{\small $B_{\text{CS}^{(\text{s})}}^{2}/{\| \bar{R}_{X} \|}^{2}_{2}$}}}
\psfrag{data5}[c][c][0.61]{\uput*{0mm}[0]{0}{\hspace*{-3.8mm}{\small $V_{\text{CS}^{(\text{s})}} /{\| \bar{R}_{X} \|}^{2}_{2}$}}}
\psfrag{x_1}[c][c][0.9]{\uput*{0.6mm}[275]{0}{\small{$\ist\ist 1$}}}
\psfrag{x_2}[c][c][0.9]{\uput*{0.6mm}[270]{0}{\small{$2$}}}
\psfrag{x_3}[c][c][0.9]{\uput*{0.6mm}[270]{0}{\small{$3$}}}
\psfrag{x_4}[c][c][0.9]{\uput*{0.6mm}[270]{0}{\small{$4$}}}
\psfrag{x_5}[c][c][0.9]{\uput*{0.6mm}[270]{0}{\small{$5$}}}
\psfrag{ylabel}[c][c][0.9]{\uput{10mm}[90]{270}{}}
\psfrag{title}[c][c][1]{\uput{2mm}[90]{0}{}}  
\psfrag{xlabel}[c][c][0.8]{\uput{5mm}[265]{0}{\vspace*{3mm} $\!\!\!\!\SS'\! /P$}}
\centering
\includegraphics[height=4.5cm,width=7.5cm]{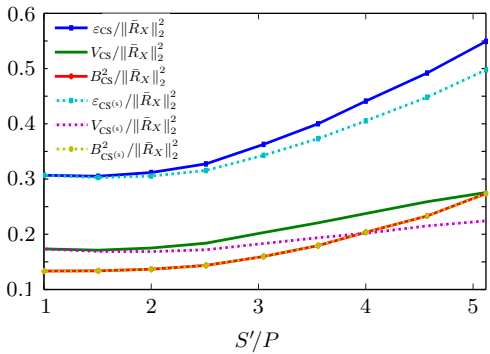}
\renewcommand{\baselinestretch}{1.15}\small\normalsize
\vspace*{1mm}
\caption{Empirical NMSE, normalized squared bias, and normalized variance of the compressive RS estimator $\hat{R}_{X,\text{CS}}[n,k]$
(solid curves) and of the symmetrized compressive RS estimator $\hat{R}_{X,\text{CS}}^{(\text{s})}[n,k]$ (dash-dotted curves) versus the
compression factor $\SS'\! /P$.}
\label{fig_CS_sweep}
\end{figure}

\begin{figure*}[t]
\vspace{3mm}
\centering
\psfrag{f}[c][c][1]{\uput{0.1mm}[160]{0}{$k$}}
\psfrag{t}[c][c][1]{\uput{3mm}[340]{0}{$n$}}
\psfrag{(a)}[c][c][1]{\uput{4.5mm}[270]{0}{\,\quad\quad{\small (a)}}}
\psfrag{(b)}[c][c][1]{\uput{4.5mm}[270]{0}{\,\quad\,{\small (b)}}}
\psfrag{(c)}[c][c][1]{\uput{4.5mm}[270]{0}{\quad\;\,{\small (c)}}}
\psfrag{(d)}[c][c][1]{\uput{4.5mm}[270]{0}{\quad$\rmv${\small (d)}}}
\psfrag{(e)}[c][c][1]{\uput{4.5mm}[270]{0}{\,\;\;\;\;{\small (e)}}}
\psfrag{(f)}[c][c][1]{\uput{4.5mm}[270]{0}{\;\;\;\;\;\,{\small (f)}}}
\psfrag{(g)}[c][c][1]{\uput{4.5mm}[270]{0}{\,\;\;\,{\small (g)}}}
\centering
\hspace*{.5mm}
\includegraphics[height=6cm,width=17cm]{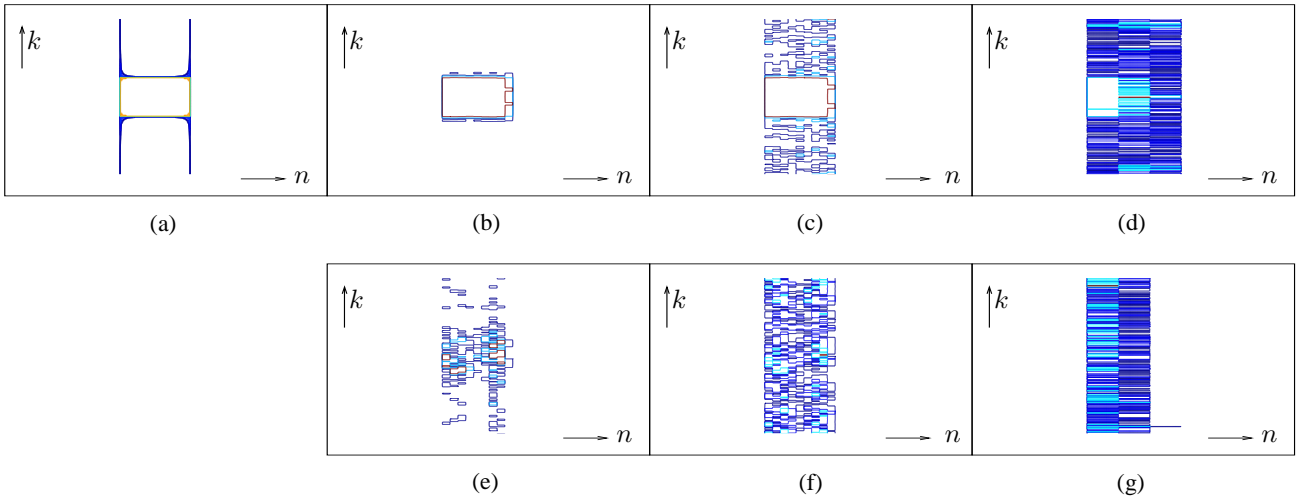}
\vspace*{3.5mm}
\renewcommand{\baselinestretch}{1.20}\small\normalsize
\caption{Averages and single realizations of the reference estimator $\hat{R}_{X}^{\text{(ref)}}[n,k]$:
(a) real part of the RS of the OFDM process $X[n]$;
(b), (c), and (d) magnitude of the average of $\hat{R}_{X}^{\text{(ref)}}[n,k]$ for $c\rmv=\rmv 1$, $c\rmv=\rmv 2$, and $c\rmv\approx\rmv 5$, respectively;
(e), (f), and (g) magnitude of a realization of $\hat{R}_{X}^{\text{(ref)}}[n,k]$ for $c\rmv=\rmv 1$, $c\rmv=\rmv 2$, and $c\rmv\approx\rmv 5$, respectively.
All TF functions are shown for $(n,k) \in  \{-150,\ldots,361\} \times \{-N/2,\ldots,N/2-1\}$, with $N \!= \rmv 512$.
}
\label{fig_TF_Leus}
\vspace{-1mm}
\end{figure*}

For a quantitative analysis of the degradation caused by the compression, we show in Fig.\ \ref{fig_CS_sweep} the empirical normalized MSE (NMSE)
of the compressive estimator $\hat{R}_{X,\text{CS}}[n,k]$ versus the compression factor $\SS'\! /P$.
The NMSE is an empirical, normalized version of the MSE $\varepsilon_{\text{CS}} = {\rm E} \big\{ \big\| \hat{R}_{X,\text{CS}} \rmv-\rmv \bar{R}_{X} \big\|^{2}_{2} \big\}$,
with the expectation replaced by the sample average over the 1000 process realizations and with normalization by
$\big\| \bar{R}_{X} \big\|^{2}_{2}$. In the same figure, we also show the empirical normalized versions of the squared
bias term $B_{\text{CS}}^2 = \big\| \ist{\rm E} \{\hat{R}_{X,\text{CS}} \} \rmv-\rmv \bar{R}_{X} \big\|_2^{2}$
and of the variance $V_{\text{CS}} = {\rm E} \big\{ \big\| \hat{R}_{X,\text{CS}} \rmv- {\rm E} \{ \hat{R}_{X,\text{CS}} \} \big\|_2^{2} \big\}$,
again with normalization by $\big\| \bar{R}_{X} \big\|^{2}_{2}$. (Recall that $\varepsilon_{\text{CS}} = B_{\text{CS}}^{2} + V_{\text{CS}}$.)
These results demonstrate a ``graceful degradation'' with increasing compression factor $\SS'\! /P$. Again, the result for $\SS'\! /P \!=\! 1$
corresponds to the basic RS estimator $\hat{R}_{X,\text{MVU}}[n,k]$. Fig.\ \ref{fig_CS_sweep}
also shows the NMSE, normalized squared bias term, and normalized variance of the symmetrized compressive estimator $\hat{R}_{X,\text{CS}}^{(\text{s})}[n,k]$.
It is seen that the variance and MSE are reduced by the symmetrization. We did not plot the MSE bounds derived in Section \ref{sec_perf_nonstat} 
because they are much larger than the empirical MSE. As mentioned in Section \ref{sec_nonstat_est_comb_msebound},
this lack of tightness is mostly due to the notoriously loose \cite{rud06} CS error bound used in \eqref{equ_err_bound_bp_nonstat_recall} 
(combined with \eqref{equ_P_cond_nonstat}).

\vspace{1mm}

\subsubsection{Comparison with a Reference Method}  \label{sec_simu_ofdm_compar}
Next, we compare our compressive nonstationary spectral estimator $\hat{R}_{X,\text{CS}}[n,k]$
with the compressive spectral estimation method proposed in \cite{Leus2011}, hereafter termed ``reference estimator.'' 
The reference estimator was devised for estimating the power spectral density of a \emph{stationary} random process; the underlying
stationarity assumption allows the use of long-term averaging. However, for the nonstationary processes considered in this paper, long-term averaging
is not an option and hence a deteriorated performance must be expected. We nevertheless chose the reference estimator for a performance comparison because
we are not aware of any previously proposed compressive spectral estimation method for general nonstationary processes.

The reference estimator uses as its input an observed realization $\mathbf{x}$ of a block of a stationary discrete random process $X[n]$ and  
calculates a reduced number of compressive measurements, for some compression factor $c$. 
From these measurements, it derives an estimate $\hat{P}_X(\omega)$ of the power spectral density 
$P_X(\omega) \triangleq \sum_{m \in \mathbb{Z}} r_X[m] \, e^{-j \omega m}\rmv$ (here, $r_X[m] \triangleq {\rm E} \ist\{X[m] \ist X^{*}[0]\}$ 
is the autocorrelation function of $X[n]$). In our case, however, $X[n]$ is the nonstationary OFDM process of length $N$ 
defined in \eqref{equ_def_OFDM_process}. In order to impart a time dependence (time resolution) to the reference estimator, 
we consecutively apply it to a sequence of overlapping length-$N_{\text{b}}$ blocks
$\mathbf{x}^{(b)} \triangleq  (x[b\ist\Delta N] \;\, x[b\ist\Delta N+\rmv 1] \,\cdots\, x[b\ist\Delta N+N_{\text{b}}-\rmv 1])^T\rmv$, $b \in \{0,\ldots,B \!-\! 1\}$ of the observed realization $x[n]$ of 
$X[n]$. Here, $B = \big\lfloor \frac{N-N_{\text{b}}+1}{\Delta N} \big\rfloor+1$ and $N_{\text{b}} \ge \Delta N$; note that $N_{\text{b}} \!-\! \Delta N$ is the overlap length.
For each block $\mathbf{x}^{(b)}\rmv$, we thus obtain a (discrete-frequency) local power spectrum 
estimate\footnote{We 
note the following details of our implementation of the reference estimator (cf.\ \cite{Leus2011} for background and notation).
The maximum correlation lag was chosen as $L \!=\! 1$. 
The second-order statistics (cross-correlation functions) $r_{y_i,y_j}[k]$ were estimated by time-averages over blocks of length $L+1 \rmv=\rmv 2$. 
The weights $c_{i}[n]$ were randomly drawn from the set $\{-1,1\}$ with equal 
probabilities.}
$\hat{P}_X^{(b)}(2 \pi k /N)$, $k \rmv\in\rmv [N]$. From the sequence of local power spectrum estimates $\hat{P}_X^{(b)}(2 \pi k /N)$, $b=0,\ldots,B \!-\! 1$, 
we then construct a \emph{time-dependent} (more specifically, piecewise constant) compressive power spectral density estimate
$\hat{R}_{X}^{\text{(ref)}}[n,k]$ by setting $\hat{R}_{X}^{\text{(ref)}}[n,k] \triangleq \hat{P}_X^{(b)}(2 \pi k /N)$ 
for $n \in \{b\ist\Delta N, b\ist\Delta N+\rmv 1, \ldots, (b+\rmv 1)\ist\Delta N-\rmv 1\}$, with $b \in \{0,\ldots,B \!-\! 1\}$.

In Fig.\ \ref{fig_TF_Leus}, we show the average of $\hat{R}_{X}^{\text{(ref)}}[n,k]$ 
obtained for 1000 realizations of the OFDM process $X[n]$ as well as a single realization of $\hat{R}_{X}^{\text{(ref)}}[n,k]$
for compression factors $c \rmv=\rmv 1$, $2$, and approximately $5$. We used block length $N_{\text{b}} \!=\! 32$ and time increment $\Delta N \!=\! 16$ 
for $c \!=\! 1$ and $c \!=\! 2$, 
and\footnote{These  
different choices of $N_{\text{b}}$ and $\Delta N$ for $c \rmv=\rmv 1$, $2$ and for $c \approx 5$ are due to
the condition $N_{\text{b}}\geq 2 \ist (2c \rmv-\! 1) \ist c$ that is required by the reference estimator 
\cite{Leus2011}.}
$N_{\text{b}} \!=\! 128$ and $\Delta N \!=\! 64$ for $c \rmv\approx\rmv 5$. For convenience,
the true RS is again re-displayed in part (a). A comparison of Fig.\ \ref{fig_TF_Leus} with Fig.\ \ref{fig_TF} shows that the proposed estimator 
$\hat{R}_{X,\text{CS}}[n,k]$ clearly outperforms the reference estimator $\hat{R}_{X}^{\text{(ref)}}[n,k]$, especially when single realizations
are considered and in the compressive case ($c \rmv>\! 1$), which are the cases of greatest relevance in our context.
This result is not surprising and should not be interpreted as evidence of poor performance of the estimator proposed in \cite{Leus2011}. In fact, 
as noted previously, that estimator was devised for stationary random processes where long-term averaging can be used,
and it was not intended for our straightforward and somewhat na\"{i}ve adaptation to nonstationary processes.

\begin{figure*}[t!]
\vspace{-2mm}
\psfrag{(a)}[c][c][0.9]{\uput*{2mm}[270]{0}{{\normalsize (a)}}}
\psfrag{(b)}[c][c][0.9]{\uput*{2mm}[270]{0}{(b)}}
\psfrag{RHS}[c][c][0.9]{\uput*{2mm}[90]{0}{}} 
\psfrag{EAF}[c][c][0.9]{\uput*{2mm}[90]{0}{}} 
\psfrag{x_m_10}[c][c][0.9]{\uput*{2mm}[270]{0}{$-10$}}
\psfrag{x_m_5}[c][c][0.9]{\uput*{2mm}[270]{0}{$-5$}}
\psfrag{x_10}[c][c][0.9]{\uput*{2mm}[270]{0}{$10$}}
\psfrag{m}[c][c][0.9]{\uput*{4.5mm}[270]{0}{$m$}}
\psfrag{n}[c][c][0.9]{\uput*{4.5mm}[270]{0}{$n$}}
\psfrag{l}[c][c][0.9]{\uput*{3mm}[90]{0}{$l$}}
\psfrag{k}[c][c][0.9]{\uput*{3mm}[90]{0}{$k$}}
\psfrag{n_0}[c][c][0.9]{\uput*{.5mm}[270]{0}{\small{$0$}}}
\psfrag{n_100}[c][c][0.9]{\uput*{.5mm}[270]{0}{\small{$100$}}}
\psfrag{n_200}[c][c][0.9]{\uput*{.5mm}[270]{0}{\small{$200$}}}
\psfrag{n_300}[c][c][0.9]{\uput*{.5mm}[270]{0}{\small{$300$}}}
\psfrag{n_400}[c][c][0.9]{\uput*{.5mm}[270]{0}{\small{$400$}}}
\psfrag{n_500}[c][c][0.9]{\uput*{.5mm}[270]{0}{\small{$500$}}}
\psfrag{x_0}[c][c][0.9]{\uput*{.5mm}[270]{0}{\small{$0$}}}
\psfrag{x_m_100}[c][c][0.9]{\uput*{.40mm}[270]{0}{\small{$\!\!\!-100$}}}
\psfrag{x_100}[c][c][0.9]{\uput*{.5mm}[270]{0}{\small{$100$}}}
\psfrag{x_m_150}[c][c][0.9]{\uput*{2mm}[270]{0}{}}
\psfrag{x_150}[c][c][0.9]{\uput*{2mm}[270]{0}{}}
\psfrag{x_50}[c][c][0.9]{\uput*{2mm}[270]{0}{}}
\psfrag{x_m_50}[c][c][0.9]{\uput*{2mm}[270]{0}{}}
\psfrag{x_m_200}[c][c][0.9]{\uput*{.5mm}[270]{0}{\small{$\!\!\!-200$}}}
\psfrag{x_200}[c][c][0.9]{\uput*{.5mm}[270]{0}{\small{$200$}}}
\psfrag{y_0}[r][r][0.9]{\uput*{1mm}[180]{0}{$0$}}
\psfrag{y_m_100}[r][r][0.9]{\uput*{1mm}[180]{0}{\small{$-100$}}}
\psfrag{y_100}[r][r][0.9]{\uput*{1mm}[180]{0}{\small{$100$}}}
\psfrag{y_m_150}[r][r][0.9]{\uput*{1mm}[180]{0}{}}
\psfrag{y_150}[r][r][0.9]{\uput*{1mm}[180]{0}{}}
\psfrag{y_50}[r][r][0.9]{\uput*{1mm}[180]{0}{}}
\psfrag{y_m_50}[r][r][0.9]{\uput*{1mm}[180]{0}{}}
\psfrag{y_m_200}[r][r][0.9]{\uput*{1mm}[180]{0}{\small{$-200$}}}
\psfrag{y_200}[r][r][0.9]{\uput*{1mm}[180]{0}{\small{$200$}}}
\centering
\includegraphics[height=5cm,width=11cm]{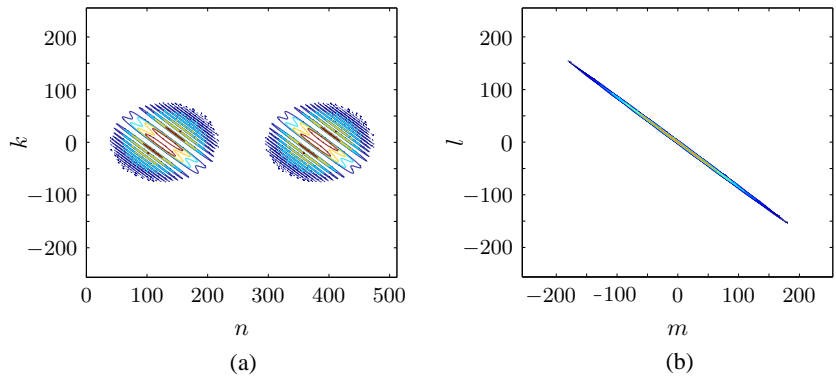}
\psfrag{x_250}[c][c][0.9]{\uput*{1mm}[270]{0}{$250$}}
\renewcommand{\baselinestretch}{1.2}\small\normalsize
\vspace{2mm}
\caption{TF representation of the two-component chirp process $X[n]$:
(a) Real part of RS $\bar{R}_X[n,k]$, displayed for $(n,k) \in [N] \times  \{-N/2,\ldots,N/2-1\}$, with $N \!= \rmv 512$;
(b) magnitude of EAF $\bar{A}_{X}[m,l]$, displayed for $(m,l) \in \{-N/2,\ldots,N/2-1\}^2$.
}
\label{fig_EAF_chirp}
\vspace{0mm}
\end{figure*}

\begin{figure*}[t]
\vspace{6mm}
\centering
\psfrag{t0}[c][c][1]{\uput{0mm}[0]{0}{$t_0$}}
\psfrag{f}[c][c][1]{\uput{0.1mm}[160]{0}{$k$}}
\psfrag{t}[c][c][1]{\uput{3mm}[340]{0}{$n$}}
\psfrag{(a)}[c][c][1]{\uput{3.5mm}[270]{0}{\,\quad\quad{\small (a)}}}
\psfrag{(b)}[c][c][1]{\uput{3.5mm}[270]{0}{\,\quad\,{\small (b)}}}
\psfrag{(c)}[c][c][1]{\uput{3.5mm}[270]{0}{\quad\;\,{\small (c)}}}
\psfrag{(d)}[c][c][1]{\uput{3.5mm}[270]{0}{\quad$\rmv${\small (d)}}}
\psfrag{(e)}[c][c][1]{\uput{3.5mm}[270]{0}{\,\;\;\;\;{\small (e)}}}
\psfrag{(f)}[c][c][1]{\uput{3.5mm}[270]{0}{\;\;\;\;\;\,{\small (f)}}}
\psfrag{(g)}[c][c][1]{\uput{3.5mm}[270]{0}{\,\;\;\,{\small (g)}}}
\psfrag{(h)}[c][c][1]{\uput{3.5mm}[270]{0}{\,\;\;\,{\small (h)}}}
\psfrag{(i)}[c][c][1]{\uput{3.5mm}[270]{0}{\,\;\;\,{\small (i)}}}
\centering
\hspace*{.5mm}\includegraphics[height=8.5cm,width=17cm]{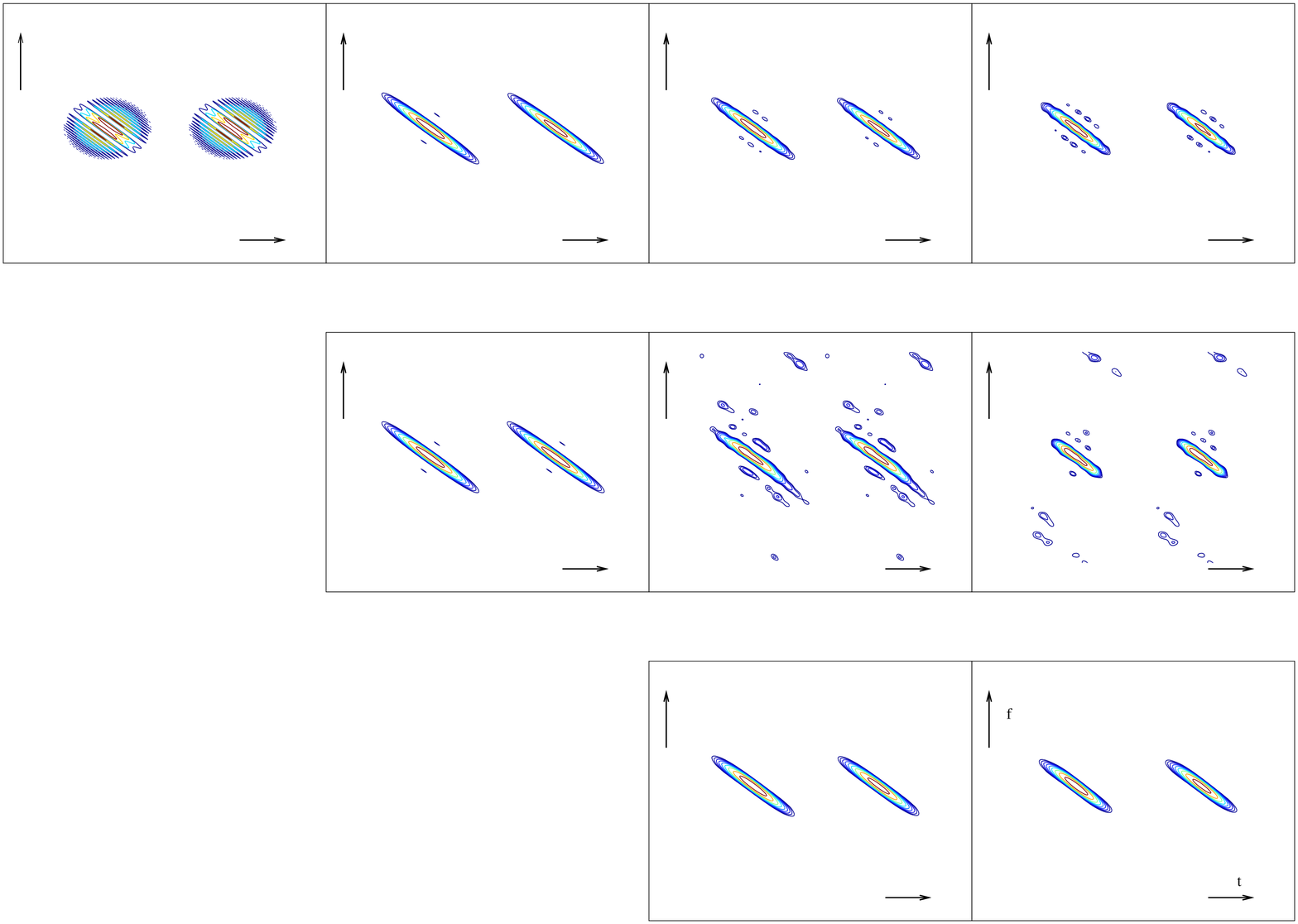}
\vspace*{3mm}
\renewcommand{\baselinestretch}{1.20}\small\normalsize
\caption{Averages and single realizations of RS estimators:
(a) RS of the chirp process $X[n]$;
(b) average of the noncompressive estimator $\hat{R}_{X,\text{MVU}}[n,k]$ (compression factor $\SS'\! /P\rmv=\!1$);
(c) and (d) average of the compressive estimator $\hat{R}_{X,\text{CS}}[n,k]$ for $\SS'\! /P\rmv\approx\rmv 5$ and $10$, respectively;
(e) realization of $\hat{R}_{X,\text{MVU}}[n,k]$;
(f) and (g) realization of $\hat{R}_{X,\text{CS}}[n,k]$ for $\SS'\! /P\rmv\approx\rmv 5$ and $10$, respectively;
(h) and (i) average of the symmetrized compressive estimator $\hat{R}_{X,\text{CS}}^{(\text{s})}[n,k]$ for $\SS'\! /P\rmv\approx\rmv 5$ and $10$, respectively.
The real parts of all TF functions are shown for $(n,k) \in [N] \times \{-N/2,\ldots,N/2-1\}$, with $N \!= \rmv 512$.}
\label{fig_TF_chirp}
\vspace{-1mm}
\end{figure*}

\vspace{-1mm}

\subsection{Chirp Process}  
 
Next, we apply our compressive RS estimator to a two-component chirp process.
Chirp signals arise, e.g., in the context of engine diagnosis \cite{Stankovic_99,scb98,aeu-detect99},
system identification and radar \cite{Xia97,Salous_et_al_1998,Rih69,Skolnik_1984}, and the study of bat echolocation \cite{Grenier86,Flandrin-bat_99}.
In our simulation, we construct a finite-length, nonstationary, discrete-time process as $X[n] \triangleq X(n T_{\text{s}})$, $n \rmv\in\rmv [512]$, where $T_{\text{s}}$ is some sampling period.
The continuous-time process $X(t)$ is given 
\vspace{-1mm}
by
\[
 X(t) \eq a_{1} \ist s(t \rmv-\rmv t_1) + a_2 \ist s(t \rmv-\rmv t_2) \,,
\vspace{.5mm}
\]
where $t_1 = 128 \, T_{\text{s}}$ and $t_2 = 384 \, T_{\text{s}}$;
$a_{1}$ and $a_{2}$ are independent zero-mean Gaussian random variables with unit variance;
and $s(t)$ is a chirp pulse defined as $s(t) =$\linebreak 
$\exp\!\big(\!\rmv\rmv-\rmv\rmv (t/T_{0})^2/2 \big) \exp(-j  \pi \beta t^2 )$,
with pulse width parameter $T_{0} = 60 \, T_{\text{s}}$ and chirp rate $\beta = 1/(600 \, T_{\text{s}}^{2})$.
The RS and EAF of the discrete-time process $X[n]$ are shown in Fig.\ \ref{fig_EAF_chirp}.
We see that the process $X[n]$ is only moderately TF sparse and not very underspread, i.e., the underspread approximation used, e.g., in
Section \ref{sec_nonstat_est_basic} can be hardly justified.

We implemented the compressive RS estimator $\hat{R}_{X,\text{CS}}[n,k]$ in \eqref{equ_MVU_nonstat_est_CS}
as well as the symmetrized compressive estimator $\hat{R}_{X,\text{CS}}^{(\text{s})}[n,k]$ in \eqref{equ_modified_compressive_RS_estimator_explicit_expr}
using $M \!=\! L \!=\! 15$ and $\Delta M \!=\! \Delta L \!=\! 32$. This corresponds to the effective EAF support
$\mathcal{A} = {\{- 15,\ldots,15 \}}_{512}\rmv \times {\{- 15,\ldots, 15 \}}_{512}$, of size $\SS \equiv (2M \rmv+\! 1)(2L\rmv+\! 1) = 961$.
The size of the extended effective EAF support $\mathcal{A}'$ is $\SS' \equiv \Delta M \ist\Delta L = 1024$.
Fig.\ \ref{fig_TF_chirp} shows the average of 1000 realizations of $\hat{R}_{X,\text{CS}}[n,k]$ and $\hat{R}_{X,\text{CS}}^{(\text{s})}[n,k]$
(obtained for 1000 realizations of $X(t)$) as well as a single realization of $\hat{R}_{X,\text{CS}}[n,k]$ for compression factors 
$\SS'\! /P\rmv \approx \rmv 5$ and $10$ or, equivalently, $P\rmv=\rmv 204$ and $102$ randomly located AF measurements.
We see that already in the noncompressive case $\SS'\! /P \!=\! 1$, where $\hat{R}_{X,\text{CS}}[n,k]$ and $\hat{R}_{X,\text{CS}}^{(\text{s})}[n,k]$ coincide with
the basic RS estimator $\hat{R}_{X,\text{MVU}}[n,k]$, there are noticeable deviations from the true RS; 
these differences are again due to the smoothing employed by $\hat{R}_{X,\text{MVU}}[n,k]$.
However, the proposed compressive RS estimator $\hat{R}_{X,\text{CS}}[n,k]$ 
still performs well in the sense that it indicates the main characteristics of the two chirp signal components---the TF locations and the chirp rate---up 
to a compression factor of 10, i.e., based on the observation of a significantly reduced number of AF samples. 
In this sense, our estimator appears to be robust to deviations from the assumed properties of approximate TF sparsity and underspreadness.
More specifically, the main deviation from the true RS is due to the fact that the oscillatory structures (\emph{inner interference terms} \cite{hla-interfer}) 
contained in the RS are suppressed by the smoothing; this result is in fact desirable in most applications.
It is furthermore seen that the average results of the symmetrized estimator $\hat{R}_{X,\text{CS}}^{(\text{s})}[n,k]$ are similar to those of $\hat{R}_{X,\text{CS}}[n,k]$.

\section{Conclusion}

For estimating a time-dependent spectrum of a nonstationary random process, long-term averaging cannot be used as this would smear out the
time-dependence of the spectrum. However, if the spectrum as a function of time and frequency is sufficiently smooth, which amounts to an \emph{underspread}
assumption, a local TF smoothing can be used. In particular, the RS of an underspread nonstationary process can be estimated by a local
smoothing of a TF distribution known as the RD.

In this paper, we have considered the practically relevant case of underspread processes that are approximately TF sparse in the sense that
only a moderate percentage of the RS values are significantly nonzero. For such processes, we have proposed a ``compressive'' RS estimator
that exploits the TF sparsity structure for a significant reduction of the number of measurements required for good estimation performance.
The measurements are values of the AF of the observed signal at
randomly chosen time lag/frequency lag positions. Our overall approach is advantageous if dedicated hardware units for computing values of the
AF from the original continuous-time signal are employed, and/or if the AF values have to be transmitted over low-rate links or stored in a memory.
The proposed compressive RS estimator extends a conventional RS estimator for underspread processes (a smoothed RD
using an MVU design of the smoothing function) by a CS reconstruction technique. 
For the latter, we used the Basis Pursuit because it is supported by a convenient
performance guarantee (a bound on the $\ell_{2}$-norm of the reconstruction error); however, other CS reconstruction
techniques can be used as well.

We provided upper bounds on the MSE of both the MVU RS estimator and its compressive extension.
The MSE bound for the compressive estimator is based on the error bound of the Basis Pursuit, which is known to be quite loose.
Therefore, the MSE bound for the compressive estimator is usually quite pessimistic.
However, it is still useful theoretically, since it reveals the asymptotic dependence of the estimation accuracy on the underspreadness and TF sparsity properties
of the process. Numerical experiments demonstrated the good performance of our compressive estimator for two typical scenarios.

We considered the RS because in the discrete setting used, it is the simplest time-dependent spectrum from a computational viewpoint.
However, for underspread processes, the RS is very close to other important time-dependent spectra such as the Wigner-Ville spectrum
and the evolutionary spectrum.
Therefore, the proposed RS estimator can also be used for estimating other time-dependent spectra if the process is sufficiently underspread.
Finally, the proposed RS estimator can also be used for estimating the EAF and the autocorrelation function, which are related to the RS via DFTs.

\section*{Appendix A: MSE of the Symmetrized Compressive RS Estimator}\label{sec_mse-symm}

\vspace{1mm}

We will prove the MSE inequality \eqref{equ_lower_mse_modified_compr_RS_est}. Let us define the symmetrization operator corresponding to \eqref{equ_modified_compressive_est_EAF}, i.e.,
\[
\mathbf{P}_{\!\text{s}}A[m,l] \eq \frac{1}{2} \ist \big[ A[m,l] + A^*[-m,-l] \,e^{-j \frac{2 \pi}{N} ml} \ist \big] 
\]
and note that (see \eqref{equ_symmetry_discrete_EAF}, \eqref{equ_modified_compressive_est_EAF})
\be
\mathbf{P}_{\!\text{s}}\bar{A}_{X}[m,l] \ist=\ist \bar{A}_{X}[m,l] \,, \quad
\mathbf{P}_{\!\text{s}}\hat{A}_{X,\text{CS}}[m,l] \ist=\ist \hat{A}_{X,\text{CS}}^{(\text{s})}[m,l] \,.
\label{equ_projections}
\ee

Furthermore, let us consider the estimation error of the compressive EAF estimator $\hat{A}_{X,\text{CS}}[m,l]$,
$E[m,l] \triangleq \hat{A}_{X,\text{CS}}[m,l] \rmv- \bar{A}_{X}[m,l]$. Using the triangle inequality \cite{RudinBook},
${\| \mathbf{P}_{\!\text{s}} E \|}_2 = \frac{1}{2} \ist {\| E[m,l] + E^*[-m,-l] \ist e^{-j \frac{2 \pi}{N} ml} \|}_2
\leq \frac{1}{2} \ist \big[ {\| E \|}_2 + {\| E \|}_2 \ist \big] = {\| E \|}_2$. 
Using $\mathbf{P}_{\!\text{s}} E = \mathbf{P}_{\!\text{s}} \hat{A}_{X,\text{CS}} \rmv-\rmv \mathbf{P}_{\!\text{s}} \bar{A}_{X}$
and \eqref{equ_projections}, it is seen that the above inequality is equivalent to
\be
\big\| \hat{A}_{X,\text{CS}}^{(\text{s})} \rmv-\rmv \bar{A}_{X} \big\|^{2}_{2}
  \,\le\, \big\| \hat{A}_{X,\text{CS}} \rmv-\rmv \bar{A}_{X} \big\|^{2}_{2} \,.
\label{equ_ineq-symm}
\ee
Furthermore, since $\bar{A}_{X}[m,l]$, $\hat{A}_{X,\text{CS}}[m,l]$, and $\hat{A}_{X,\text{CS}}^{(\text{s})}[m,l]$ are related to
$\bar{R}_X[n,k]$, $\hat{R}_{X,\text{CS}}[n,k]$, and $\hat{R}_{X,\text{CS}}^{(\text{s})}[n,k]$, respectively via the 2D DFT in
\eqref{equ_accompanying_RS_est_AF_est_compressive}, which is norm-preserving, 
\pagebreak 
the inequality \eqref{equ_ineq-symm}
implies that
\[
\big\| \hat{R}_{X,\text{CS}}^{(\text{s})} \rmv-\rmv \bar{R}_{X} \big\|^{2}_{2} \,\le\, \big\| \hat{R}_{X,\text{CS}} \rmv-\rmv \bar{R}_{X} \big\|^{2}_{2} \,.
\]
Finally, taking the expectation on both sides yields the MSE inequality \eqref{equ_lower_mse_modified_compr_RS_est}.

\section*{Appendix B:\, TF Shift Matrices}\label{sec_shift-matrices}

We consider the family of (scaled) discrete TF shift matrices ${\{ \mathbf{J}_{m,l} \}}_{m,l \in [N]}$ of size $N \!\times\! N$
whose action on $\mathbf{x} \!\in\! \mathbb{C}^N$ is given by
\begin{equation}
\label{equ_action_discrete_TF_shift}
{(\mathbf{J}_{m,l} \ist\ist \mathbf{x})}_{n+1} \eq \frac{1}{\sqrt{N}} \, ( \mathbf{x} )_{{ (n-m)}_{N} \ist+\ist 1} \, e^{ j \frac{2 \pi}{N} l n} \ist , \quad n \in [N] \,,
\end{equation}
with ${(n)}_{N} \triangleq n \,\ist {\rm mod} \, N$. These matrices can be written $\mathbf{J}_{m,l} = \frac{1}{\sqrt{N}} \,\mathbf{M}_{l} \mathbf{T}_{m}$,
where $\mathbf{M}_{l}$ is the diagonal $N \times N$ matrix with diagonal elements
$1, \ist e^{j \frac{2 \pi}{N} l}, \ldots, e^{j \frac{2 \pi}{N} l (N-1)}$ and $\mathbf{T}_{m}$ is the circulant $N \!\times\! N$ matrix
whose entries ${( \mathbf{T}_{m} )}_{n,n'}$ are given by $1$ if ${(n\^{E}\!-\!n')}_{N} \rmv= {(m)}_{N}$ and $0$ otherwise.
It can be easily verified that the set ${\{ \mathbf{J}_{m,l} \}}_{m,l \in [N]}$ forms an orthonormal basis for the linear space of matrices $\mathbb{C}^{N \times N}\!$ 
equipped with inner product $\langle \mathbf{A}, \mathbf{B} \rangle = {\rm tr} \{ \mathbf{A} \mathbf{B}^{H} \}$, i.e.,
\begin{equation}
\label{equ_condition_ONB_TF_shifts}
\langle \mathbf{J}_{m,l}, \mathbf{J}_{m'\!,l'} \rangle \ist=\ist \delta{[m \rmv\rmv-\rmv\rmv m']}_{N} \, \delta{[l \rmv\rmv-\rmv\rmv l']}_{N}
\vspace{-2mm}
\end{equation}
and
\vspace{.5mm}
\begin{equation}
\label{equ_condition_ONB_TF_shifts_A}
\mathbf{A} \ist=\rmv \sum_{m,l \in [N]} \!\langle \mathbf{A}, \mathbf{J}_{m,l} \rangle \, \mathbf{J}_{m,l}  \,,
  \quad \text{for all} \;\, \mathbf{A} \rmv\rmv\in\rmv\rmv \mathbb{C}^{N \times N} .
\end{equation}
It can furthermore be shown that the EAF in \eqref{equ_def_discrete_EAF} and the AF in \eqref{equ_AF} can be written as
\begin{align}
\bar{A}_{X}[m,l] &\eq \sqrt{N} \, \langle \mathbf{\Gamma}_{\!X} , \mathbf{J}_{m,l} \rangle \nonumber \\[.7mm]
A_{X}[m,l] &\eq \sqrt{N} \, \langle \mathbf{x}\ist \mathbf{x}^{H} \!, \mathbf{J}_{m,l} \rangle \,,
  \label{equ_AF_J}
\end{align}
where $\mathbf{\Gamma}_{\!X} = {\rm E} \ist\{ \mathbf{x} \mathbf{x}^H \}$ with $\mathbf{x} = (X[0] \,\cdots\, X[N\!-\!1])^T$.
Thus, according to \eqref{equ_condition_ONB_TF_shifts_A}, we have the expansions
\begin{align}
\mathbf{\Gamma}_{\!X} &\eq \frac{1}{\sqrt{N}} \rmv\sum_{m,l \in [N]} \!\bar{A}_{X}[m,l] \, \mathbf{J}_{m,l}
  \label{equ_decomposition_R_x_discrete_TF_shift} \\[.5mm]
\mathbf{x} \ist \mathbf{x}^{H}\rmv\rmv &\eq \frac{1}{\sqrt{N}} \rmv\sum_{m,l \in [N]} \!A_{X}[m,l] \, \mathbf{J}_{m,l} \,. \nonumber 
\end{align}
Finally, from \eqref{equ_action_discrete_TF_shift}, one can deduce the following relations:
\begin{align}
\mathbf{J}_{m,l} \ist\ist \mathbf{J}_{m'\!,l'}\rmv &\eq \frac{1}{\sqrt{N}} \, \mathbf{J}_{m\ist+\ist m'\!,\ist l\ist+\ist l'} \,e^{-j\frac{2 \pi}{N} ml' }
  \label{equ_composition_discrete_TF_shift} \\[.5mm]
\mathbf{J}_{m,l}^{H} &\eq \mathbf{J}_{-m,-l} \, e^{-j \frac{2 \pi}{N} ml} \,,
  \label{equ_Hermitian_discrete_TF_shift} \\[-7mm]
&\nonumber
\end{align}
and, in turn,
\begin{equation}
\label{equ_composition_3_factors_discrete_TF_shift}
 \mathbf{J}_{n,k} \ist\ist \mathbf{J}_{m,l} \ist\ist \mathbf{J}^{H}_{n,k} \,=\, \frac{1}{N} \, \mathbf{J}_{m,l} \, e^{-j\frac{2 \pi}{N} (nl - km)} \,.
\end{equation}

\section*{Appendix C:\, Derivation of Expressions \eqref{equ_var_A_non} and \eqref{equ_chi_def_1_0}}

\vspace{1mm}

We will derive \eqref{equ_var_A_non} and \eqref{equ_chi_def_1_0} from \eqref{equ_variance_nonstat_equ_s_x_mvu}.
Our derivation will be based on expansions of $\mathbf{C}^{(\text{R})}_{n,k}$ and $\mathbf{C}^{(\text{I})}_{n,k}$ into the TF shift matrices $\mathbf{J}_{m,l}$.
Using \eqref{eq:C_Re-Im}, \eqref{equ_def_C_n_k}, and \eqref{equ_Hermitian_discrete_TF_shift}, we have
\begin{align}
 \mathbf{C}^{(\text{R})}_{n,k} & \rmv\stackrel{\eqref{eq:C_Re-Im},\eqref{equ_def_C_n_k}}{\eq}\rmv \frac{1}{2\sqrt{N}} 
   \Bigg[ \sum_{m=-M}^{M} \sum_{l=-L}^{L} \! e^{-j \frac{2 \pi}{N}(km - nl)} \ist\ist\mathbf{J}_{m,l}^H \nonumber \\[.5mm]
&\rule{28mm}{0mm}\,+\! \sum_{m=-M}^{M} \sum_{l=-L}^{L} \! e^{j \frac{2 \pi}{N}(km - nl)} \ist\ist\mathbf{J}_{m,l} \Bigg] \nonumber \\[1mm]
& \;\;\ist\ist\stackrel{\eqref{equ_Hermitian_discrete_TF_shift}}{\eq}  \frac{1}{2\sqrt{N}} 
  \Bigg[ \sum_{m=-M}^{M} \sum_{l=-L}^{L} \! e^{-j \frac{2 \pi}{N}(km - nl)} \ist\ist\mathbf{J}_{-m,-l} \, e^{-j\frac{2\pi}{N} ml} \nonumber \\[.5mm]
&\rule{28mm}{0mm}\,+\! \sum_{m=-M}^{M} \sum_{l=-L}^{L} \! e^{j \frac{2 \pi}{N}(km - nl)} \ist\ist\mathbf{J}_{m,l} \Bigg] \nonumber \\[1mm]
&\;\;\,\eq \frac{1}{2\sqrt{N}} \!\sum_{m=-M}^{M} \sum_{l=-L}^{L} \rmv e^{j \frac{2 \pi}{N}(km - nl)} \ist\ist \big(e^{-j\frac{2\pi}{N} ml} + 1 \big) \,\mathbf{J}_{m,l} \nonumber \\[2mm]
&\;\;\,\eq\! \sum_{m,l \in [N]} \! e^{j \frac{2 \pi}{N}(km - nl)} \, c^{(\text{R})}_{m,l} \,\mathbf{J}_{m,l} \,, \label{equ_expansion_E_n_k} \\[-6mm]
& \nonumber
\end{align}
with
\vspace{2.5mm}
\begin{equation}
\label{equ_def_e_m_l}
c^{(\text{R})}_{m,l} \,\triangleq\, \frac{1}{2\sqrt{N}} \,\ist\ist I_{\mathcal{A}}[m,l] \, \big(e^{-j\frac{2\pi}{N} ml} +\rmv 1 \big) \,.
\vspace{1mm}
\end{equation}
In a similar way, we obtain from \eqref{eq:C_Re-Im}
\be
 \mathbf{C}^{(\text{I})}_{n,k} \eq\! \sum_{m,l \in [N]} \! e^{j \frac{2 \pi}{N}(km - nl)} \, c^{(\text{I})}_{m,l} \,\mathbf{J}_{m,l} \,,
\label{equ_expansion_O_n_k}
\vspace{-2.5mm}
\ee
with
\vspace{1.5mm}
\begin{equation}
\label{equ_def_o_m_l}
c^{(\text{I})}_{m,l}  \,\triangleq\, \frac{1}{2j\sqrt{N}} \,\ist\ist I_{\mathcal{A}}[m,l] \, \big(e^{-j\frac{2\pi}{N} ml} \rmv-\rmv\rmv 1 \big) \,.
\vspace{1mm}
\end{equation}

The first term in \eqref{equ_variance_nonstat_equ_s_x_mvu} can then be written as
\begin{equation}
\sum_{n,k \in [N]} \rmv\! {\rm tr} \big\{ \mathbf{C}^{(\text{R})}_{n,k} \mathbf{\Gamma}_{\!X} \mathbf{C}^{(\text{R})}_{n,k} \mathbf{\Gamma}_{\!X} \rmv\big\}
  \eq {\rm tr} \big\{ \mathbf{D} \mathbf{\Gamma}_{\!X} \rmv\big\}
  \eq {\rm tr} \big\{ \mathbf{D} \mathbf{\Gamma}_{\!X}^H \rmv\big\} \,,
  \label{equ_variance_firstterm}
\vspace{-1.5mm}
\end{equation}
with
\begin{align}
\mathbf{D} &\,\triangleq\ist\ist \sum_{n,k \in [N]} \rmv\! \mathbf{C}^{(\text{R})}_{n,k} \mathbf{\Gamma}_{\!X}  \mathbf{C}^{(\text{R})}_{n,k} \nonumber \\[.5mm]
&\eq \sum_{n,k \in [N]} \rmv\! \mathbf{C}^{(\text{R})}_{n,k} \mathbf{\Gamma}_{\!X}  \mathbf{C}^{(\text{R})H}_{n,k} \nonumber \\[.5mm]
&\stackrel{\eqref{equ_expansion_E_n_k}}{\eq} \sum_{m,l,m'\!,l' \in [N]} \! c^{(\text{R})}_{m,l} \, c^{(\text{R})*}_{m'\!,l'} \nonumber \\[1mm]
&\rule{15mm}{0mm} \times \ist\underbrace{\Bigg[ \sum_{n,k \in [N]} \! e^{j\frac{2 \pi}{N}[k(m-m')-n(l-l')]} \Bigg]}_{N^{2} \, \delta{[m - m']}_N \ist {\delta[l - l']}_N}
  \ist\mathbf{J}_{m,l} \ist\mathbf{\Gamma}_{\!X}  \mathbf{J}^{H}_{m'\!,l'} \nonumber \\[1mm]
&\eq N^{2} \!\rmv \sum_{m,l \in [N]} \! \big|c^{(\text{R})}_{m,l}\big|^{2} \, \mathbf{J}_{m,l} \ist \mathbf{\Gamma}_{\!X} \mathbf{J}^{H}_{m,l}\nonumber \\[.5mm]
&\stackrel{\eqref{equ_decomposition_R_x_discrete_TF_shift}}{\eq} N \sqrt{N} \!\sum_{m,l,m'\!,l' \in [N]}
  \!\big|c^{(\text{R})}_{m,l}\big|^{2} \ist\ist \bar{A}_{X}[m'\!,l'] \, \mathbf{J}_{m,l} \ist \mathbf{J}_{m'\!,l'} \ist \mathbf{J}^{H}_{m,l} \nonumber \\ 
&\stackrel{\eqref{equ_composition_3_factors_discrete_TF_shift}}{\eq} \sqrt{N} \! \sum_{m,l,m'\!,l' \in [N]}
  \! \big|c^{(\text{R})}_{m,l}\big|^{2} \ist\ist \bar{A}_{X}[m'\!,l'] \, \mathbf{J}_{m'\!,l'} \nonumber \\[-2mm]
&\rule{47mm}{0mm} \times e^{-j\frac{2 \pi}{N} (ml' \rmv- lm')}
  \label{equ_variance_D} \\[-7mm]
&\nonumber
\end{align}
and
\vspace*{1.5mm}
\be
\mathbf{\Gamma}_{\!X}^H \stackrel{\eqref{equ_decomposition_R_x_discrete_TF_shift}}{\eq} \frac{1}{\sqrt{N}} \rmv \sum_{m,l \in [N]} \!\bar{A}^*_{X}[m,l] \, \mathbf{J}^{H}_{m,l} \,.
  \label{equ_variance_RH}
\ee
Inserting \eqref{equ_variance_D} and \eqref{equ_variance_RH} into \eqref{equ_variance_firstterm} then 
\vspace*{1mm}
yields
\begin{align}
&\sum_{n,k \in [N]} \rmv\! {\rm tr} \big\{\mathbf{C}^{(\text{R})}_{n,k} \mathbf{\Gamma}_{\!X} \mathbf{C}^{(\text{R})}_{n,k} \mathbf{\Gamma}_{\!X} \rmv\big\} \nonumber \\[0mm]
&\rule{2mm}{0mm}\eq {\rm tr} \Bigg\{ \sum_{m,l,m'\!,l'\!,m''\!,l'' \in [N]}
  \! \big|c^{(\text{R})}_{m,l}\big|^{2} \ist\ist \bar{A}_{X}[m'\!,l'] \, \mathbf{J}_{m'\!,l'} \nonumber \\[-1.5mm]
&\rule{30mm}{0mm} \times e^{-j\frac{2 \pi}{N} (ml' \rmv- lm')} \ist\ist \bar{A}^*_{X}[m''\!,l''] \, \mathbf{J}^{H}_{m''\!,l''}\Bigg\} \nonumber \\[0mm]
&\rule{2mm}{0mm}\eq \rmv\sum_{m,l,m'\!,l'\!,m''\!,l'' \in [N]}
  \! \big|c^{(\text{R})}_{m,l}\big|^{2} \ist\ist \bar{A}_{X}[m'\!,l'] \ist  \bar{A}^*_{X}[m''\!,l''] \nonumber \\[.5mm]
&\rule{22mm}{0mm} \times e^{-j\frac{2 \pi}{N} (ml' \rmv- lm')}
  \hspace*{-12mm} \underbrace{{\rm tr} \big\{\mathbf{J}_{m'\!,l'} \ist \mathbf{J}^{H}_{m''\!,l''} \rmv\big\}}_{\langle \mathbf{J}_{m'\!,l'},\ist \mathbf{J}_{m''\!,l''} \rangle
  \,\stackrel{\eqref{equ_condition_ONB_TF_shifts}}{=}\,\ist \delta{[m' \rmv- m'']}_{N} \ist \delta{[l' \rmv- {l}'']}_{N}} \nonumber \\[2.5mm]
&\rule{2mm}{0mm}\eq \rmv\sum_{m,l,m'\!,l' \in [N]}
  \! \big|c^{(\text{R})}_{m,l}\big|^{2} \ist\ist \big|\bar{A}_{X}[m'\!,l']\big|^{2} \, e^{-j\frac{2 \pi}{N} (ml' \rmv- lm')}\nonumber \\[1mm]
&\rule{2mm}{0mm}\eq \rmv\sum_{m,l,m'\!,l' \in [N]}
 \! \big|c^{(\text{R})}_{m,l}\big|^{2} \ist\ist \big|\bar{A}_{X}[m'\!,l']\big|^{2} \, e^{j\frac{2 \pi}{N} (ml' \rmv- lm')} \,,
  \label{equ_sum_variance_Even_term_TF_shifts}
\end{align}
where the symmetry relation \eqref{equ_symmetry_discrete_EAF} has been used in the last step.

In a similar manner, using \eqref{equ_expansion_O_n_k}, we obtain for the second term in \eqref{equ_variance_nonstat_equ_s_x_mvu}
\begin{align}
\label{equ_sum_variance_Odd_term_TF_shifts}
&\sum_{n,k \in [N]} \rmv\! {\rm tr} \big\{ \mathbf{C}^{(\text{I})}_{n,k} \mathbf{\Gamma}_{\!X}  \mathbf{C}^{(\text{I})}_{n,k} \mathbf{\Gamma}_{\!X} \rmv\big\}\nonumber \\[-.5mm]
&\rule{4mm}{0mm}\eq \! \sum_{m,l,m'\!,l' \in [N]} \! \big|c^{(\text{I})}_{m,l}\big|^{2} \ist\ist \big|\bar{A}_{X}[m'\!,l'] \big|^{2} \, e^{j\frac{2 \pi}{N} (ml' \rmv- lm')} \,.
\end{align}
Inserting \eqref{equ_sum_variance_Even_term_TF_shifts} and \eqref{equ_sum_variance_Odd_term_TF_shifts}
into \eqref{equ_variance_nonstat_equ_s_x_mvu} then gives 
\vspace*{1mm}
\eqref{equ_var_A_non}:
\begin{align*}
V &= \sum_{m,l,m'\!,l' \in [N]}
  \! \big|c^{(\text{R})}_{m,l} \big|^{2} \ist\ist \big|\bar{A}_{X}[m'\!,l']\big|^{2} \, e^{j\frac{2 \pi}{N} (ml' \rmv-\ist lm')}\nonumber \\[.5mm]
&\rule{9mm}{0mm}\,+\rmv \sum_{m,l,m'\!,l' \in [N]}
  \! \big|c^{(\text{I})}_{m,l} \big|^{2} \ist\ist \big|\bar{A}_{X}[m'\!,l']\big|^{2} \, e^{j\frac{2 \pi}{N} (ml' \rmv-\ist lm')} \nonumber\\[1.5mm]
&= \sum_{m'\!,l' \in [N]} \! \big| \bar{A}_{X}[m'\!,l'] \big|^{2} \, \chi[m'\!,l'] \,, \\[-6mm]
&\nonumber
\end{align*}
with
\vspace{1mm}
\be
\chi[m,l] \,\triangleq \sum_{m'\!,l' \in [N]} \!\rmv \big( \big|c^{(\text{R})}_{m'\!,l'}\big|^{2} + \big|c^{(\text{I})}_{m'\!,l'}\big|^{2} \big) \, e^{j\frac{2 \pi}{N} (m'l - l'm)} \,.
\label{equ_chi_def}
\ee
Using \eqref{equ_def_e_m_l} and \eqref{equ_def_o_m_l}, we 
\vspace*{1mm}
have
\begin{align*}
&\big|c^{(\text{R})}_{m,l}\big|^{2} + \big|c^{(\text{I})}_{m,l}\big|^{2}\nonumber \\[0mm]
&\rule{5mm}{0mm} \eq \frac{1}{N} \, I_{\mathcal{A}}[m,l] \, 
   \Bigg[ \ist\ist\bigg| \frac{e^{-j \frac{2 \pi}{N} ml} \ist+\ist 1}{2} \bigg|^{2}
   \rmv\rmv+\ist\ist \bigg| \frac{e^{-j \frac{2 \pi}{N} ml}-1}{2j} \bigg|^{2} \ist\ist\Bigg] \nonumber\\[0mm]
&\rule{5mm}{0mm}\eq \frac{1}{N} \, I_{\mathcal{A}}[m,l] \, \Big[ \cos^2 \!\Big(\frac{ \pi}{N} ml \Big) \ist+\ist\ist \sin^2 \!\Big(\frac{ \pi}{N} ml \Big) \Big]  \nonumber\\[2mm]
&\rule{5mm}{0mm}\eq \frac{1}{N} \, I_{\mathcal{A}}[m,l] \,.
\end{align*}
Inserting this into \eqref{equ_chi_def} yields \eqref{equ_chi_def_1_0}.

\section*{Appendix D:\, Derivation of Expression \eqref{equ_approx_P_p_q_sum_quad}}

\vspace{.5mm}

We will derive \eqref{equ_approx_P_p_q_sum_quad} from \eqref{equ_power_non_exact}.

\vspace{1mm}

\subsubsection{Expansions of $\mathbf{T}^{(\text{\em R})}_{\rmv p,q}$ and $\mathbf{T}^{(\text{\em I})}_{\rmv p,q}$} \label{sec_expr-hpq_expans}
Our derivation will be based on the underspread assumption and on expansions of $\mathbf{T}^{(\text{R})}_{\rmv p,q}$ and
$\mathbf{T}^{(\text{I})}_{\rmv p,q}$ into the TF shift matrices $\mathbf{J}_{m,l}$. Inserting \eqref{equ_def_T_p_q} into the definition
of $\mathbf{T}^{(\text{R})}_{\rmv p,q}$ in \eqref{equ_def_T_p_q_RI} yields
\begin{align}
\mathbf{T}^{(\text{R})}_{\rmv p,q} &\ist\eq\ist \frac{\sqrt{N}}{2} \ist \Bigg[
  \sum_{m=-M}^{M} \sum_{l=-L}^{L} \! e^{-j 2 \pi \big(\frac{qm} {\lendelay} \ist-\ist \frac{pl}{\lendoppler} \big)} \,\mathbf{J}_{m,l}^H\nonumber\\[0mm]
  &\rule{25mm}{0mm}\,+\rmv \sum_{m=-M}^{M} \sum_{l=-L}^{L} \! e^{j 2 \pi \big(\frac{qm} {\lendelay} \ist-\ist \frac{pl}{\lendoppler} \big)} \,\mathbf{J}_{m,l}  \Bigg] \nonumber\\[.5mm]
&\ist\stackrel{\eqref{equ_Hermitian_discrete_TF_shift}}{\eq}\ist
\frac{\sqrt{N}}{2} \ist \Bigg[ \sum_{m=-M}^{M} \sum_{l=-L}^{L} \! e^{-j 2 \pi \big(\frac{qm} {\lendelay} \ist-\ist \frac{pl}{\lendoppler} \big)} 
  \,\mathbf{J}_{-m,-l} \, e^{-j \frac{2 \pi}{N} ml}\nonumber\\[0mm]
  &\rule{25mm}{0mm}\,+\rmv \sum_{m=-M}^{M} \sum_{l=-L}^{L} \! e^{j 2 \pi \big(\frac{qm} {\lendelay} \ist-\ist \frac{pl}{\lendoppler} \big)} \,\mathbf{J}_{m,l} \Bigg] \nonumber\\[.5mm]
&\ist\ist\eq\ist \frac{\sqrt{N}}{2} \!\! \sum_{m=-M}^{M} \sum_{l=-L}^{L} \! e^{j 2 \pi \big(\frac{qm} {\lendelay} \ist-\ist \frac{pl}{\lendoppler} \big)}
  \big( e^{-j \frac{2 \pi}{N} ml} +\rmv 1 \big) \,\mathbf{J}_{m,l} \nonumber\\[.5mm]
&\ist\ist\eq\ist\ist \sum_{m,l \in [N]} \! e^{j 2 \pi \big(\frac{qm} {\lendelay} \ist-\ist \frac{pl}{\lendoppler} \big)} \, t^{(\text{R})}_{m,l} \, \mathbf{J}_{m,l} \,, \label{equ_def_T_p_q_real}
 \\[-7mm]
&\nonumber
\end{align}
with
\vspace{1mm}
\be
t^{(\text{R})}_{m,l} \eq \frac{\sqrt{N}}{2} \, I_{\mathcal{A}}[m,l] \,
   \big( e^{-j \frac{2 \pi}{N} ml} +\rmv 1 \big) \,.
\label{equ_sf_matrices_C_R}
\ee
In a similar manner, we obtain the expansion
\be
\mathbf{T}^{(\text{I})}_{\rmv p,q} \eq\! \sum_{m,l \in [N]} \! e^{j 2 \pi \big(\frac{qm} {\lendelay} \ist-\ist \frac{pl}{\lendoppler} \big)} \, t^{(\text{I})}_{m,l} \, \mathbf{J}_{m,l} \,,
  \label{equ_def_T_p_q_imag}
\vspace{-2mm}
\ee
with
\vspace{1.5mm}
\be
t^{(\text{I})}_{m,l} \eq \frac{\sqrt{N}}{2j} \, I_{\mathcal{A}}[m,l] \,
   \big( e^{-j \frac{2 \pi}{N} ml} \rmv-\rmv 1 \big) \,.
\label{equ_sf_matrices_C_I}
\ee

For an underspread process $X[n]$, the effective EAF support $\mathcal{A} \equiv {\{- M,\ldots,M \}}_N\rmv \times {\{- L,\ldots, L \}}_N$
is a small region about the origin of the $(m,l)$ plane (plus its periodically continued replicas, which are irrelevant to our argument and will hence be disregarded).
Looking at the expressions of $t^{(\text{R})}_{m,l}$ and $t^{(\text{I})}_{m,l}$ in \eqref{equ_sf_matrices_C_R} and \eqref{equ_sf_matrices_C_I},
we can then conclude from the presence of the factor $I_{\rmv \mathcal{A}}[m,l]$ that $t^{(\text{R})}_{m,l}$ and $t^{(\text{I})}_{m,l}$
can be nonzero only for $|ml| \ll N$. This means that in \eqref{equ_sf_matrices_C_R} and \eqref{equ_sf_matrices_C_I},
we can approximate $e^{-j\frac{2 \pi}{N}ml}$ by $1$, yielding
\begin{align}
t^{(\text{R})}_{m,l} &\,\approx\, \sqrt{N} \, I_{\rmv \mathcal{A}}[m,l] \label{equ_sf_matrices_C_R_approx} \\[1mm]
t^{(\text{I})}_{m,l} &\,\approx\, 0 \,. \label{equ_sf_matrices_C_I_approx}
\end{align}
Using \eqref{equ_sf_matrices_C_I_approx} in \eqref{equ_def_T_p_q_imag} yields
\be
\mathbf{T}^{(\text{I})}_{\rmv p,q} \ist\approx\ist \mathbf{0} \,,
\label{equ_T_I_zero}
\ee
and thus \eqref{equ_power_non_exact} approximately simplifies 
\pagebreak 
to
\be
\HH_ {p,q} \,\approx\, {\rm tr} \big\{\mathbf{T}^{(\text{R})}_{\rmv p,q} \mathbf{\Gamma}_{\!X} \mathbf{T}^{(\text{R})}_{\rmv p,q}  \mathbf{\Gamma}_{\!X} \big\}
         +\ist \big| {\rm tr} \{ \mathbf{\Gamma}_{\!X} \mathbf{T}^{H}_{\rmv p,q} \} \big|^{2} \ist.
\label{equ_power_non_exact_1}
\ee

\vspace{1mm}

\subsubsection{First term in \eqref{equ_power_non_exact_1}} \label{sec_expr-hpq_first}
We will now develop the two terms on the right-hand side of \eqref{equ_power_non_exact_1}. The first term can be written
\vspace{-1mm}
as
\be
{\rm tr} \big\{ \mathbf{T}^{(\text{R})}_{\rmv p,q} \mathbf{\Gamma}_{\!X} \mathbf{T}^{(\text{R})}_{\rmv p,q} \mathbf{\Gamma}_{\!X} \big\}
  \eq \big\langle \mathbf{T}^{(\text{R})}_{\rmv p,q} \mathbf{\Gamma}_{\!X} \ist , \big( \mathbf{T}^{(\text{R})}_{\rmv p,q} \mathbf{\Gamma}_{\!X} \big)^{\rmv H} \big\rangle \,.
\label{equ_power_non_exact_firstterm}
\ee
In order to find an approximation for this inner product, we use the following general result for
the product $\mathbf{C} = \mathbf{A} \mathbf{B}$ of two $N \!\times\! N$ matrices $\mathbf{A}$ and $\mathbf{B}$. 
The matrices $\mathbf{A}$, $\mathbf{B}$, and $\mathbf{C}$ can be expanded into
the orthonormal basis ${\{ \mathbf{J}_{m,l} \}}_{m,l \in [N]}$, with respective expansion coefficients $a_{m,l}$, $b_{m,l}$, and $c_{m,l}$,
e.g., $\mathbf{A} = \sum_{m,l \in [N]} a_{m,l} \, \mathbf{J}_{m,l}$.
Then the $c_{m,l}$ are related to the $a_{m,l}$ and $b_{m,l}$ by the ``twisted convolution'' \cite{koz-diss,GM-phd,folland89,groebook}
\vspace{-1mm}
\begin{equation}
\label{equ_conv_spreading}
c_{m,l} \eq\rmv \frac{1}{\sqrt{N}} \!\sum_{m'\!,l' \in [N]} \!\rmv a_{m'\!,l'} \ist b_{m-m'\!,l-l'} \,e^{-j \frac{2 \pi}{N} m'(l - l') } \,.
\end{equation}
This expression can be verified using \eqref{equ_composition_discrete_TF_shift}.
Let us apply it to the matrix product $\mathbf{T}^{(\text{R})}_{\rmv p,q} \mathbf{\Gamma}_{\!X}$. We have the expansion
\be
 \mathbf{T}^{(\text{R})}_{\rmv p,q} \mathbf{\Gamma}_{\!X} \rmv\rmv\eq\rmv\! \sum_{m,l \in [N]} \! d_{p,q;m,l} \, \mathbf{J}_{m,l} \,.
\vspace{-1mm}
\label{equ_approx_prod_TR}
\ee
The expansion coefficients of $\mathbf{T}^{(\text{R})}_{\rmv p,q}$ are $e^{j 2 \pi \big(\frac{qm} {\lendelay} \ist-\ist \frac{pl}{\lendoppler} \big)} \ist t^{(\text{R})}_{m,l}$
(see \eqref{equ_def_T_p_q_real}); those of $\mathbf{\Gamma}_{\!X}$ are $\frac{1}{\sqrt{N}} \, \bar{A}_{X}[m,l]$ (see \eqref{equ_decomposition_R_x_discrete_TF_shift}).
Inserting these expressions into \eqref{equ_conv_spreading} yields
\begin{align}
d_{p,q;m,l} &\eq \frac{1}{\sqrt{N}} \! \sum_{m'\!,l' \in [N]} \!\Big[ e^{j 2 \pi \big(\frac{qm'} {\lendelay} \ist-\ist \frac{pl'}{\lendoppler} \big)} \ist t^{(\text{R})}_{m'\!,l'} \Big] \nonumber \\[.5mm]
&\rule{12mm}{0mm} \times \bigg[ \frac{1}{\sqrt{N}} \, \bar{A}_{X}[m \rmv\rmv-\rmv\rmv m'\!,l \rmv\rmv-\rmv\rmv l'] \bigg] \ist e^{-j \frac{2 \pi}{N} m'(l - l') } \nonumber\\[1mm]
&\,\approx \frac{1}{\sqrt{N}} \! \sum_{m'\!,l' \in [N]} \! e^{j 2 \pi \big(\frac{qm'} {\lendelay} \ist-\ist \frac{pl'}{\lendoppler} \big)} \ist I_{\rmv \mathcal{A}}[m'\!,l'] \nonumber \\[.5mm]
&\rule{12mm}{0mm} \times \bar{A}_{X}[m \rmv\rmv-\rmv\rmv m'\!,l \rmv\rmv-\rmv\rmv l'] \, e^{-j \frac{2 \pi}{N} m'(l - l') } \,,
\label{equ_approx_prod_TR_d}
\end{align}
where the approximate expression \eqref{equ_sf_matrices_C_R_approx} was used in the last step. For an underspread process, because of the support of
$I_{\rmv \mathcal{A}}[m,l]$ and the effective support of $\bar{A}_{X}[m,l]$, the terms in the sum \eqref{equ_approx_prod_TR_d} are
significantly nonzero only for $|m'(l-l')| \ll N$. We can thus use the approximation $e^{-j \frac{2 \pi}{N} m'(l - l') } \approx 1$, which yields
\begin{align}
\hspace*{-1mm}d_{p,q;m,l} &\,\approx \frac{1}{\sqrt{N}} \! \sum_{m'\!,l' \in [N]} \! I_{\rmv \mathcal{A}}[m'\!,l'] \, 
  \bar{A}_{X}[m \rmv\rmv-\rmv\rmv m'\!,l \rmv\rmv-\rmv\rmv l'] \nonumber \\[-2mm]
&\rule{39mm}{0mm} \times e^{j 2 \pi \big(\frac{qm'} {\lendelay} \ist-\ist \frac{pl'}{\lendoppler} \big)} .
\label{equ_approx_prod_TR_d_2}\\[-8mm]
&\nonumber
\end{align}
Next, we consider
\begin{align}
\big( \mathbf{T}^{(\text{R})}_{\rmv p,q} \mathbf{\Gamma}_{\!X} \big)^{\rmv H}
& \ist\ist\stackrel{\eqref{equ_approx_prod_TR}}{\eq}\! \sum_{m,l \in [N]} \! d^*_{p,q;m,l} \, \mathbf{J}^{H}_{m,l} \nonumber \\
& \,\ist\ist\stackrel{\eqref{equ_Hermitian_discrete_TF_shift}}{\eq} \sum_{m,l \in [N]} \! d^{*}_{p,q;m,l} \, \mathbf{J}_{-m,-l} \, e^{-j \frac{2 \pi}{N} ml}  \nonumber \\
& \,\ist\ist\eq \ist\sum_{m,l \in [N]}  d^{*}_{p,q;-m,-l} \, \mathbf{J}_{m,l} \, e^{-j \frac{2 \pi}{N} ml} \,,
\label{equ_approx_prod_TR_H}
\end{align}
where the $N$-periodicity of $d_{p,q;m,l}$ with respect to $m$ and $l$ was used in the last step.
For an underspread process, 
\pagebreak 
again because of the support of $I_{\rmv \mathcal{A}}[m,l]$ and the effective support of $\bar{A}_{X}[m,l]$,
it follows from \eqref{equ_approx_prod_TR_d_2} that the coefficients $d_{p,q;m,l}$ are significantly nonzero only for $|ml| \rmv\ll\rmv N$.
Hence, we can set $e^{-j \frac{2 \pi}{N} ml} \approx 1$ in \eqref{equ_approx_prod_TR_H}, which gives
\be
\big( \mathbf{T}^{(\text{R})}_{\rmv p,q} \mathbf{\Gamma}_{\!X} \big)^{\rmv H} \approx \!\sum_{m,l \in [N]} \! d^{*}_{p,q;-m,-l} \, \mathbf{J}_{m,l} \,.
\label{equ_approx_prod_TR_H_2}
\ee
In a similar way, we obtain from \eqref{equ_symmetry_discrete_EAF} the following approximation:
\begin{equation}
\label{equ_symmetry_discrete_EAF_underspread_approx}
\bar{A}_{X}^{*}[-m,-l] \ist\approx\ist \bar{A}_{X}[m,l] \,, \quad \text{for} \; |ml| \rmv\ll\rmv N \,.
\end{equation}

We now insert \eqref{equ_approx_prod_TR} and \eqref{equ_approx_prod_TR_H_2} into \eqref{equ_power_non_exact_firstterm}, and obtain
\begin{align}
&{\rm tr} \big\{ \mathbf{T}^{(\text{R})}_{\rmv p,q} \mathbf{\Gamma}_{\!X} \mathbf{T}^{(\text{R})}_{\rmv p,q} \mathbf{\Gamma}_{\!X} \big\}\nonumber \\[.5mm]
&\rule{1mm}{0mm} \,\approx\, \Bigg\langle \sum_{m,l \in [N]} \! d_{p,q;m,l} \, \mathbf{J}_{m,l} \, ,
  \sum_{m'\!,l' \in [N]} \! d^{*}_{p,q;-m'\!,-l'} \, \mathbf{J}_{m'\!,l'} \Bigg\rangle  \nonumber \\[.5mm]
&\rule{1mm}{0mm}\rmv\stackrel{\eqref{equ_condition_ONB_TF_shifts}}{\eq} \! \sum_{m,l \in [N]} \! d_{p,q;m,l} \, d_{p,q;-m,-l} \,.
\label{equ_power_non_exact_firstterm_1}
\end{align}
From the underspread approximations \eqref{equ_approx_prod_TR_d_2} and \eqref{equ_symmetry_discrete_EAF_underspread_approx},
it readily follows that $d_{p,q;-m,-l} \approx d_{p,q;m,l}^{*}$. Indeed,
\begin{align}
d_{p,q;-m,-l} & \stackrel{\eqref{equ_approx_prod_TR_d_2}}{\,\approx\,}
  \frac{1}{\sqrt{N}} \! \sum_{m'\!,l' \in [N]} \! I_{\rmv \mathcal{A}}[m'\!,l'] \, \bar{A}_{X}[-m \rmv\rmv-\rmv\rmv m'\!,-l \rmv\rmv-\rmv\rmv l']\nonumber \\[-2.5mm]
&\rule{46mm}{0mm}\times e^{j 2 \pi \big(\frac{qm'} {\lendelay} \ist-\ist \frac{pl'}{\lendoppler} \big)} \nonumber \\[0mm]
& \stackrel{\eqref{equ_symmetry_discrete_EAF_underspread_approx}}{\approx}
  \frac{1}{\sqrt{N}} \! \sum_{m'\!,l' \in [N]} \! I_{\rmv \mathcal{A}}[m'\!,l'] \, \bar{A}^{*}_{X}[m \rmv+\rmv m'\!,l \rmv+\rmv l']\nonumber \\[-2.5mm]
&\rule{46mm}{0mm}\times e^{j 2 \pi \big(\frac{qm'} {\lendelay} \ist-\ist \frac{pl'}{\lendoppler} \big)} \nonumber \\[-1mm]
& \,\stackrel{(*)}{\eq} \frac{1}{\sqrt{N}} \! \sum_{m'=-M}^{M} \sum_{l'=-L}^{L} \! \bar{A}^{*}_{X}[m \rmv+\rmv m'\!,l \rmv+\rmv l']\nonumber \\[-2.5mm]
&\rule{46mm}{0mm}\times e^{j 2 \pi \big(\frac{qm'} {\lendelay} \ist-\ist \frac{pl'}{\lendoppler} \big)} \nonumber \\[-1mm]
& \,\eq \frac{1}{\sqrt{N}} \! \sum_{m'=-M}^{M} \sum_{l'=-L}^{L} \! \bar{A}^{*}_{X}[m \rmv\rmv-\rmv\rmv m'\!,l \rmv\rmv-\rmv\rmv l']\nonumber \\[-2.5mm]
&\rule{46mm}{0mm}\times e^{-j 2 \pi \big(\frac{qm'} {\lendelay} \ist-\ist \frac{pl'}{\lendoppler} \big)} \nonumber \\[0mm]
&\,\stackrel{(*)}{\eq} \frac{1}{\sqrt{N}} \! \sum_{m'\!,l' \in [N]} \! I_{\rmv \mathcal{A}}[m'\!,l'] \, \bar{A}^{*}_{X}[m \rmv\rmv-\rmv\rmv m'\!,l \rmv\rmv-\rmv\rmv l']\nonumber \\[-2.5mm]
&\rule{46mm}{0mm}\times e^{-j 2 \pi \big(\frac{qm'} {\lendelay} \ist-\ist \frac{pl'}{\lendoppler} \big)} \nonumber \\[-1mm]
& \stackrel{\eqref{equ_approx_prod_TR_d_2}}{\approx} d_{p,q;m,l}^*\,,
\label{equ_approx_equal_d_m_l} 
\end{align}
where the periodicity of the summand with respect to $m'$ and $l'$ has been exploited in the steps labeled with $(*)$.
Using \eqref{equ_approx_equal_d_m_l} in \eqref{equ_power_non_exact_firstterm_1} then gives
\begin{equation}
{\rm tr} \big\{ \mathbf{T}^{(\text{R})}_{\rmv p,q} \mathbf{\Gamma}_{\!X} \mathbf{T}^{(\text{R})}_{\rmv p,q} \mathbf{\Gamma}_{\!X} \big\}
\,\approx \rmv\sum_{m,l \in [N]} \! |d_{p,q;m,l}|^2 \,.
\label{equ_power_non_exact_firstterm_2}
\end{equation}

\vspace{.5mm}

\subsubsection{Second term in \eqref{equ_power_non_exact_1}} \label{sec_expr-hpq_second}
Next, we consider the second term on the right-hand side of \eqref{equ_power_non_exact_1}. We have
\begin{align}
{\rm tr} \{ \mathbf{\Gamma}_{\!X} \mathbf{T}^{H}_{\rmv p,q} \}
&\stackrel{\eqref{equ_def_T_p_q_RI}}{\,\eq\,}\ist {\rm tr} \big\{ \mathbf{\Gamma}_{\!X} \mathbf{T}^{(\text{R})}_{\rmv p,q} \big\}
  \ist+\ist j\,{\rm tr} \big\{ \mathbf{\Gamma}_{\!X} \mathbf{T}^{(\text{I})}_{\rmv p,q} \big\}\nonumber \\[0mm]
&\stackrel{\eqref{equ_T_I_zero}}{\,\approx\,}\ist {\rm tr} \big\{ \mathbf{\Gamma}_{\!X} \mathbf{T}^{(\text{R})}_{\rmv p,q} \big\}\nonumber \\[1mm]
&\,\eq {\rm tr} \big\{ \mathbf{T}^{(\text{R})}_{\rmv p,q} \mathbf{\Gamma}_{\!X} \big\} \,.
\label{equ_approximation_Gamma_X_T_p_q}
\end{align}
Using $\mathbf{J}_{0,0} \rmv=\rmv \frac{1}{\sqrt{N}}\ist\mathbf{I}$, where $\mathbf{I}$ denotes the $N \rmv\times\rmv N$ identity matrix, we obtain further
\begin{align}
{\rm tr} \{ \mathbf{\Gamma}_{\!X} \mathbf{T}^{H}_{\rmv p,q} \}
&\stackrel{\eqref{equ_approximation_Gamma_X_T_p_q}}{\approx} {\rm tr} \big\{ \mathbf{T}^{(\text{R})}_{\rmv p,q} \mathbf{\Gamma}_{\!X} \mathbf{I} \big\} \nonumber \\[1mm]
&\,\eq \sqrt{N} \, {\rm tr} \big\{  \mathbf{T}^{(\text{R})}_{\rmv p,q} \mathbf{\Gamma}_{\!X} \mathbf{J}_{0,0}^{H} \big\} \nonumber \\[1mm]
&\,\eq \sqrt{N} \, \big\langle \mathbf{T}^{(\text{R})}_{\rmv p,q} \mathbf{\Gamma}_{\!X} \ist,\ist \mathbf{J}_{0,0}\^{E}\big\rangle \nonumber \\[.5mm]
&\stackrel{\eqref{equ_approx_prod_TR}}{\eq} \rmv\sqrt{N} \, \Bigg\langle \sum_{m,l \in [N]} \! d_{p,q;m,l} \, \mathbf{J}_{m,l} \, , \mathbf{J}_{0,0} \!\Bigg\rangle \nonumber \\[-.5mm]
&\,\stackrel{\eqref{equ_condition_ONB_TF_shifts}}{\eq} \ist\sqrt{N} \, \ist d_{p,q;0,0} \,.
\label{equ_power_non_exact_secondterm}
\end{align}


\subsubsection{Approximation of $\HH_ {p,q}$} \label{sec_expr-hpq_approx}
Inserting \eqref{equ_power_non_exact_firstterm_2} and \eqref{equ_power_non_exact_secondterm} into \eqref{equ_power_non_exact_1},
we obtain the following approximation of $\HH_ {p,q}$:
\[
\HH_ {p,q} \,\approx\rmv \sum_{m,l \in [N]} \! |d_{p,q;m,l}|^2 \ist+\ist N \ist | d_{p,q;0,0} |^{2} \ist.
\]
This can be expressed as
\be
\HH_ {p,q} \,\approx\rmv \sum_{n,k \in [N]} \! |\hat{d}_{p,q;n,k}|^2 \ist+\ist N \, \Bigg|\frac{1}{N}  \!\rmv \sum_{n,k \in [N]} \! \hat{d}_{p,q;n,k} \Bigg|^{2} ,
\label{equ_approx_power_non_3_FT}
\ee
where $\hat{d}_{p,q;n,k}$ is the 2D DFT of $d_{p,q;m,l}$ with respect to $(m,l)$. We have
\begin{align}
&\hat{d}_{p,q;n,k}\nonumber \\[1mm]
&\rule{2mm}{0mm} \ist\eq \frac{1}{N} \! \sum_{m,l \in [N]} \! d_{p,q;m,l} \, \twiddle[(km-nl)] \nonumber \\[0mm]
&\rule{2mm}{0mm}\rmv\rmv\rmv\stackrel{\eqref{equ_approx_prod_TR_d_2}}{\,\approx\,}\!\rmv \frac{1}{\sqrt{N}} \! \sum_{m'\!,l' \in [N]} \!\rmv I_{\rmv \mathcal{A}}[m'\!,l'] \,
  \Bigg[ \frac{1}{N} \! \sum_{m,l \in [N]} \!\rmv \bar{A}_{X}[m \rmv\rmv-\rmv\rmv m'\!,l \rmv\rmv-\rmv\rmv l'] \nonumber \\[-1mm]
  &\rule{39mm}{0mm}\times \twiddle[(km-nl)] \Bigg]  \,
  e^{j 2 \pi \big(\frac{qm'} {\lendelay} \ist-\ist \frac{pl'}{\lendoppler} \big)} \nonumber \\[-1mm]
&\rule{2mm}{0mm}\stackrel{\eqref{equ_fourier_eaf_rhs}}{\eq} \frac{1}{\sqrt{N}} \! \sum_{m'\!,l' \in [N]} \!\rmv I_{\rmv \mathcal{A}}[m'\!,l'] \, \bar{R}_X[n,k] \nonumber \\[-3mm]
  &\rule{39mm}{0mm}\times \twiddle[(km'-nl')] \, e^{j 2 \pi \big(\frac{qm'} {\lendelay} \ist-\ist \frac{pl'}{\lendoppler} \big)} \nonumber \\[0mm]
&\rule{2mm}{0mm}\eq  \frac{1}{\sqrt{N}} \, \bar{R}_X[n,k] \!\sum_{m'\!,l' \in [N]} \!\rmv I_{\rmv \mathcal{A}}[m'\!,l'] \nonumber \\[-2.5mm]
&\rule{39mm}{0mm}\times e^{-j \frac{2 \pi}{N} \rmv \big[ \big(k - \frac{N}{\lendelay}q \big)m' -\ist \big(n - \frac{N}{\lendoppler}p \big)l' \big]} \nonumber \\[.5mm]
&\rule{2mm}{0mm}\stackrel{\eqref{eq:Phi_MVU}}{\eq} \sqrt{N} \, \bar{R}_X[n,k] \,\ist \Phi_{\text{MVU}}[n \rmv-\rmv p \ist \Delta n, k \rmv-\rmv q \ist \Delta k] \,,
\label{equ_approx_prod_TR_d_RS}\\[-4.5mm]
\nonumber
\end{align}
where, as before, $\Delta n = N/\lendoppler$ and $\Delta k = N/\lendelay$.
Finally, inserting \eqref{equ_approx_prod_TR_d_RS} into \eqref{equ_approx_power_non_3_FT} yields \eqref{equ_approx_P_p_q_sum_quad}.

\section*{Acknowledgment}

The authors would like to thank the anonymous reviewers for numerous helpful comments, which have led to improvements of this paper.


\renewcommand{\baselinestretch}{1.09}\small\normalsize\small
\bibliographystyle{ieeetr}
\bibliography{/Users/ajung/Arbeit/LitAJ_ITC.bib,/Users/ajung/Arbeit/tf-zentral}


\vspace*{8mm}

\noindent {\bf Alexander Jung}
received the Diplom-Ingenieur and Dr. techn. degrees in electrical engineering from 
Vienna University of Technology, Vienna, Austria, in 2008 and 2011, respectively.
Since 2008, he has been a Research Assistant with the Institute of Telecommunications, Vienna University of Technology. 
His research interests are in statistical signal processing with emphasis on sparse estimation problems. 
He received several national awards and a Best Student Paper Award at IEEE ICASSP 2011.

\vspace*{4mm}

\noindent {\bf Georg Taub\"ock} 
(S'01--M'07) received the Dipl.-Ing.\ degree and the Dr.\ techn.\ degree (with highest honors) in electrical engineering and the Dipl.-Ing.\ degree in mathematics (with highest honors) from Vienna University of Technology, Vienna, Austria in 1999, 2005, and 2008, respectively. He also received the diploma in violoncello from the Conservatory of Vienna, Vienna, Austria, in 2000. 

From 1999 to 2005, he was with the FTW Telecommunications Research Center Vienna, Vienna, Austria, and since 2005, he has been with the Institute of Telecommunications, Vienna University of Technology, Vienna, Austria. From February to August 2010, he was a visiting researcher with the 
Communication Technology Laboratory/Communication Theory Group at ETH Zurich, Zurich, Switzerland.

His research interests include wireline and wireless communications, compressed sensing, signal processing, and information theory.

\vspace*{4mm}

\noindent {\bf Franz Hlawatsch} 
(S'85--M'88--SM'00--F'12) received the Diplom-Ingenieur, Dr. techn., and Univ.-Dozent (habilitation) degrees in electrical engineering/signal processing from Vienna University of Technology, Vienna, Austria in 1983, 1988, and 1996, respectively.

Since 1983, he has been with the Institute of Telecommunications, Vienna University of Technology, where he is currently an Associate Professor. During 1991--1992, as a recipient of an Erwin Schr\"odinger Fellowship, he spent a sabbatical year with the Department of Electrical Engineering, University of Rhode Island, Kingston, RI, USA. In 1999, 2000, and 2001, he held one-month Visiting Professor positions with INP/ENSEEIHT, Toulouse, France and IRCCyN, Nantes, France. He (co)authored a book, two review papers that appeared in the {\sc IEEE Signal Processing Magazine}, about 200 refereed scientific papers and book chapters, and three patents. He coedited three books. His research interests include signal processing for wireless communications and sensor networks, statistical signal processing, and compressive signal processing.

Prof. Hlawatsch was Technical Program Co-Chair of EUSIPCO 2004 and served on the technical committees of numerous IEEE conferences. He was an Associate Editor for the {\sc IEEE Transactions on Signal Processing} from 2003 to 2007 and for the {\sc IEEE Transactions on Information Theory} from 2008 to 2011. From 2004 to 2009, he was a member of the IEEE SPCOM Technical Committee. He is coauthor of papers that won an IEEE Signal Processing Society Young Author Best Paper Award and a Best Student Paper Award at IEEE ICASSP 2011.

\onecolumn

\clearpage\newpage

\end{document}